\documentclass[10pt]{article}

\usepackage{amsmath}
\usepackage{amssymb}
\usepackage{graphicx}
\usepackage{color} 
\usepackage[all]{xy}
\usepackage{amscd,amsfonts,amsthm}
\usepackage{graphicx}
\usepackage{comment}
\usepackage{amstext}
\usepackage[normalsize]{subfigure}
\usepackage{epsf}
\usepackage{epsfig}
\usepackage{xspace}
\usepackage[active]{srcltx}
\usepackage[all]{xy}
\usepackage[table]{xcolor}
\usepackage{hyperref}
\usepackage{tabularx,ragged2e,booktabs,caption}
\usepackage{bibunits}

\usepackage[labelfont=bf,labelsep=period,justification=raggedright]{caption}

\makeatletter
\renewcommand{\@biblabel}[1]{\quad#1.}
\makeatother

\date{}

\newcolumntype{C}[1]{>{\Centering}m{#1}}

\newcommand{\N}{\mathbb N}

\newcommand{\R}{\mathbb R}

\newtheorem{theorem}{Theorem}[section]

\newtheorem{proposition}[theorem]{Proposition}

\newtheorem{definition}[theorem]{Definition}
\newtheorem{example}[theorem]{Example}

\numberwithin{equation}{section}

\begin{document}
\begin{bibunit}[plain]

\begin{flushleft}
{\Large
\textbf{Topological strata of weighted complex networks}
}
\\
Giovanni Petri$^{1,\ast}$, 
Martina Scolamiero$^{1,2}$, 
Irene Donato$^{1,3}$ 
Francesco Vaccarino$^{1,3}$ 
\\

\bf{1} ISI Foundation, Via Alassio 11/c, 10126 Torino - Italy \\
\bf{2} Dipartimento di Ingegneria Gestionale e della Produzione, Politecnico di Torino, C.so Duca degli Abruzzi n.24, Torino, 10129, Italy \\
\bf{3} Dipartimento di Scienze Matematiche, Politecnico di Torino, C.so Duca degli Abruzzi n.24, Torino, 10129, Italy 
\\
$\ast$ E-mail:  giovanni.petri@isi.it
\end{flushleft}

\section*{Abstract}
The statistical mechanical approach to complex networks is the dominant paradigm in describing natural and societal complex systems 
The study of network properties, and their implications on dynamical processes, mostly focus on locally defined quantities of nodes and edges, 
such as node degrees, edge weights and --more recently--  correlations between neighboring nodes.
However, statistical methods quickly become cumbersome when dealing with many-body properties  and do not capture the precise mesoscopic structure of complex networks. \\
Here we introduce a novel method, based on persistent homology, to detect particular non-local structures, akin to {\it weighted holes}  within the link-weight network fabric, which are invisible to existing methods.
Their properties divide weighted networks in two broad classes: one is characterized by small hierarchically nested holes, while the second displays  larger and longer living inhomogeneities. 
These classes cannot be reduced to known local or quasilocal network properties, because of the intrinsic non-locality of homological properties, and thus yield a new classification built on high order coordination patterns. \\
Our results show that topology can provide novel insights relevant for many-body interactions in social and spatial networks. Moreover, this new method creates the first bridge between network theory and algebraic topology, which will allow to import the toolset of algebraic methods to complex systems.   


\section{Introduction}
Complex networks have become one of the prominent tools in the study of social, technological and biological systems \cite{newmanfunction, boccaletti,doro}.
In particular, weighted networks have been largely used to convey not only the presence but also the intensity of relations between nodes in a network. 
Real-world networks display however intricate patterns of redundant links with edge weights and node degrees usually ranging over various orders of magnitudes \cite{Barrat:2004bk,puppappera}.
This makes very hard to extract the significant network structure from the background \cite{motifs,barabazzo,dk,Conradi:2007jo}, especially in the case of very dense networks \cite{dante,tomaso1}. 
Alongside topological filtering methods \cite{tomaso2,kcore}, the typical approach to this problem is to choose a suitable threshold for the edge weights, e.g. global \cite{dante} or local \cite{backbone}, and study the reduced graph composed by only the edges of weight larger (smaller) than the threshold parameter.
In any case, some properties of the original graph are inevitably lost under such transformation.  \\
To avoid this pitfall, we propose to consider the set of all filtered networks, ordered by the descending thresholding weight parameter, in the spirit of {\it persistent homology} \cite{Ghrist:2008tw,(Pers),(Top)}. \\
This set, which we call {\it graph filtration}, combines link weights and connectivity structure over all weight scales.
The graph filtration proceeds on the network $\Omega$ following these steps:
\begin{itemize}
\item Rank the weights of links from $\omega_{max}$ to $\omega_{min}$: the discrete parameter $\epsilon_t$ scans the sequence.
\item  At each step $t$ of the decreasing edge ranking we consider the thresholded graph $G(\omega_{ij},\epsilon_t)$, i.e. the subgraph of $\Omega$ with links of  weight larger than $\epsilon_t$.
\end{itemize}
Figure 1a provides a schematic illustration of the rank filtration. 
This approach preserves the complete topological and weight information, allowing us to focus on special mesoscopic structures: {\it weighted network holes}, that relate the network's weight-degree structure to its homological backbone. \\
A weighted network hole of weight $\omega$ is a loop composed by $n$ nodes $i_0,i_1, i_2, ...., i_{n-1}$, where all cyclic edges $(i_l, i_{l+1})$ (with $i_0\equiv i_{n}$) have weights $\geq \omega$ , while all the other possible edges crossing the loop are strictly weaker than $\omega$.  We focus on this special class of subgraphs, because formally such weighted holes are generators of the first homology group, $H_1$, of the clique complex of the graph thresholded by weight $\omega$ (see Materials and Methods). 
The aim of this paper is to characterize the evolution of these generators along the network filtration. 
As we swipe the network from the largest to the smallest weights, network holes appear and potentially close.  \\
By unearthing  their properties, we obtain the main contribution of this paper:  the statistical features of  weighted network holes yield a classification of real-world networks in two classes,  depending on the compatibility or lack thereof with null models generated by graph randomisations. Furthermore, this classification is defined by mesoscopic homological structures that cannot be reconduced to local properties alone.\\
The method used for the classification itself, which we call {\it weighted clique rank homology}, is the second novel main contribution of this paper.
It allows to recover complete and accurate long-range information from noisy redundant network data, by building on persistent homology \cite{(Pers)}, a recent theory developed in computational topology \cite{(Top)}, which we extend to the case of networks. 

Each weighted hole $g$ is characterized by three quantities: its birth index $\beta_g$, its persistence $p_g$ and its length $\lambda_g$. 
After ranking links in a descending order according to their weights, the birth index of a hole is the rank $t$ of its weight $\omega$. 
As we proceed adding links to the filtration in ranking order, it is possible that a link with rank $t'>t$ will appear and cross the hole. 
We call this closure of the weighted hole, or {\it death} $\delta_g$. 
The persistence $p_g$ is the interval between the birth and death of $g$, $p_g =\delta_g - \beta_g = t' -t $. 
Finally, the length  $\lambda_g$ is the number of links composing $g$. \\
Similarly to stratigraphy, each step of the filtration is a topological stratum of the network, where the edge weight rank plays the role of depth. 
Intuitively, $g$ can then be thought as an underground cavity, hidden in the link-weight fabric of the network,  and $\beta_g$, $p_g$ and $\lambda_g$ as its maximal depth, vertical size and girth respectively.  \\

\section{Results}

\subsection{Homological network classes}
We applied this analysis to various social, infrastructural and biological networks (see SI for a detailed list).
In order to compare datasets, indices are normalized by the corresponding filtration length (maximal rank) $T$, so that all $\beta_{g}$, $\delta_{g}$, and thus $p_{g}$, vary in the unit interval. 
In addition, we compared each dataset with two randomized versions, obtained by weight reshuffling and edge-swapping respectively. 
While both randomisations preserve the weight and degree sequences, the first one redistributes the edge weights and is meant to destroy weight correlations, while preserving the linking patterns and thus the degree assortativity. 
The second instead randomizes the network through double-edge swaps, destroying both weight and degree correlations \cite{richclub}.  
We stress that, as the degree and weight sequences are preserved in the randomisations, they cannot account for the differences in the observed homology. \\
%
The statistical distributions obtained for the $\{ \beta_{g} \}$, $\{ p_{g} \}$ and $\{ \lambda_{g} \}$ for $H_1$ cycles highlight a natural division of the analysed networks in two broad classes (Fig.~2):
\begin{description}
\item[Class I networks:]  cycle distributions are markedly different from the randomized versions (cycles display shorter persistence times, earlier and broader birth distributions and very short lengths as compared to their randomized versions);  
\item[Class II networks:] cycle distributions are very close to their random versions ( late appearance, short persistences, long cycles). 
\end{description}
The short cycles of Class I networks nest hierarchically and appear and die over all scales while those in the randomized counterparts are born uniformly along the filtration but are more persistent, producing largely hollow network instances. 
The implications are twofold.
Since cycles represent weaker connectivity regions, this results in class I networks being more {\it solid} than the randomized versions, while class II networks resemble more closely the randomized instances. 
Second, since the cycle abundance ratio between real and random instances is the same in the two groups, 
the differences between class I and II does not depend on cycle abundance, but rather on their properties. \\
This can be seen easily by compressing the whole information within two scalar metrics which do not depend on the number of generators in a given network filtration.
We define the {\it network hollowness} $h_i$ and the {\em chain-length normalized hollowness} $\tilde{h}_i$  as:
\begin{eqnarray}
h_k = \frac{1}{N_{g_k}} \sum_{g_k } \frac{ p_{g_k}}{T} \\
 \tilde{h}_k = \frac{1}{N_{g_k}} \sum_{g_k } \frac{\lambda_{g_k}}{N} \frac{ p_{g_k}}{T}
\end{eqnarray} 
where $\{g_k\}$ is the set of generators of the $k$-th homological group $H_k$ and $N_{g_k}=\dim{H_k}$ their number. 
The first is a measure of the average persistence, while the second weights generators according to both their length and persistence.
Table \ref{table::h1} reports the values for $h_1$ and $\tilde{h}_1$.
Class I networks have lower hollowness values as compared to their randomized versions, while class II ones show comparable values.\\ 
Interestingly, the hollowness values for the $H_2$ generators mostly vanish for the randomized instances (Table \ref{table::h1}), as opposed to the case of real networks.
It appears that, while persistent one-dimensional cycles are more easily generated in the randomized instances, higher forms of network coordinations, e.g. $H_2$ generators (akin to two-dimensional surfaces bounding three-dimensional voids), do not only display different properties in comparison to the real network, but are instead wiped away.
These findings hint therefore to the presence of higher order coordination mechanisms in real world networks. \\

Naturally, the two network classes do not represent a binary taxonomy and should be considered as two extremes of a range over which networks distribute. 
For example, we find networks that interpolate between these classes, e.g. the online messages network has short persistence intervals, but also late cycle appearances and short length cycles.
However, classes do not appear to display uniform behavior for local and two-body quantities: degree- and weight- distributions and correlations are mixed within the same group and do not provide a direct answer for the nature of the two classes. 
Similarly, a recently proposed measure of structural organisation, {\it integrativeness} \cite{Pajevic:2012ew}, which measures the neighborhood overlap around strong links, does not provide insights to explain class I, since within the latter one finds both integrative and dispersive networks. \\
Finally, the classes do not show a consistent pattern in {\it assortativity}: for example, class I includes the gene network (assortative) and the airport networks (disassortative), while class II  includes the assortative co-authorship networks and the disassortative Twitter data. Therefore, assortativity cannot be the discriminating factor between classes. 

\subsection{Higher order organization}
Because homology is essentially a non-local property, it was expectable that the local measures just mentioned would not be able to explain the observed homological patterns. 
Network homology can be seen in fact as the weighted complement to the {\it perturbative} $dK$-series approach \cite{dk}: the latter proceeds by successive bottom-up constraints on $k$-body correlations, rapidly becoming very cumbersome, while our method returns the complete superposition of the network's degree and weight correlation layers in a non-perturbative (top-down) fashion.   

A simple artificial network helps illustrating this point: Random Geometric Graphs (RGG) have been recently shown to display long-range many-body correlations \cite{barthe,antonioni}. 
We find also that they have homological structures reminding of class I networks (Fig.~2a, b and c) ) and the same relation to their randomized versions. 
Class I networks are the result of high-order coordination in a similar way.  
This is supported also by the presence in real networks and RGGs of higher homology generators, which  require elaborate coordination patterns in order to appear. 
While these cycles almost disappear in randomized versions of real-world networks, they are present in the case of RGGs. \\ 
For the latter and the airports, this organisation can be thought as the result of the non-local constraint imposed by the metric of the underlying space \cite{barrat2005}. 
Although spatial constraints are harder to fathom for social and genetic systems,  alternative explanations are possible: for example, the homological structure of the observed online communication and gene networks can be thought as stemming from group interactions among people (e.g. mailing lists, multi-user mails) and biological functions (e.g. pathways ) respectively, which provide an underlying non-local mechanism for the emergence of homological patterns.   \\

Further evidence of this behavior can be found by zooming on specific cycles which convey information about underlying constrains hidden in the network weight-link connectivity patterns. 
For example, the cycle structure of the air passenger network detects the expected reduced connectivity over oceans --in the form of strong persistent cycles-- and the strong backbone of US airport hubs, which is then filled by the local (intra-community) links (Fig.~1b).  
Another example can be found in the school children's face-to-face contact network.
As expected we find the most significant cycles to link together different school classes (yellow and pink cycles in Fig.~1c). However, we also find that a school class (green nodes), despite being both a network community and 3-clique component \cite{Palla:2005cja}, is characterized by a strong internal $H_1$ generator, which might be reflecting peculiar social dynamics coming from different seating arrangements or schedules for part of the class \cite{juliette}. \\
%
%

\subsection{Spectral correlates of homology classes}
At the opposite extreme of local quantities lie the spectral properties of networks. 
It is very important therefore to investigate whether it is possible to highlight peculiar spectral signatures of the two classes.
Network eigenvalues, especially those of the Laplacian matrix, figure prominently in a number of applications, ranging from
spectral clustering \cite{spectral_clust} to the propensity to synchronize of a set of oscillators distributed on the nodes \cite{bocca_synch}.
Given a graph $G$, we denote its adjacency matrix $A(G)$ and its Laplacian matrix as $L(G) = D-A(G)$, where $d_{ij}=\delta_{ij }\sum_k a_{ik}$. 
For a symmetric network with $N$ nodes, $A(G)$ has a set of real eigenvalues $\lambda_1 \geq \lambda_2 \geq \dots \lambda_{N-1} \geq \lambda_N$. 
The spectral gap $\Delta \lambda_A = \lambda_1 - \lambda_2$, and its normalized version,  $R_{A} = \frac{\lambda_1 - \lambda_2}{\lambda_2 - \lambda_N}$, effectively measure how far the leading eigenvalue lies in comparison to the bulk of the eigenvalue distribution \cite{farkas}. \\
Interestingly, we find that class I networks have significantly larger spectral gaps  ($p<0.05$ comparing the distributions) than class  II networks (panel IV in Fig.~2a).  
Despite being somewhat neglected in the complex networks literature, $\Delta \lambda_A$ has been linked to the notion of {natural connectivity} \cite{jun}: it encodes spectral information about network redundancy in terms of the number of closed paths and is defined as $\bar{\lambda} = \log \left [ \frac{1}{N} \sum_{i=1}^N e^{\lambda_i} \right]$. 
Rewriting $\bar{\lambda} = \lambda_1 + \log \left [ \frac{1}{N} (1 + \sum_{i=2}^N e^{\lambda_i - \lambda_1})  \right]$, it is easy to see that for large gaps all the terms in the sum are exponentially suppressed and therefore $\bar{\lambda}$ is essentially dominated by the leading adjacency eigenvalue modulo a size effect, $\bar{\lambda} \sim \lambda_1 - \log N$.  
This result is consistent with the nested cycle structure that we highlighted in class I. 
More importantly, we find a difference between the two classes in the topological constraints to synchronization processes . 
For the Laplacian $L(G)$, label the set of eigenvalues $0=\lambda_1<\lambda_2^L \leq \lambda_3^L \leq \dots \leq \lambda_N^L$ and define the Laplacian eigenratio $R_{L} = \frac{\lambda_N^L}{\lambda_2^L}$.
Barahona and Pecora \cite{pecora} showed that a set of dynamical systems, placed on the network's nodes and coupled according to the graph adjacency with a global coupling $\sigma$, has a linearly stable synchronous state if 
\begin{align}
R_L < \beta 
\end{align}
where $\beta$ is a purely dynamical parameter. 
This inequality implies that networks displaying very large $R_L$ are hard (or impossible) to synchronize. 
Panel IVb of Fig.~2 shows again a significant difference between the two classes: class I networks have much larger eigenratios, making them 
hardly synchronizable. \\
Our results show therefore a deep connection between the homological network structure, the network spectral properties and their implications on network dynamics.  
Indeed, the role of mesoscopic structures in the stability and evolution of dynamical systems on networks is gradually emerging, as shown for example by recent work based on the concepts of basic symmetric subgraphs and their legacy eigenvalues in the global network spectrum \cite{mac1}, and is indeed being shaped by algebraic methods, well suited to capture the geometric information hidden within the network fabric.

\section{Conclusions}  
Hitherto, the homological structure of weighted networks could not be systematically studied. 
Our method, grounded in computational topology, allows to probe multiple layers of organized structure.
It highlighted two classes of network distinguished by their homological features, which we interpreted as caused by differences in the higher order networks organisations that are not captured by (quasi)local approaches. \\
Among the many possible applications,  two very relevant ones for social and infrastructural networks are the study of the weighted rich club's geometry beyond the aggregate measure \cite{richclub,rich}, and the generalisation of network embedding models to include homological information \cite{Boguna:2010bc}.
Furthermore, the two classes displayed also a marked difference in their spectral gap distributions and in particular in the values of the algebraic connectivity, implying that the different homological structures are correlated with different synchronizability thresholds. \\
This work therefore provides a stepping stone towards understanding the coupling between network dynamical processes and the network's homology. \\ 
Finally, the filtration's construction rule is flexible and can be readily adapted to other problems. Similarly to changing goggles, different edge metrics can be used (e.g. betweenness or salience \cite{salience}), the thresholding method varied (e.g. local thresholding \cite{backbone}) or the filtration promoted to a filtering on two quantities (e.g. edge weight and time in a temporal network) using {\it multi-persistent} homology \cite{multipersistence}.  \\


\section*{Methods and Materials}

\subsection*{Persistent homology}
The method we use to uncover weighted holes is persistent homology of the weight clique rank filtration. 
In this section we will briefly explain persistent homology and its realization through the weight rank clique filtration.\\
Persistent homology is a technique from computational algebraic topology that can be viewed as parametrized version of simplicial homology\cite{Simp}. The two definitions needed for simplicial homology are those of {\it simplicial complex} and {\it homology}. 
A {\it simplicial complex} is a non empty family $X$ of finite subsets, called faces, of a vertex set with the two constraints:
\begin{enumerate}
\item[-] a subset of a face in $X$ is a face in $X$,
\item[-] the intersection of any two faces in $X$ is either a face of both or empty.
\end{enumerate}
We assume that the vertex set is finite and totally ordered. A face of $n+1$ vertices is called $n-$face and denoted by $[p_0,\ldots, p_n]$.
The interpretation of low dimensional  faces is intuitive: a $0-$face is a vertex, a $1-$face is a segment, a $2-$face is a full triangle, a $3-$face is a full tetrahedron.
The dimension of a simplicial complex is the highest dimension of the faces in the complex. \\
Morphism between simplicial complexes are called simplicial maps.
A simplicial map is a map between simplicial complexes with the
property that the image of a vertex is a vertex and the image of a $n-$face is face of dimension $\leq n$.\\
{\it Simplicial Homology} with coefficients in a field is a functor from the category of simplicial complexes to 
the category of vector spaces \cite{Simp}.
Homology of dimension $n$ assigns to each simplicial complex $X$, the vector space $H_n(X)$ of $n$-cycles modulo boundaries and to every simplicial map
 $X\stackrel{f}\rightarrow Y$ the linear map $H_n(f):H_n(X)\rightarrow H_n(Y)$.\\

The construction that leads to the vector space $H_n$ is the following.
Given a simplicial complex $X$ of dimension $d$, consider the vector spaces $C_n$ on the set of $n-$faces in $X$ for $0\leq n \leq d$. Elements in $C_n$ are called $n-$chains.
The linear maps sending a $n-$face to the alternate sum of its $(n-1)-$faces.

\begin{eqnarray*}
\partial_n: C_n & \longrightarrow & C_{n-1}\\
\text{[}p_{0},\ldots, p_{n}\text{]} & \rightarrow & \sum_{i=0}^n(-1)^i\text{[}p_{0},
\ldots, p_{i-1}, p_{i+1},\ldots, p_{n}\text{]}.
\end{eqnarray*}

shares the property $\partial_{n-1}\circ\partial_n=0$. \\
The subspace $ker \,\partial_n$ of $C_n$ is called the vector space of $n-$cycles and denoted by $Z_n$.
The subspace  $Im \, \partial_{n+1}$ of $C_n$, is called the vector space of $n-$boundaries and denoted by $B_n$.
Note that from $\partial_{n-1}\circ\partial_n=0$ it follows that $B_n\subseteq Z_n$ for all $n$. \\
The $n-$th simplicial homology group of $X$, with coefficients in $k$, is the vector space $H_n:=Z_n/B_n$.\\

%
{\it Persistent homology} is the homology of a {\it filtration}, i.e. an increasing sequence of simplicial complexes
$$X_0\subset X_1\subset \ldots \subset X_n=X,$$ as opposed to that of a single simplicial complex. \\
It assigns to a filtration the homology groups of the simplicial complexes $H_n(X_v)$ and the linear maps $i_{v,w}: H_n(X_v)\rightarrow H_n(X_{w})$ induced in homology by the inclusions $X_v\hookrightarrow X_w$ for all $v\leq w$. 
Note that the linear maps $i_{v,v+1}$ are not always injective, meaning that some homological features can disappear along the filtration. 
These features are encoded by the persistent homology generators: an element $g \in H_n(X_v)$ such that there is no $h \in H_n(X_w)$ for $ w<v$ with the property that $i_{w,v-w}h=g.$
Two indices completely determine a generator $g\in H_n(X)$, namely its birth, $\beta_{g}$ and its death $\delta_{g}$. The index $\beta_{g}$ traces the first index such that $g$ is in the filtration and $\delta_{g}$ is the index of the simplicial complex in which the cycle becomes a boundary (i.e. disappears homologically). The persistence (lifetime) of a generator is measured by $p_g:=\delta_g-\beta_g$.
The length of a cycle, that is the number of faces composing it, is denoted by $\lambda_g$. \\
For each homology group, the information about the filtration is collected in a barcode: the set of intervals $[\beta_{g}; \delta_{g}]$ for all generators $g\in H_n$, which constitutes a handy complete invariant of $H_n$ \cite{(Pers)}.
An alternative way to represent the persistent homology of a filtration is through persistence diagrams \cite{(Pers), Pers3}, which we use extensively in the SI. 
A persistence diagram is a set of points in the plane counted with multiplicity. It can be recovered from the barcode considering the points $(\beta_g,\delta_g)\in \R^2$ with multiplicity given by the number of generators with the same persistence interval. In the SI, the reader can find $H_1$ persistent diagrams of the real world datasets examined for the classification, together with the explicit comparison to the results for their relevant randomized versions. \\

\subsection*{Filtrations}
In classical applications, the filtration is obtained from a point cloud using the Rips-Vietoris complex and persistent homology used to uncover robust topological features of the point cloud.
We instead use the clique weight rank filtration to uncover properties deriving from the topology and weighted structure of weighted networks.\\
Recalling that an $n-$clique is a complete subgraph on $n+1$ vertices, the {\it clique complex} is a simplicial complex built from the cliques of a graph. Namely there is a $n-$face in the simplicial complex for every $(n+1)-$clique in the graph. The compatibility relations are satisfied because subsets of cliques and intersection of cliques are cliques themselves.\\

The {\it Weight Rank Clique filtration} on a weighted network $\Omega$ combines the clique complex construction with a thresholding on weights following three main steps.
\begin{itemize}
\item{} Rank the weights of links from $\omega_{max}$ to $\omega_{min}$: the discrete parameter $\epsilon_t$ indexes the sequence.
\item{} At each step $t$ of the decreasing edge ranking we consider the thresholded graph $G(\omega_{ij},\epsilon_t)$, i.e. the subgraph of $\Omega$ with links of  weight larger than $\epsilon_t$.
\item{} For each graph $G(\omega_{ij},\epsilon_t)$ we build the clique complex $K(G,\epsilon_t)$.
\end{itemize}
The clique complexes are nested along the growth of $t$ and determine the weight rank clique filtration.
Note that this construction is in fact the clique complex of each element in the graph filtration. \\
In particular, persistent one dimensional cycles in the weight rank clique filtration represent weighted loops with much weaker internal links. \\
There is a conceptual difference in interpreting $H_1$ persistent homology of data with the Rips-Vietoris filtration and $H_1$ persistent homology of weighted networks with the weight rank clique filtration.
While in the first case persistent generators are relevant and considered features of the data, short cycles are more interesting for networks.
This is because random networks, or randomisations of real networks, display one dimensional persistent generators at all scales, while short lived generators testify the presence of local organisation properties on different scales.

\section*{Acknowledgments}
The authors acknowledge M. Rasetti for stimulating discussions.

\clearpage
\begin{figure*}
\centering
\includegraphics[width=0.8\textwidth]{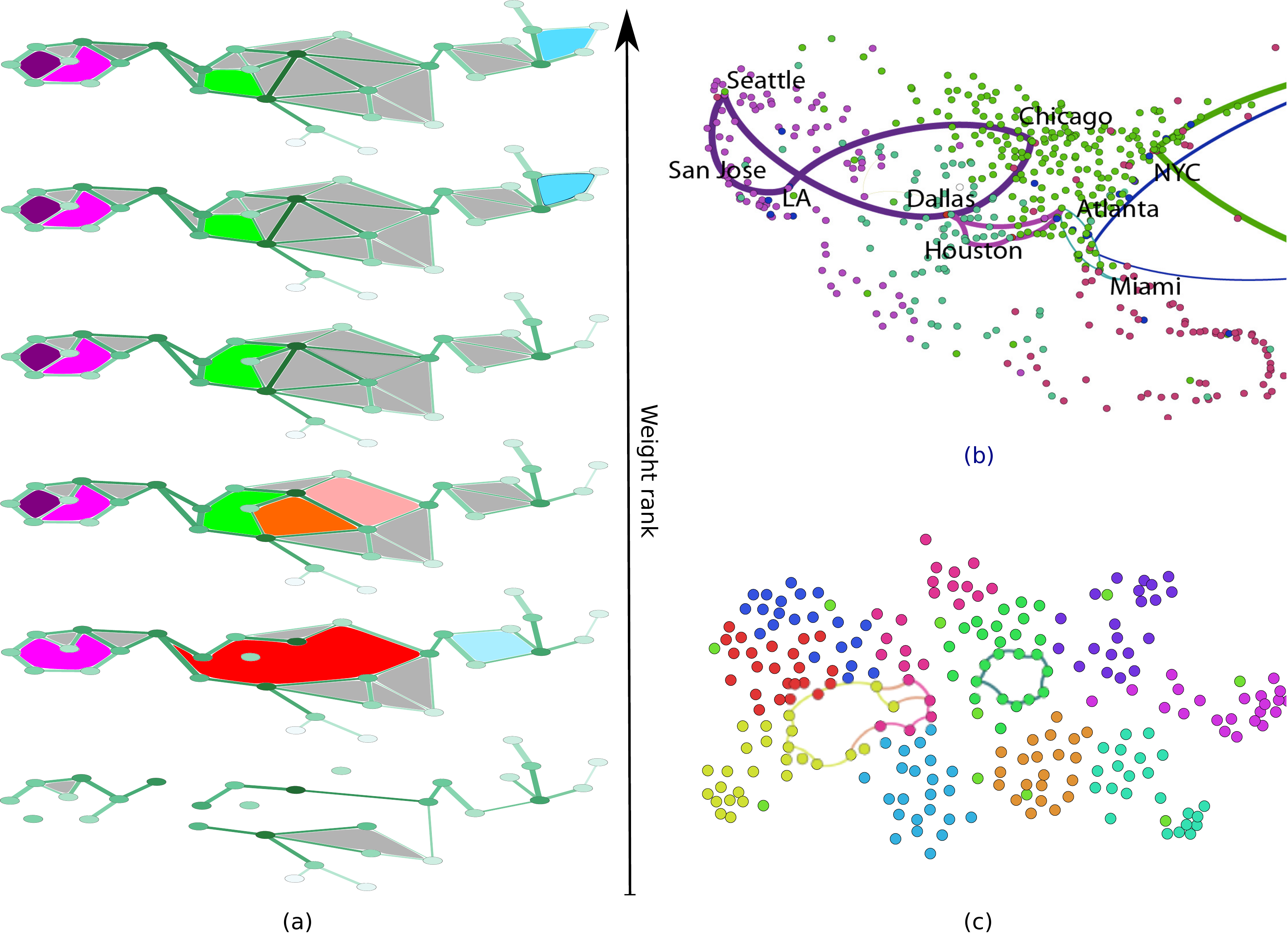}

\caption{{\bf Weight rank clique filtration and homology of networks.} 
{\it (a)} The weight rank filtration proceeds from the bottom up. Weighted holes (colored) and cliques (gray) appear as links are added. Weighted holes can branch into smaller holes, which have then independent evolution, persisting or dying along the filtration as links close them by 3-cliques. The cartoon shows two very long-persistence holes (violet and purple) appearing quite early and living until the end, while the largest hole (red) branches into three smaller holes, of only one survives to the end of the filtration (green). 
{\it (b)} A selection of weighted holes from the US air passenger network (year 2000). The node colors represent the best modularity partition of the entire network. The cycles are all long-persistence one, chosen to represent different behaviors: for example, the Chicago-Los Angeles-San Jose-Seattle cycle spans a large spatial distance, implying weaker connectivity across the cycle and within the region encompassed by the cycle, while the cycle going east from New York connects the east coast to three large European network and its persistence is due to the reduced connectivity due to the Atlantic Ocean. %
(c) A selection of the strongest cycles in the face-to-face contact network in a primary school (see SI for details on dataset). Node colors represent different classes in the school. Cycles are often found across communities, since by definition they probe the presence of holes among network regions. However, this is not the only information they convey. The cycle contained in a single community (green) testify the presence of peculiar contact geometries even within dense community structures. }\label{fig::filtration}
\end{figure*}

\begin{figure*}
\centering
\subfigure[]{
\includegraphics[width=.46\textwidth]{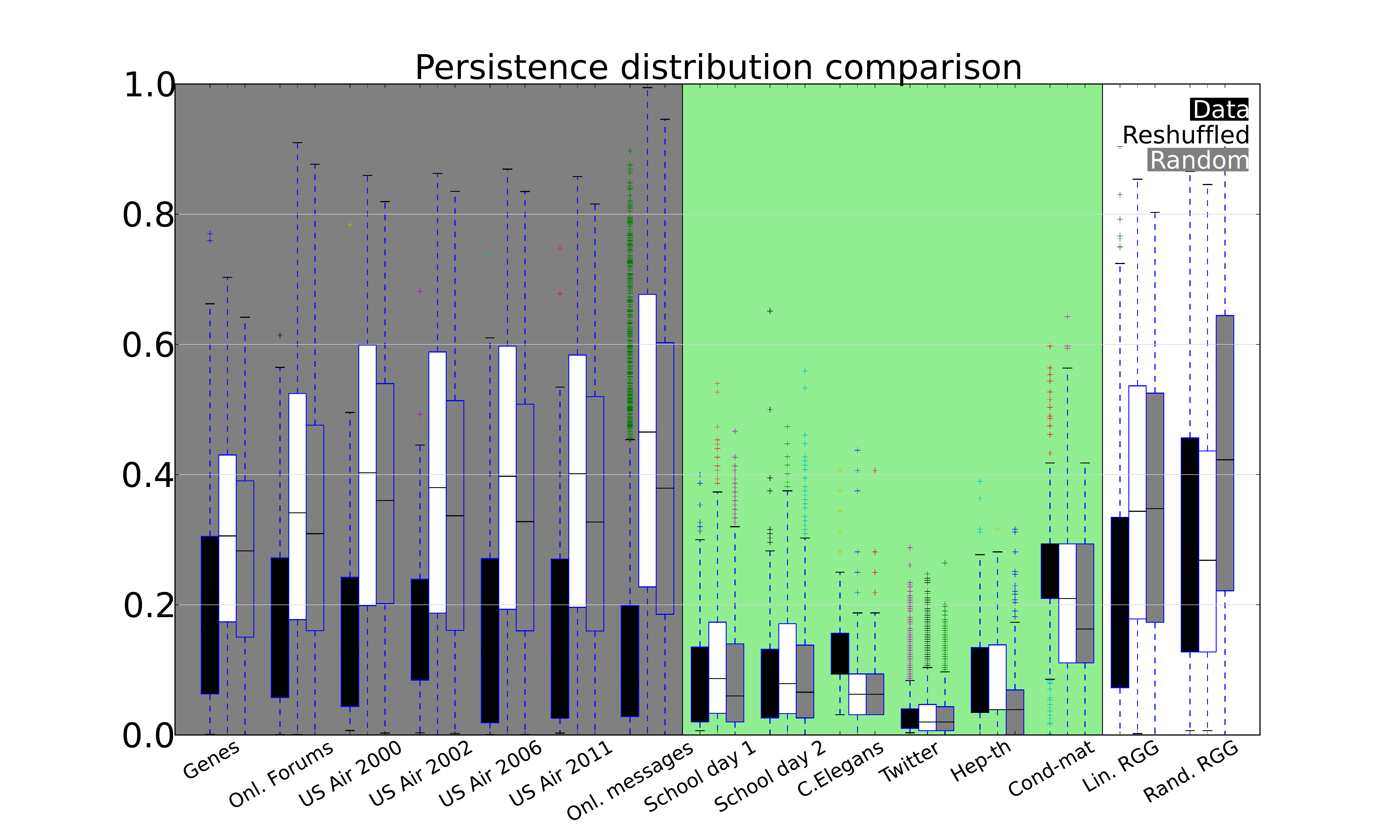}
}
\subfigure[]{
\includegraphics[width=.46\textwidth]{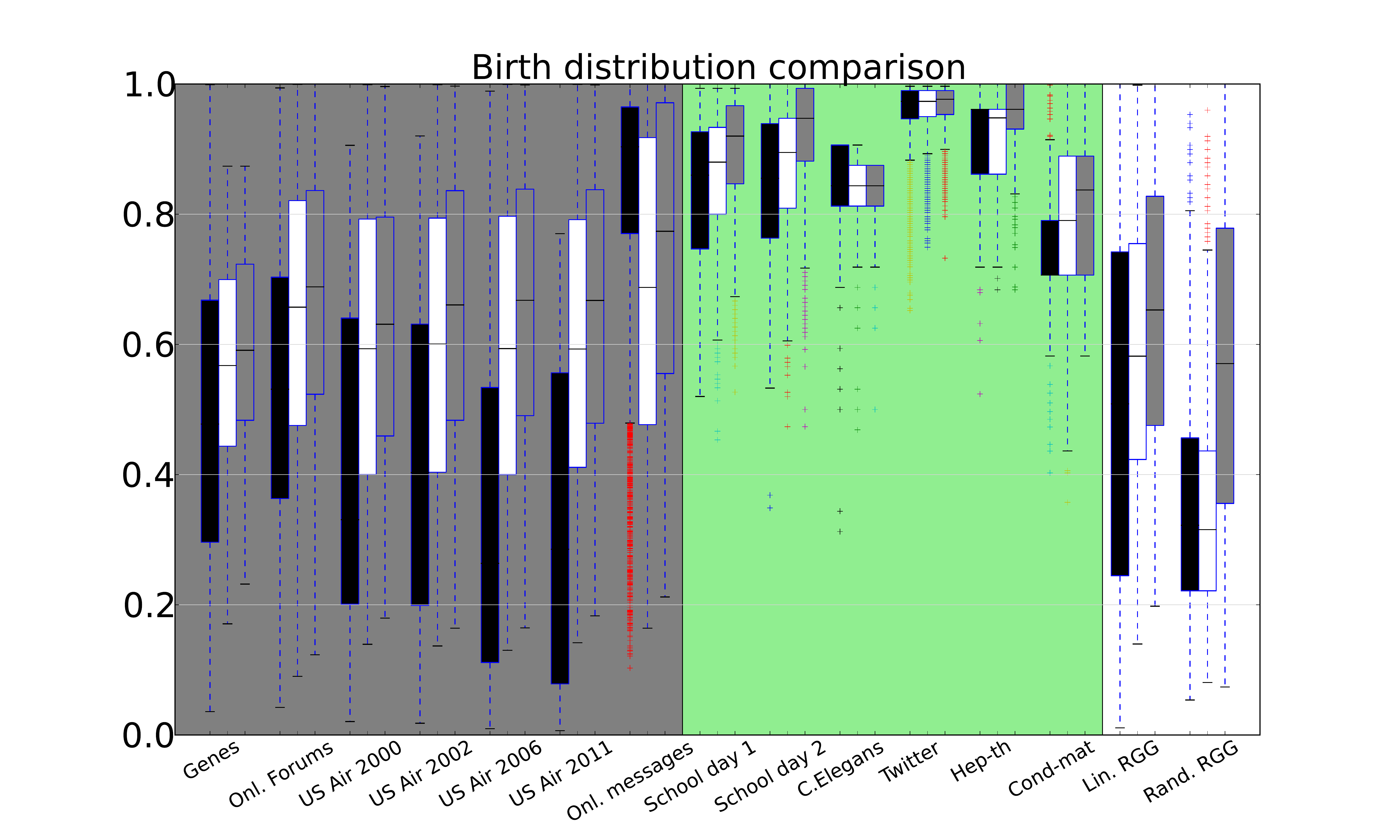}
}
\subfigure[]{
\includegraphics[width=.46\textwidth]{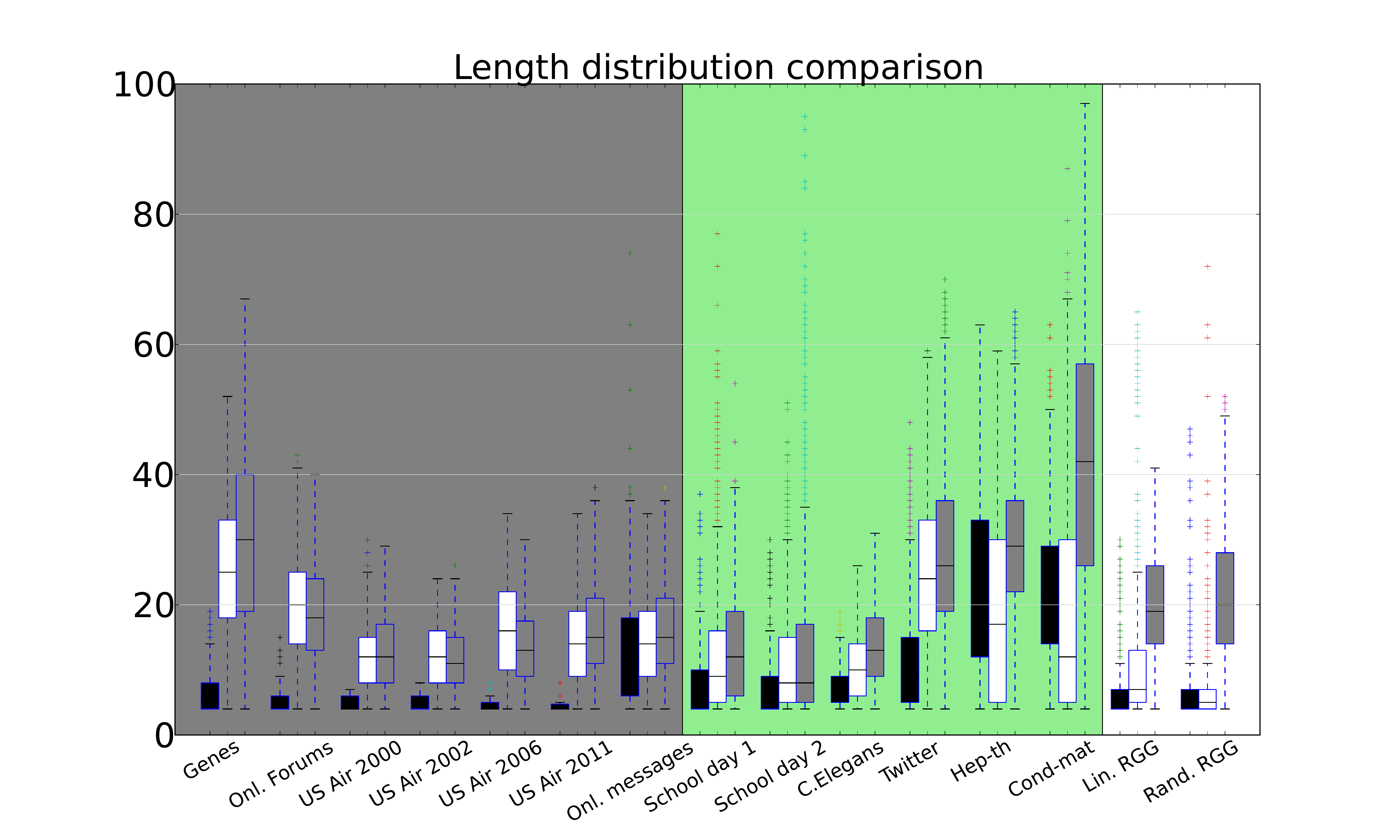}
}
\subfigure[]{
\includegraphics[width=0.23\textwidth]{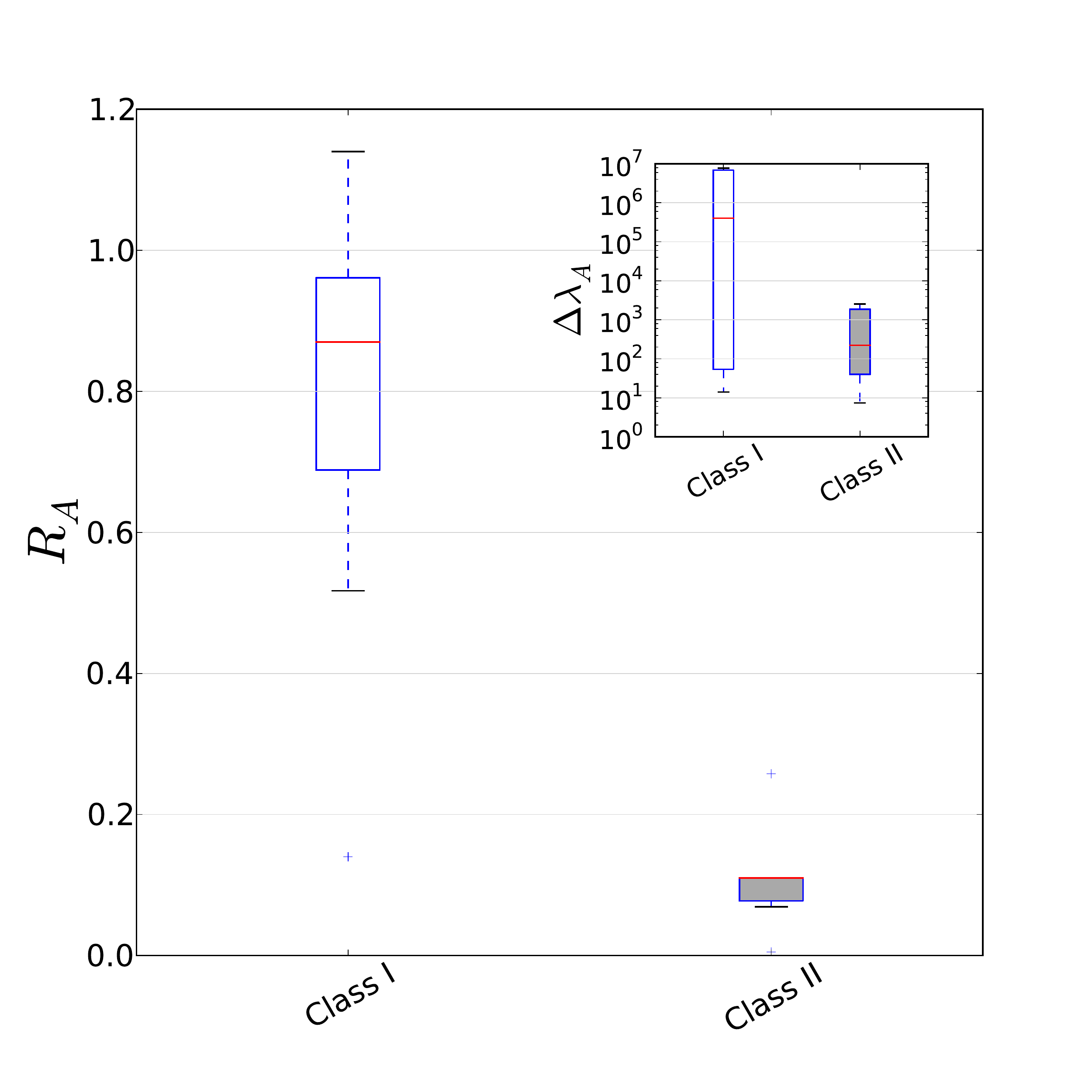}
\includegraphics[width=0.23\textwidth]{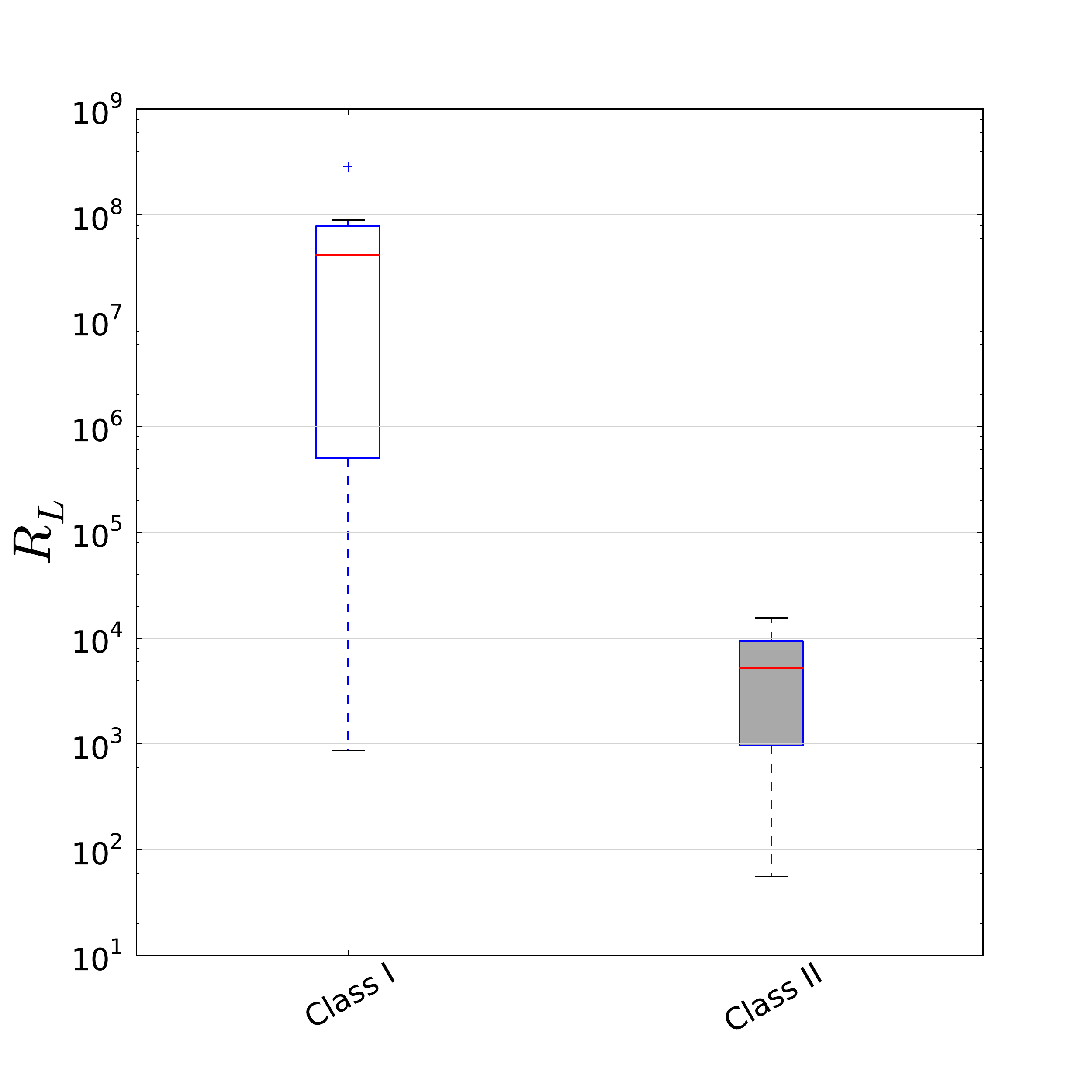}
}
\caption{{\bf Figure 2. Statistical and spectral properties of $H_1$ generators.}
Box plots of the distributions of persistences $\{p_g\}$ (panel $a)$), births $\{\beta_g\}$ (panel $b)$) and lengths $\{ \lambda_g \}$ (panel $c)$) for the 1d cycles ($H_1$ generators) of real networks ({\it black}), reshuffled ({\it white}) and randomized ({\it gray}).   The gray and green shaded areas identify the two network classes described in the main text: class I is significantly different from the random expectations, with shorter, less persistent cycles that appear across the entire filtration; class II networks are not significantly different from the random versions, with long cycles and late birth times in the filtration. The characteristics of class I networks imply a stratification of cycles that betrays the presence of large, non-local organisation in the network structure, which is not present in class II networks. For comparison, an example of RGG network (600 nodes in the unitary disk, linking distance 0.01), known to have higher order degree correlations, had edge weights set according to $\omega_{ij}\propto (k_i k_j)^\theta$, with $\theta=1$ (linearly correlated weight RGG) and $\theta=0$ (random weight RGG). In both cases, the distributions of cycles' properties resemble closely those of class I networks. Panel $d)$ finally reports the distribution of adjacency spectral gaps $\Delta \lambda_A$ and $R_A$ (left plot) and the Laplacian eigenratio $R_L$ (right plot). All the quantities show significant ($p<0.05$) differences between the two classes, implying that the homological structure affect the dynamical properties of networks, e.g. the synchronizability threshold.}

\end{figure*}

\clearpage

\begin{table}
\begin{center}
\tiny
\begin{tabular}{|c|c|c|c|c|c|c|c|c|}
\toprule[1.5pt]
Dataset (class) & $h_1$ & $\tilde{h}_1$ & $h^{sh}_1$ & $\tilde{h}^{sh}_1$ & $h^{rnd}_1$ & $\tilde{h}^{rnd}_1$ & $h_2$ & $\tilde{h}_2$ \\
\hline 
Genes(I)   & 0.515 & 0.003 & $0.020 \pm 0.001$ & $ 0.0007 \pm 0.00001$ & $0.0151 \pm 0.0004$ & $0.00023 \pm 0.00005$ & 0.35 & 0.006 \\
Online forums(I) & $0.175$ & $0.001$ & $0.355\pm 0.005$ & $0.007\pm 0.001$& $0.325\pm 0.005$ & $0.007\pm 0.001$ & 0.02 & 0.0003\\
US Air 2000(I) & 0.160 & 0.001 & $ 0.405 \pm 0.005$  & $0.0065 \pm 0.0007$ & $0.358 \pm 0.006$ & $0.0060 \pm 0.0005$ & 0.02 & 0.0003  \\
US Air 2002(I) & 0.186  & 0.0008  & $0.39 \pm 0.01$ &$ 0.0037 \pm 0.0003$ & $0.34 \pm 0.01$  & $ 0.0034 \pm 0.0003 $ &0.23  & 0.002 \\ 
US Air 2006 (I)& 0.167 & 0.0005  &  $ 0.398 \pm 0.005 $ & $0.0036 \pm 0.0005$ & $0.348 \pm 0.008$  & $ 0.0032 \pm 0.0003 $   & 0.165  & 0.001 \\ 
US Air 20011(I)& 0.181 & 0.0006 & $0.41 \pm 0.01$ & $0.0034 \pm 0.0002$ & $0.35 \pm 0.01$  & $0.0033 \pm 0.0003 $ &0.076  & 0.0007 \\ 
Online messages(I)& 0.21 & 0.0014 & $0.190 \pm 0.002$ & $0.0017 \pm 0.0001$ & $ 0.185  \pm 0.002$& $0.0015\pm 0.0001$ & 0.02 & 0.0003  \\ 
School day  1 (II)& 0.088 & 0.0034 & $ 0.113 \pm 0.002$ & $0.007\pm 0.001$ & $0.093 \pm 0.002$ &  $0.006 \pm 0.001$ & 0.015 & 0.0012 \\   
School day 2 (II)& 0.090 & 0.0033  & $0.115\pm 0.002$ & $0.0065 \pm 0.0005$  & $0.098 \pm 0.003$ & $0.0089 \pm 0.0008$ & 0.01412 & 0.00095 \\
C. elegans (II)& 0.0784 & 0.002 &   $0.0745\pm 0.0017$ & $0.001 \pm 0.0001$  & $0.0896\pm 0.0023$ & $0.0041\pm 0.0005$ & 0.058 &  0.002 \\
Twitter (II)&  0.03 & 0.0001 & $ 0.030 \pm 0.001 $    & $0.0002 \pm 0.0001$  &  $0.029 \pm 0.001$ & $0.0002 \pm 0.0001$  & 0.01 & 0.0001\\
Hep-th (II)& 0.08 & 0.0002 & $0.075 \pm 0.001$ & $0.0002 \pm 0.0001$ & $0.0508 \pm 0.0003$ &$0.0002 \pm 0.0001$ & - & -  \\ 
Cond-mat (II)& 0.26 & 0.0004 & $ 0.20\pm 0.003$ & $0.0002\pm 0.0001$ & $0.180 \pm 0.002$ & $0.0005 \pm 0.0001 $  & - & -\\ 
Lin. RGG & 0.227 & 0.003 & $0.368 \pm 0.005$ & $ 0.006\pm 0.001 $ & $0.355  \pm 0.002$ & $ 0.012 \pm 0.001$   & 0.28 & 0.006 \\  
Ran. RGG & 0.3 & 0.0041 & $0.299 \pm 0.005$ & $0.0045\pm 0.0002$ & $0.649 \pm 0.40$ & $0.015\pm 0.001$  & 0.115 & 0.003 \\
\bottomrule[1.25pt]

\end{tabular}\par
\end{center}
\bigskip

\caption{\normalsize {\bf Summary of hollowness values.} For each dataset, we report the values of the {\it hollowness} $h_1$ and {\it cycle-length normalized hollowness} $\tilde{h}_1$ for $H_1$ cycles for real networks and their randomisations ($sh$ and $rnd$). Most networks (class I in particular) show lower values than for their randomized versions. We also report the values of the {\it hollowness} $h_2$ and {\it cycle-length normalized hollowness} $\tilde{h}_2$ for $H_2$ cycles for real networks. The values for the randomized networks are not reported as --strikingly-- the randomisations do not display any higher homology, while almost all real networks display positive values of the $H_2$ hollowness.}\label{table::h1}
\end{table}



\end{bibunit}

\clearpage
\begin{bibunit}[plain]

\renewcommand{\thefigure}{S.\arabic{figure}}

\section*{Supplementary Information for ''Topological strata of weighted complex networks''}

The Supplementary Information is organized in four sections. Section I contains definitions and references concerning persistent homology, our main tool.
In section II some constructions for filtrations are presented, in particular the weight rank clique filtration is introduced. In section III we describe the datasets tested for our main result, the classification of networks based on persistent $H_1$ generators. In section IV the reader can find more detailes on the classification, and the plots supporting the result.

\subsection*{Persistent homology}

This section is devoted to the mathematical framework supporting persistent homology \cite{(Top)} \cite{(Pers)} \cite{Pers2}. Persistent homology can be viewed as parametrized version of simplicial homology, that requires the definitions of simplicial complex and homology, for detailed information we refer to \cite{Simp}.

\begin{definition}
A \textit{simplicial complex} is a non empty family $X$ of finite
subsets, called faces, of a vertex set with the two constraints:
\begin{enumerate}
\item[-] a subset of a face in $X$ is a face in $X$,
\item[-] the intersection of any two faces in $X$ is a face of both.
\end{enumerate}
\end{definition}

We assume that the vertex set is finite and totally ordered. A face of $n+1$ vertices is called $n-$face and denoted by $[p_0,\ldots, p_n]$.
A $0-$face is a vertex, a $1-$face is a segment, a $2-$face is a full triangle, a $3-$face is a full tetrahedron.
The dimension of a simplicial complex is the highest dimension of the faces in the complex. \\
\begin{example}
The clique complex is a simplicial complex constructed from a graph. There is a $n-$face in the simplicial complex for every $(n+1)-$clique in the graph, i.e a complete subgraph on $n+1$ vertices.
The compatibility relations are satisfied because subsets of cliques and intersection of cliques are cliques themselves.\\
\end{example} 

Morphism between simplicial complexes are called simplicial maps.
\begin{definition}
A simplicial map is a map between simplicial complexes with the
property that the image of a vertex is a vertex and the image of a $n-$face is face of dimension $\leq n$.
\end{definition}

Fixed a field $k$, in the following by vector space we intend a $k-$vector space and $k[t]$ is the polynomial ring in one variable with coefficients in $k$.
Given a simplicial complex $X$ of dimension $d$, consider the vector spaces $C_n$ on the set of $n-$faces in $X$ for $0\leq n \leq d$. Elements in $C_n$ are called $n-$chains.
The linear maps sending a $n-$face to the alternate sum of it's $(n-1)-$faces.

\begin{eqnarray*}
\partial_n: C_n & \longrightarrow & C_{n-1}\\
\text{[}p_{0},\ldots, p_{n}\text{]} & \rightarrow & \sum_{i=0}^n(-1)^i\text{[}p_{0},
\ldots, p_{i-1}, p_{i+1},\ldots, p_{n}\text{]}.
\end{eqnarray*}

shares the property $\partial_{n-1}\circ\partial_n=0$.

The subspace $ker \,\partial_n$ of $C_n$ is called the vector space of $n-$cycles and denoted by $Z_n$.
The subspace  $Im \, \partial_{n+1}$ of $C_n$, is called the vector space of $n-$boundaries and denoted by $B_n$.
Note that from $\partial_{n-1}\circ\partial_n=0$ it follows that $B_n\subseteq Z_n$ for all $n$.
\begin{definition}
The $n-$th simplicial homology group of $X$, with coefficients in $k$, is the vector space $H_n:=Z_n/B_n$.\\
The rank of $H_n$ is called the $n$-th Betti number of $X$. 
\end{definition}

The first Betti numbers of $X$ have an easy intuitive meaning: the $0$-th Betti number is the number of connected components of $X$, the first Betti number is the number of two dimensional (poligonal) holes, the third Betti number is the number of three dimensional holes (convex polyhedron).\\

It is fundamental to note that homology is a functor, this implies the following proposition.

\begin{proposition}
Let $X$ and $Y$ be two simplicial complexes, a simplicial map $f: X \rightarrow Y$ determines a linear map between the homology groups $H_i (f): H_i(X)\rightarrow H_i(Y)$ for all $i$.
\end{proposition}

The starting point in persistent homology is a filtration.
As in \cite{(Pers)}, we call a simplicial complex $X$
filtered if we are given a family of subspaces
$\{X_v\}$ parametrized by $\N$, such that $X_v\subseteq X_w$
whenever $v\leq w.$ The family $\{X_v\}$ is called a
\textit{filtration}. There are many ways to construct a filtration from a point cloud or a network, some relevant ones are explained in section II.

\begin{definition}
The \textit{persistent homology module} of a filtration is given by the homology groups of the simplicial complexes $H_n(X_v)$ and the linear maps $i_{v,w}: H_n(X_v)\rightarrow H_n(X_{w})$ induced in homology by the inclusions $X_v\hookrightarrow X_w$ for all $v\leq w$. 
\end{definition}
Following \cite{(Pers)}, this system is called a module because  the vector space $H_n=\oplus_v H_n(X_v)$ can actually be endowed with a $k[t]-$module structure, defining $t\cdot m:=i_{v,v+1}(m)$ for $m\in H_n(X_v)$. Note that the linear maps $i_{v,v+1}$ are not always injective.
A persistent homology generator is a generator of $H_n$ according to the $k[t]-$structure, i.e an element $g \in H_n(X_v)$ such that there is no $h \in H_n(X_w)$ for $ w<v$ with the property that $t^{v-w}h=g.$
By the structure theorem on modules over PID, $k[t]-$modules are completely determined by the degree of each generator $g$ (birth of the generator $\beta_{g}$) and the degree in which the generator is annihilated by the module action (death of the generator $\delta_{g}$). The persistence (lifetime) of a generator is measured by $p_g:=\delta_g-\beta_g$.
The length of a cycle, number of faces composing it, is denoted by $\lambda_g$. \\
The barcode of a filtration is the set of intervals $[\beta_{g}; \delta_{g}]$ for all generators $g\in H_n$, this is a handy complete invariant of $H_n$, \cite{(Pers)}.
By persistent topological features we intend generators of $H_n$ such that the interval $[\beta_g; \delta_g]$ is large with respect to the filtration length.\\
An alternative way to represent persistent homology modules is the persistence diagram \cite{Pers2}, \cite{Pers3}.
A persistence diagram is a set of points in the plane counted with multiplicity, it can be recovered from the barcode considering the points $(\beta_g,\delta_g)\in \R^2$ with multiplicity given by the number of generators with the same persistence interval.\\
Persistent homology modules can be computed using the libraries {\it javaPlex} (Java) or {\it Dionysus} (C++), which are both available from the Stanford's CompTop group website (\url{http://comptop.stanford.edu/}), and presented using the barcode or the persistence diagram.
We developed a Python module to wrap the {\it javaPlex} library, consisting of a number of scripts able to preprocess complex networks and store the resulting homological information in a manageable form. 

\subsection*{Filtrations}

In this section we will go through some basic constructions that generate a filtration starting from a point cloud or a complex network.\\

The most popular filtration for data analysis is the \textit{Rips-Vietoris filtration} \cite{(Pers)}. The Rips-Vietoris complex is a simplicial complex associated to a set of points in a metric space in the following way: every point $p$ is the center of a radius $\epsilon$ ball $D(p,\epsilon)$ and $n+1$ points $\{p_0,\ldots, p_n\}$ determine a $n-$face in the Rips-Vietoris complex if the corresponding radius $\epsilon$ balls intersect two by two, i.e $D(p_i,\epsilon)\cap D(p_j,\epsilon)\ne \emptyset$ for all $ i\ne j \in \{0\ldots n\}$.
Clearly the Rips-Vietoris complex depends on the parameter $\epsilon$ and if $\epsilon_1<\epsilon_2$ the complex with $\epsilon_1$ radius balls is contained in the complex with $\epsilon_2$ radius balls.
To the growth of $\epsilon$ we obtain an increasing sequence of simplicial complexes, a filtration, the Rips-Vietoris filtration. In this context persistent topological features of the filtration are considered as features of the point cloud.\\

For unweighted networks, the \textit{Clique filtration} is used in \cite{Horak:2009bi} to analyse the difference between the barcodes of random networks, networks with exponential connectivity distribution and scale-free networks.
The $k-$skeleton $X_k$ of a simplicial complex $X$ is the subcomplex of $X$ containing all the faces of dimension smaller or equal to $k$.
Consider a complex network and the corresponding clique complex $X$, the clique filtration is obtained by filtering the clique complex according to the dimension of the skeleton:
\[
X_0\subseteq X_1 \subseteq X_2 \subseteq \ldots \subseteq X.
\] 
Note that persistent features of the Clique filtration are generators of the homology groups of the clique complex. These generators can be directly calculated from the clique complex of the graph, thus the filtration gives no extra information. This is not the case for the following filtration we have introduced for weighted networks in which persistent features cannot be determined from a single simplicial complex in the family but instead reveal the intricate multiscale relation between weights and links in a weighted indirect network.  

The \textit{Weight Rank Clique filtration} on a weighted network $\Omega$ combines the clique complex construction with a thresholding on weights.
The fist step is to rank the weights of links from $\omega_{max}$ to $\omega_{min}$: the discrete parameter $\epsilon_t$ scans the sequence.
At each step $t$ of the decreasing edge ranking we consider the thresholded graph $G(\omega_{ij},\epsilon_t)$, i.e. the subgraph of $\Omega$ with links of  weight larger than $\epsilon_t$.
For each graph $G(\omega_{ij},\epsilon_t)$ we build the clique complex $K(G,\epsilon_t)$.
The clique complexes are nested to the growth of $t$ and determine the weight rank clique filtration.
Persistent one dimensional cycles represent weighted loops with much weaker internal links. 

\subsection*{Datasets}
The dataset analysed in this paper cover a broad range of fields, spanning  social, infrastructural and biological networks. 
In detail they are:
\begin{description}
\item[US Air passenger networks] The networks refer to the years 2000, 2002, 2006 and 2011. The years were chosen to provide snapshots of the air traffic situation at 4-5 years intervals, plus one extra (year 2000) just before the events of 9/11 which significantly affected the air transportation industry. 
The data used are publicly available from the website of the Bureau of Transportation Statistics (\url{http://www.transtats.bts.gov/}). Individual flights between airports were aggregated on routes as defined by origin and destination cities. The weight reported is the yearly aggregated passenger traffic. 

\item[C. Elegans] The network is available at \url{http://cdg.columbia.edu/cdg/datasets} and reports a weighted, directed representation of the C. Elegans's neuronal network \cite{watts}. The network was symmetrized by summing the weights present on edges between the same nodes (given $\omega_{ij}$ and $\omega_{ji}$, $\omega^{symm}_{ij} = \omega^{symm}_{ji} =\omega_{ij} + \omega_{ji}$). 

\item[Online Messages and Forums] The online messages network consists of messages in a student online community at University of California \cite{opsahl1}. The online forum network refers to the same online community, but focuses on the activity of users in public forums, rather than on private messages \cite{opsahl2}. Both networks are publicly available online at Tore Opsahl's website (\url{http://toreopsahl.com/datasets/}).

\item[Gene network] The gene interaction network used in the paper is a sampling of the complete human genome dataset available from  the University of Florida Sparse Matrix Collection. Each node is an individual gene, while the edges correlates the expression level of a gene with that of the genes (using a NIR score \cite{NIR}). The node set of  the analysed network was obtained by randomly choosing an origin node, then adding its neighborhood to the node set; the neighborhoods of the newly added nodes were then added to the node set recursively until a given number of nodes was obtained (in the case used the target number of nodes was $N=1300$). Then all the edges present in the original network between the nodes in the node set were added, effectively taking a connected subgraph of the original network. 
To reduce the computational complexity due to the large density of the graph, the weighted clique filtration was stopped at an edge weight of $0.09$ (similarly to the choice made in \cite{Pajevic:2012ew}). 

\item[Twitter] The dataset consists of a network of mentions and retweet between Twitter users and is available online on the Gephi dataset page (\url{http://wiki.gephi.org/index.php/Datasets}). Weights are proportional to the number of interactions between a pair of users. 

\item[School face-to-face contact network] The dataset contains two days of recorded face-to-face interactions in a primary school. 
Each node represents a child, with the edge weight between two nodes being proportional to the amount of time the two children spent face to face. 
We analysed the two days separately, yielding two networks. The dataset has been collected by the Sociopattern project (\url{http://www.sociopatterns.org/}) and analysed in \cite{juliette}. 

\item[Co-authorship networks] The networks analysed are the weighted co-authorship networks of the Condensed Matter E-print Archive between 1995 and 1999  (cond-mat) and the High-Energy Theory E-print Archive between 1995 and 1999 (hep-th) \cite{newman1}. 

\end{description}

Finally, for comparison we use Random Geometric Graphs (RGG) \cite{bart,penrose}, which are simple models of spatial networks: a RGG is generated by sprinkling $N$ of nodes randomly on a metric space that acts as a substrate (usually a disk of unitary radius or a square with identified edges), and then linking 
nodes that are closer than a given linking distance $d$. \\ 

Finally, the networks analysed in this article are undirected and weighted, because the weighted clique filtration finds a natural application in such case. However, schemes for directed networks can be easily devised and tailored to specific case studies, e.g. one could adopt the definition used 
in the directed clique percolation method \cite{dcpm} in order to associate network structures to simplices.

\subsection*{Results for weight rank clique filtration}

\begin{figure*}[h]
\centering
\subfigure[]{\includegraphics[width=0.32\textwidth]{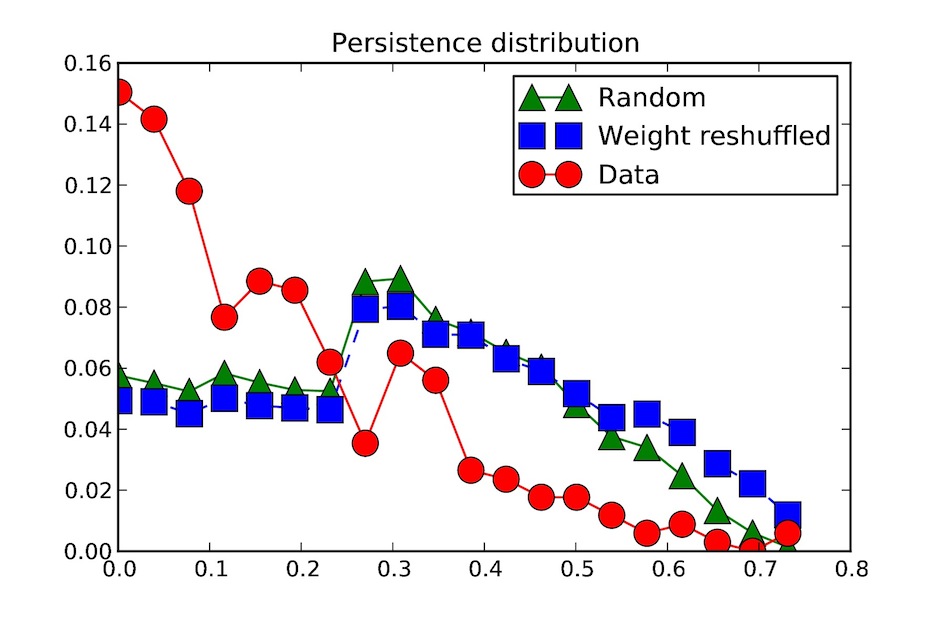}}	
\subfigure[]{\includegraphics[width=0.32\textwidth]{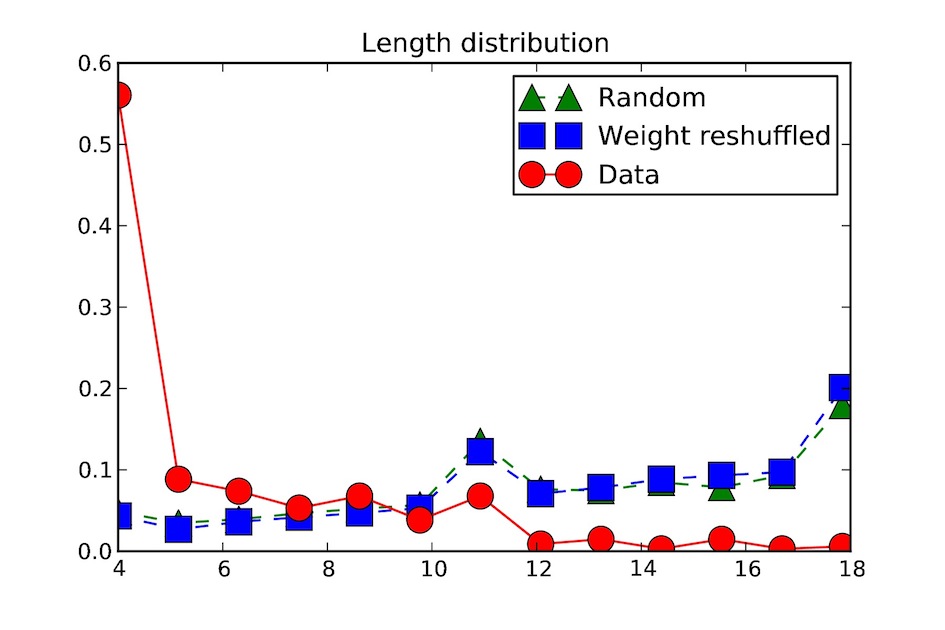}} 
\subfigure[]{\includegraphics[width=0.32\textwidth]{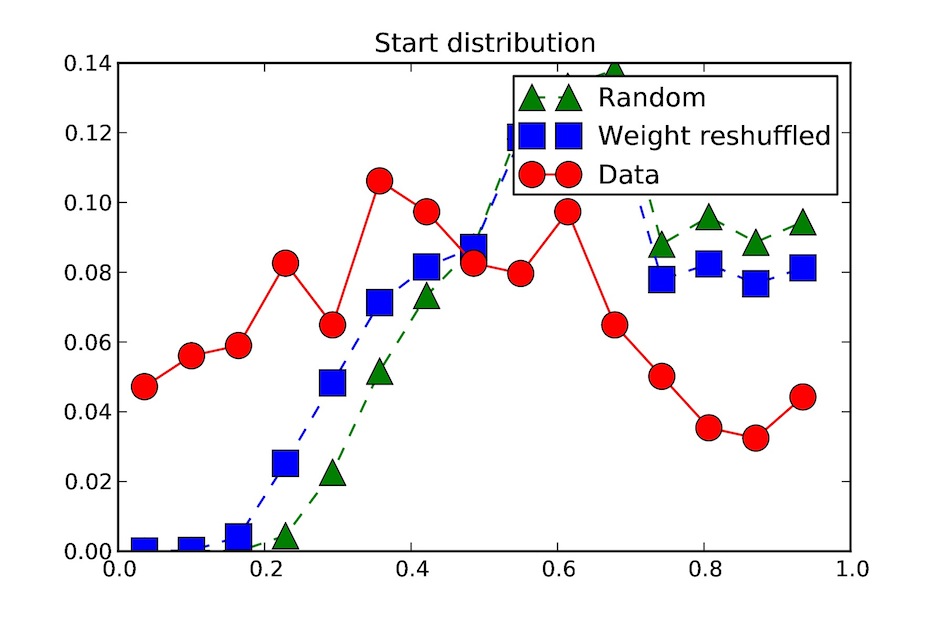}}
\\
\subfigure[]{\includegraphics[width=0.32\textwidth]{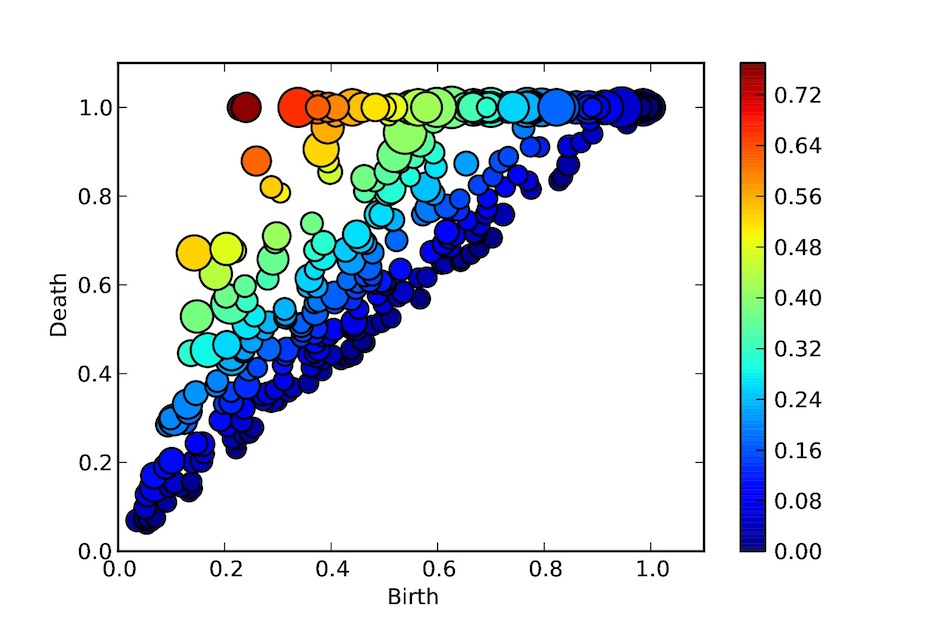}
}
\subfigure[]{\includegraphics[width=0.32\textwidth]{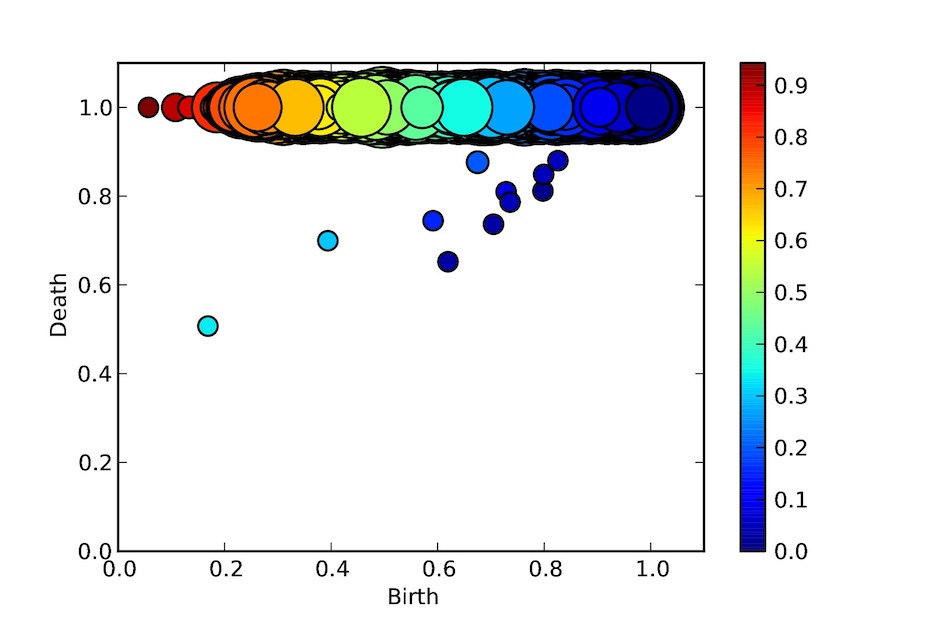}
}
\subfigure[]{\includegraphics[width=0.32\textwidth]{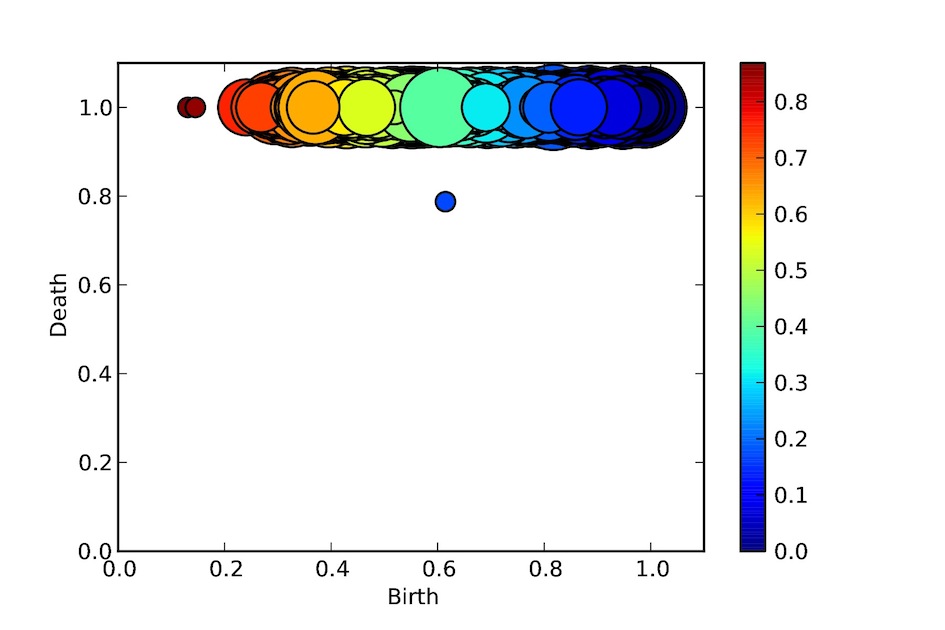}
}
\caption{{\bf Summary of $H_1$ persistent homology results for the human gene interaction network 2 (Class I).}}\label{fig::gene_random_comparison}
\end{figure*}

\begin{figure*}[h]
\centering
\subfigure[]{\includegraphics[width=0.31\textwidth]{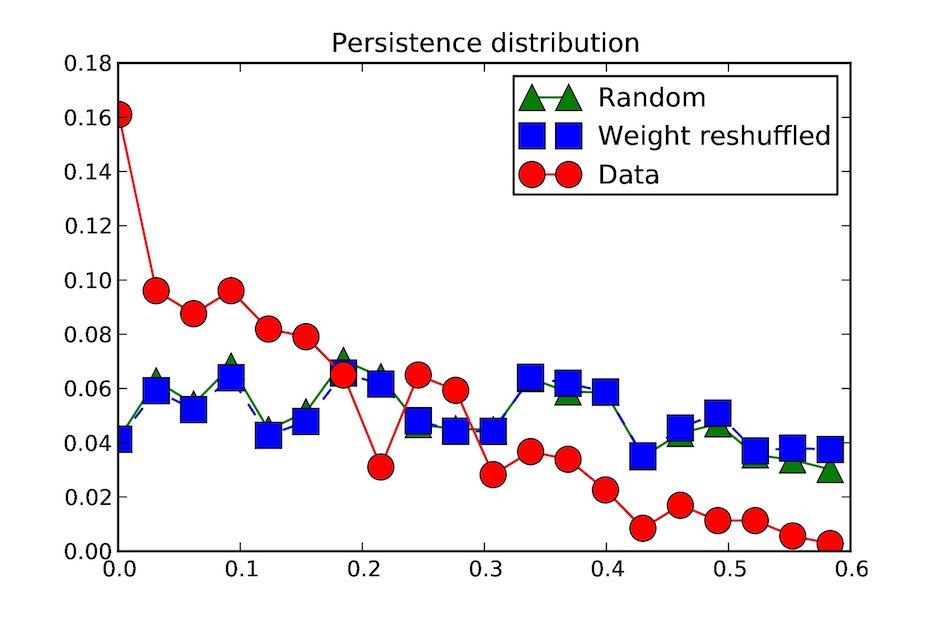}}
\subfigure[]{\includegraphics[width=0.31\textwidth]{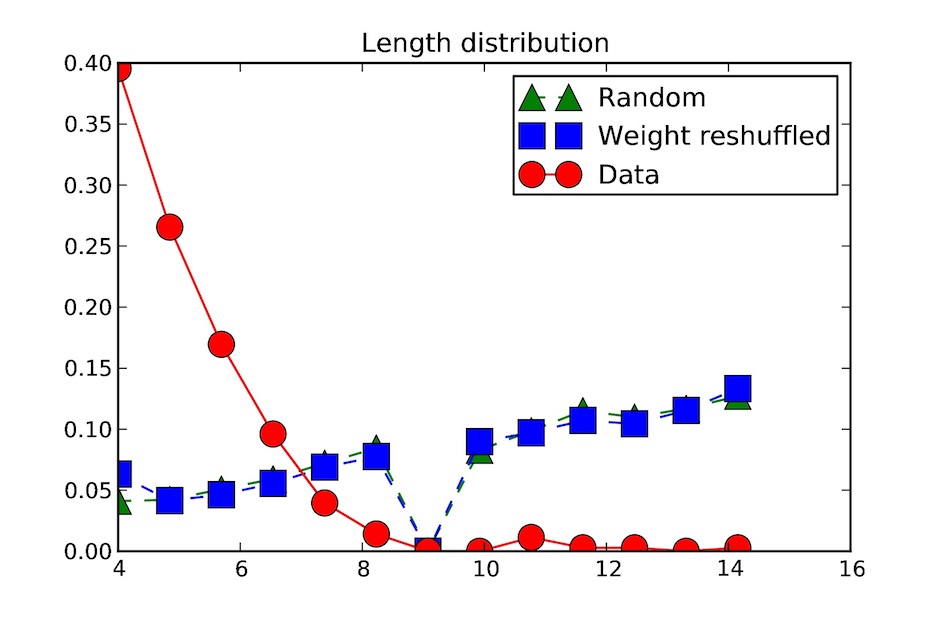}}
\subfigure[]{\includegraphics[width=0.31\textwidth]{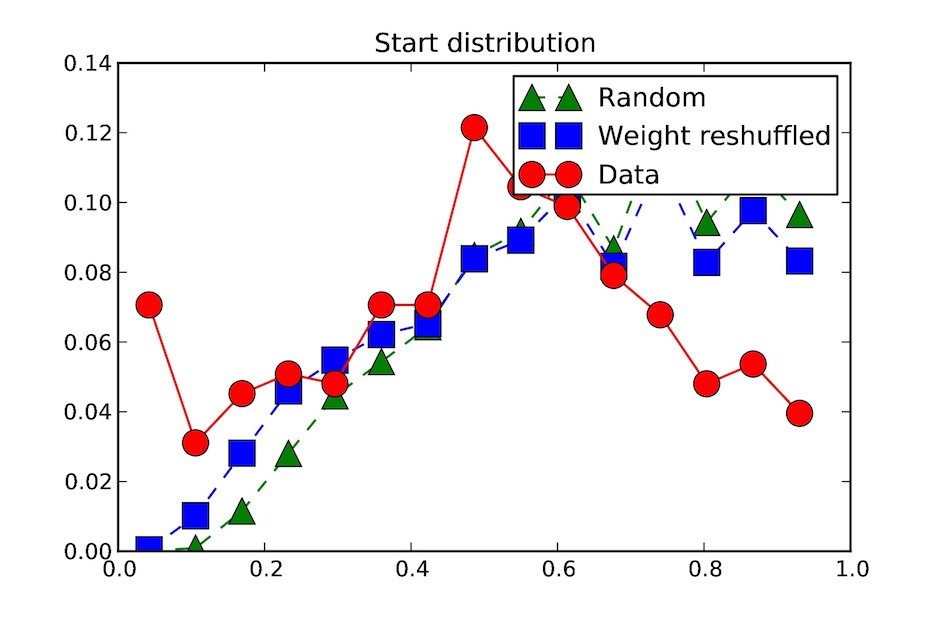}}
\\
\subfigure[]{\includegraphics[width=0.32\textwidth]{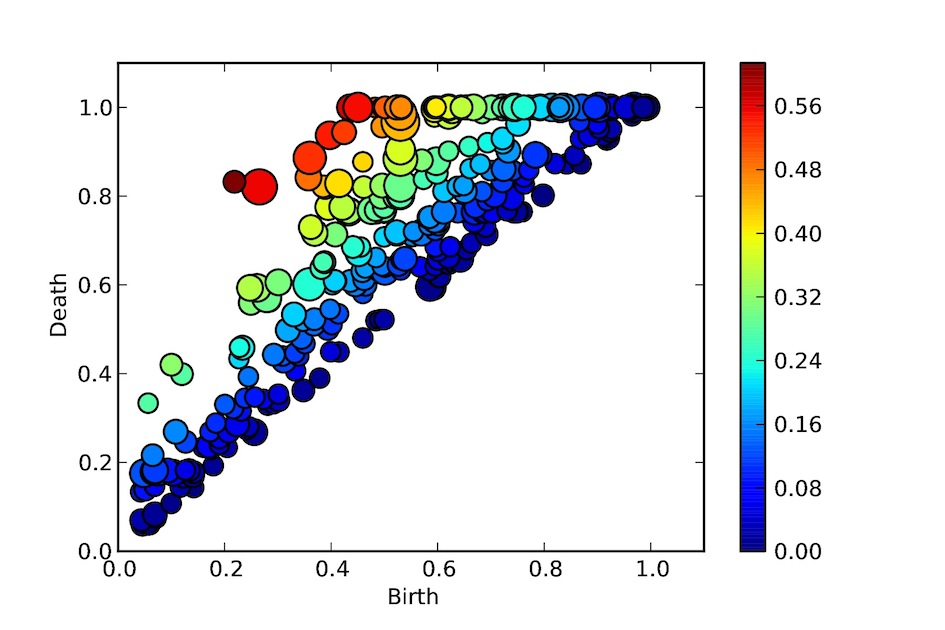}}
\subfigure[]{\includegraphics[width=0.32\textwidth]{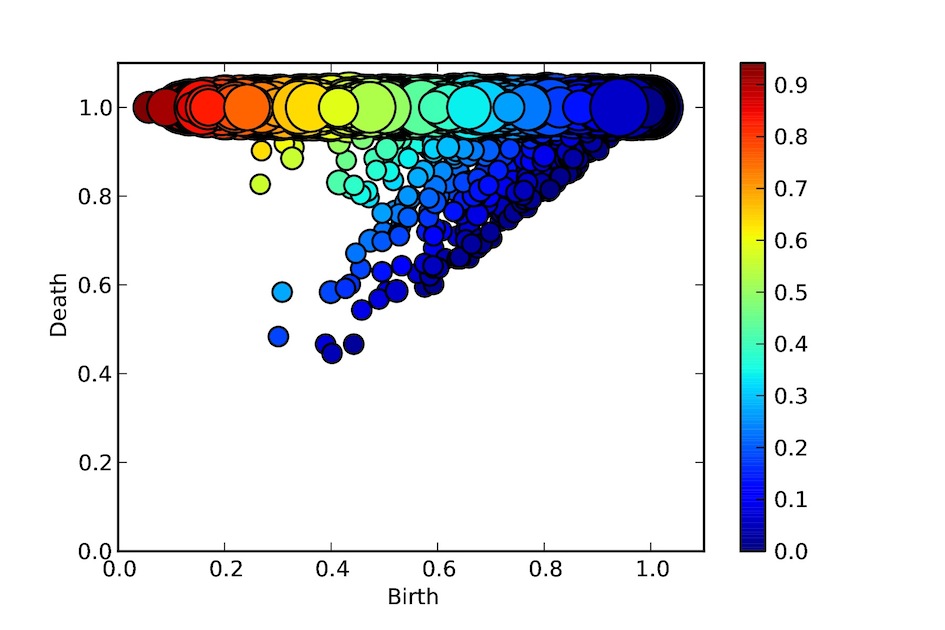}}
\subfigure[]{\includegraphics[width=0.32\textwidth]{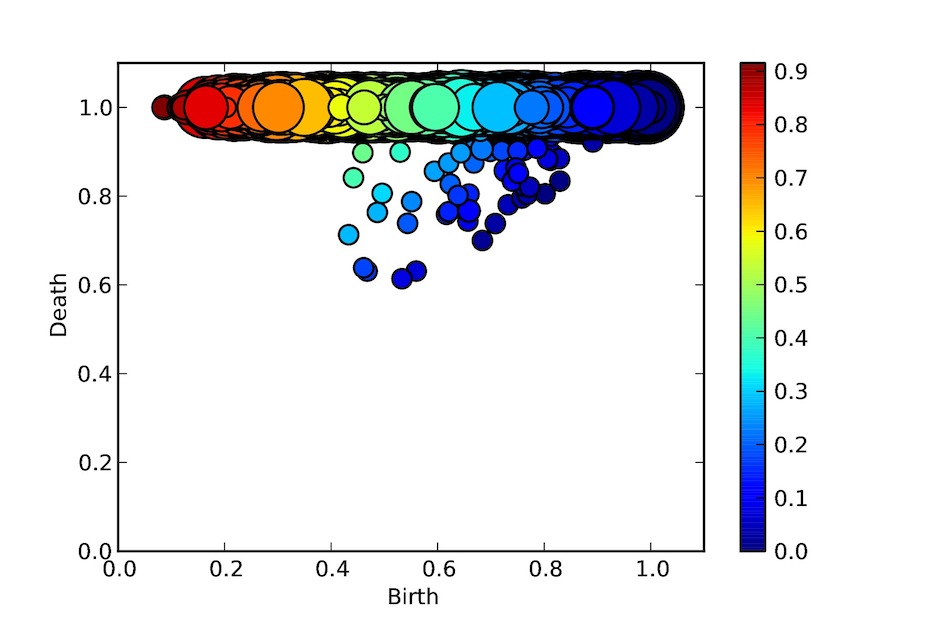}}
\caption{{\bf Summary of $H_1$ persistent homology results for online forum network of \cite{opsahl2} (Class I).} \label{fig::tore_random_comparison}}
\end{figure*}

\begin{figure*}[h]
\centering
\subfigure[]{\includegraphics[width=0.32\textwidth]{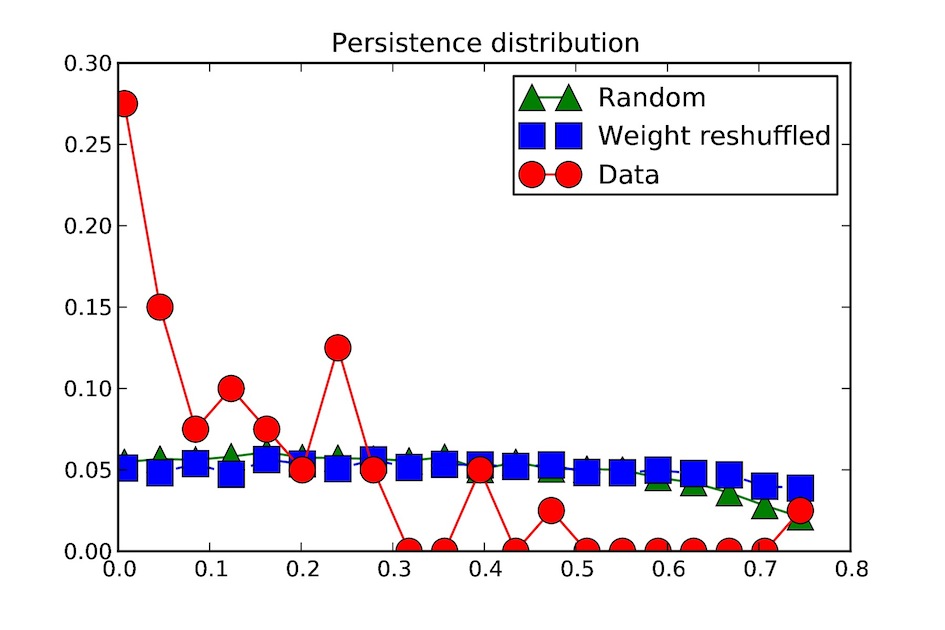}}
\subfigure[]{\includegraphics[width=0.32\textwidth]{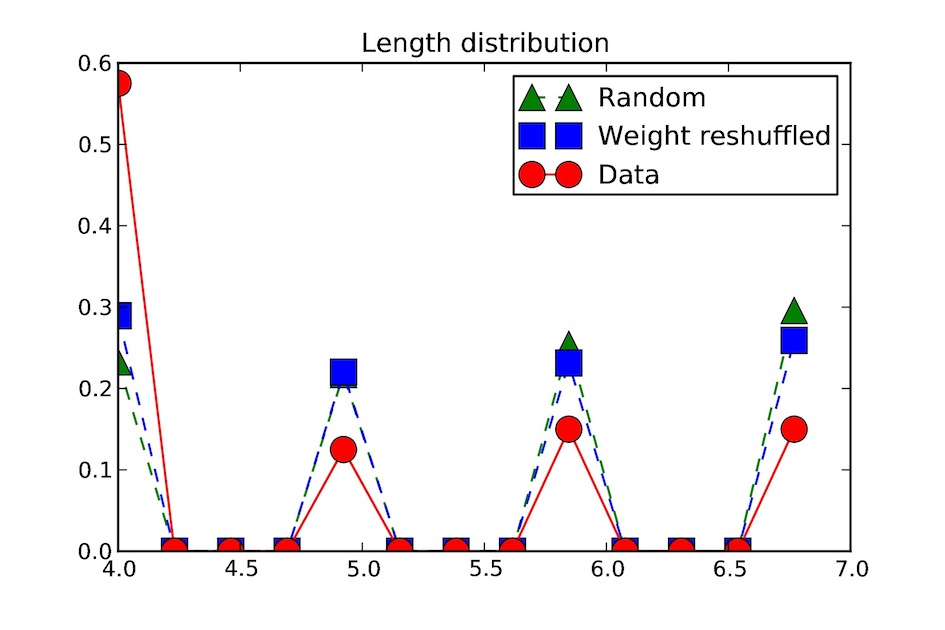}}
\subfigure[]{\includegraphics[width=0.32\textwidth]{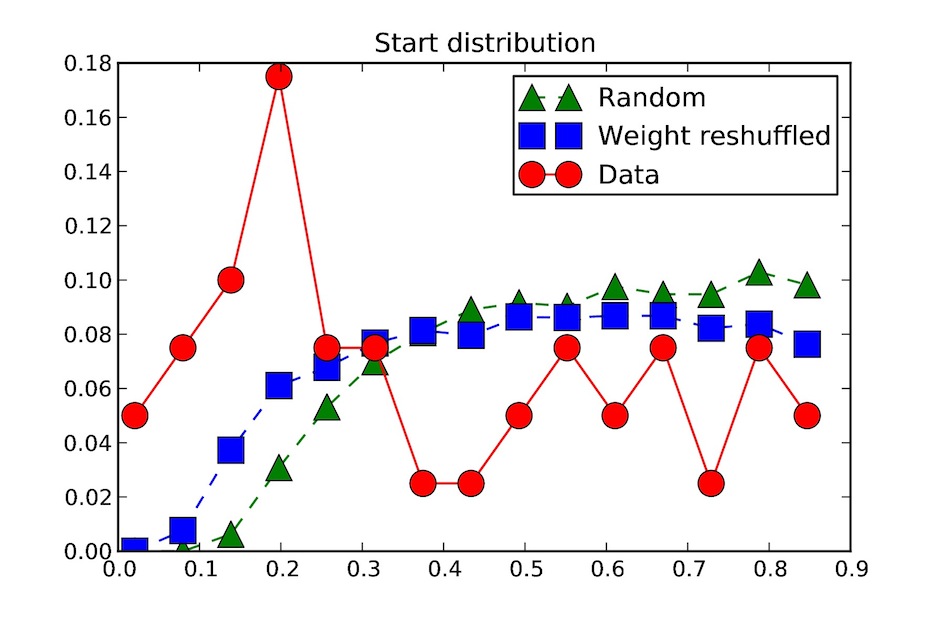}}
\\
\subfigure[]{\includegraphics[width=0.32\textwidth]{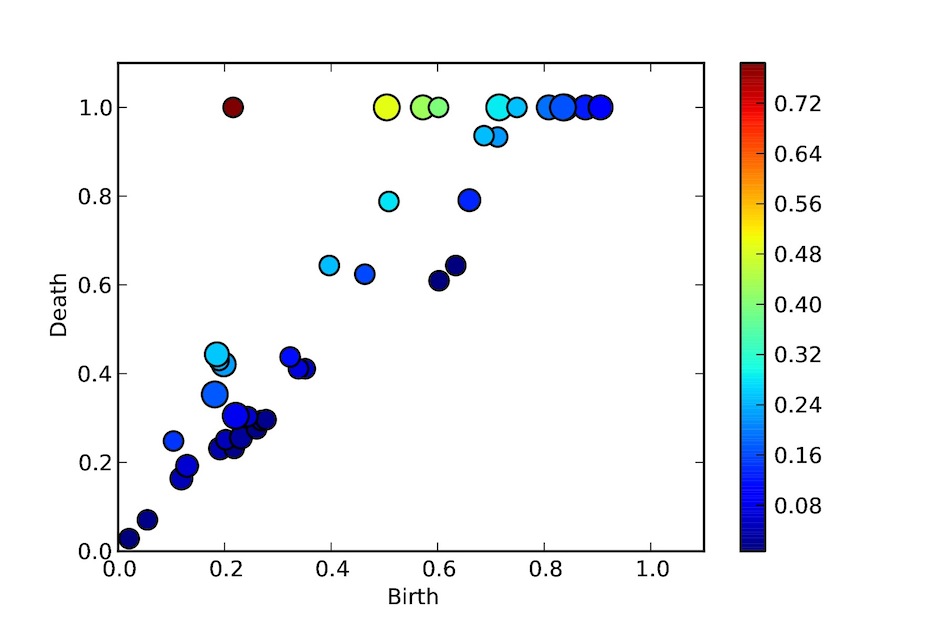}}
\subfigure[]{\includegraphics[width=0.32\textwidth]{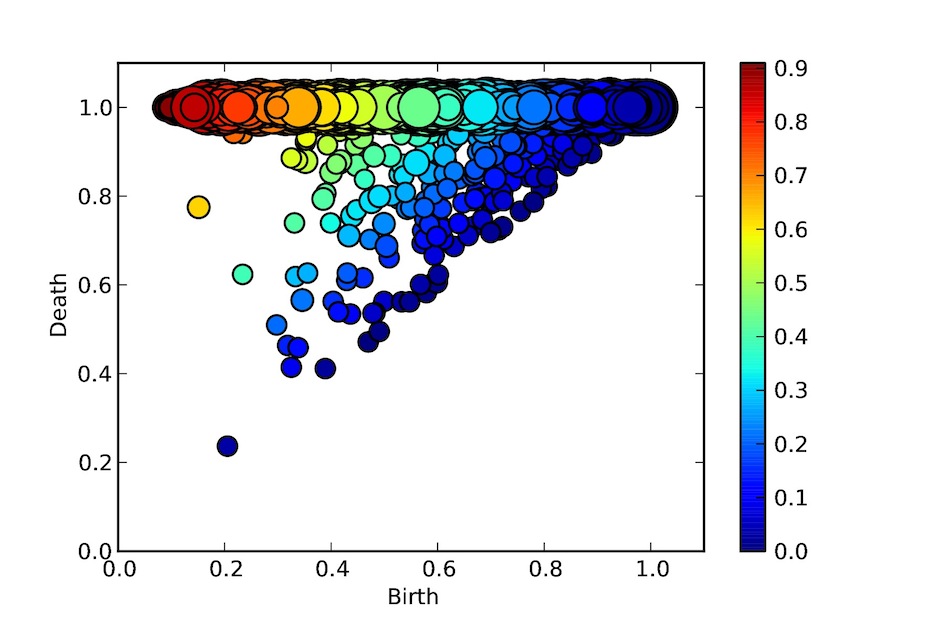}}
\subfigure[]{\includegraphics[width=0.32\textwidth]{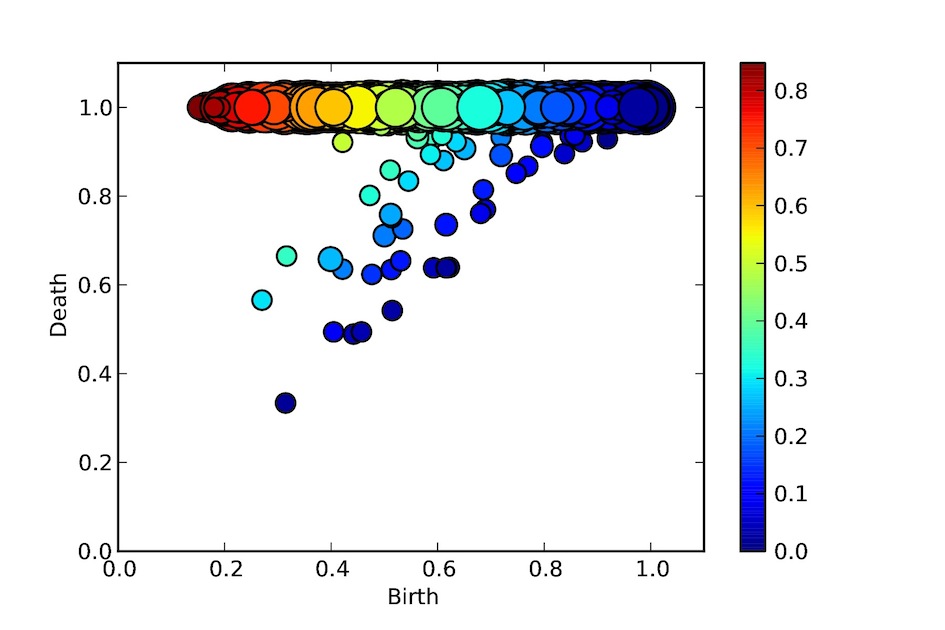}}
\caption{{\bf Summary of $H_1$ persistent homology results for the US airways passenger network for 2000 (Class I).} \label{fig::US2000_random_comparison}}

\end{figure*}

\begin{figure*}[h]
\centering
\subfigure[]{\includegraphics[width=0.32\textwidth]{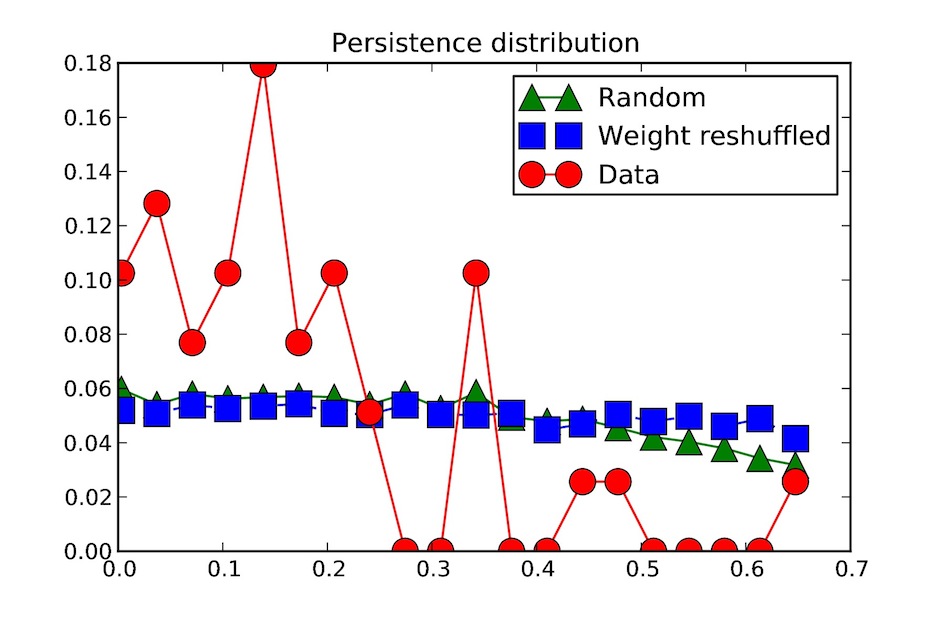}}
\subfigure[]{\includegraphics[width=0.32\textwidth]{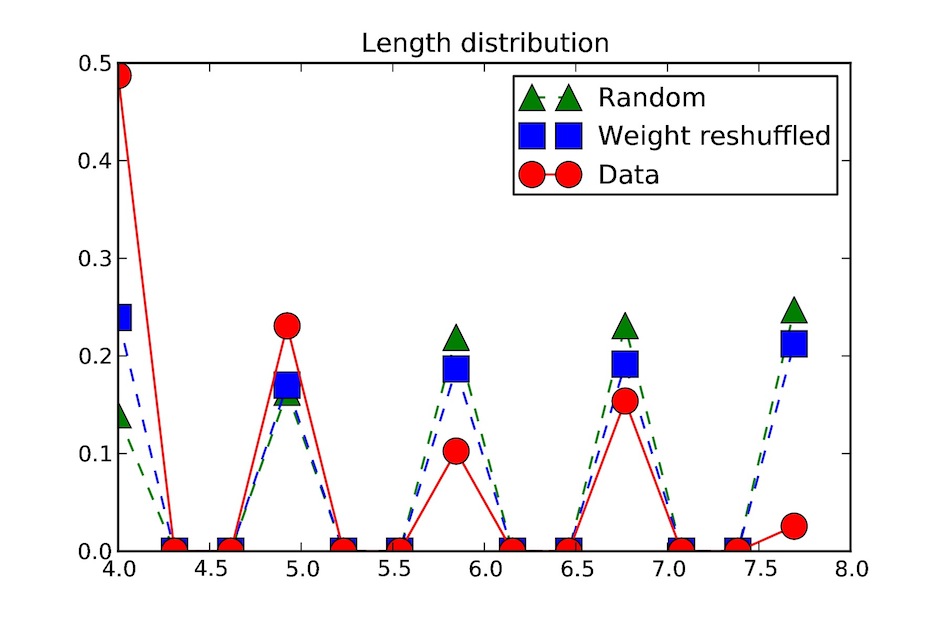}}
\subfigure[]{\includegraphics[width=0.32\textwidth]{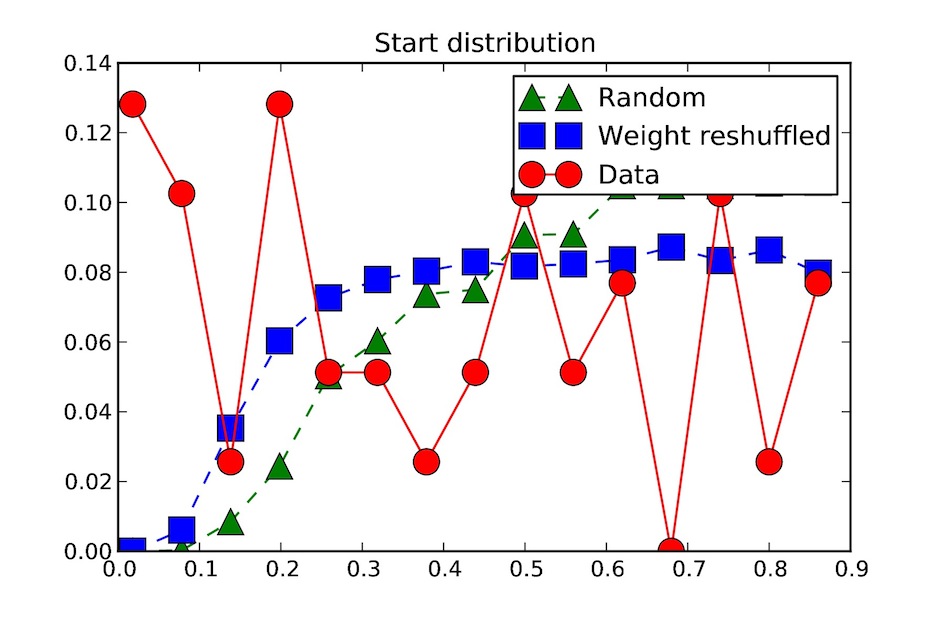}}
\\
\subfigure[]{\includegraphics[width=0.32\textwidth]{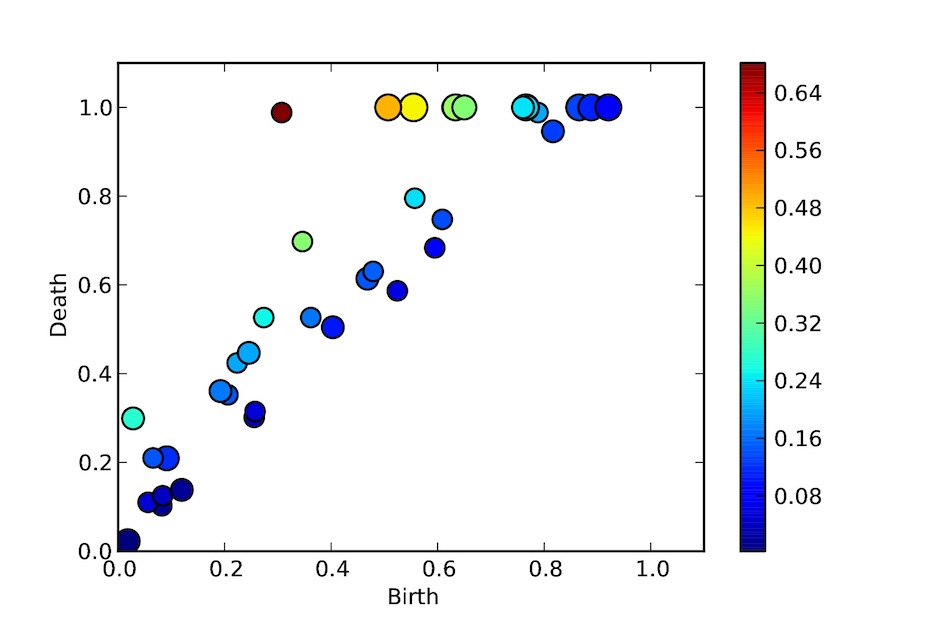}}
\subfigure[]{\includegraphics[width=0.32\textwidth]{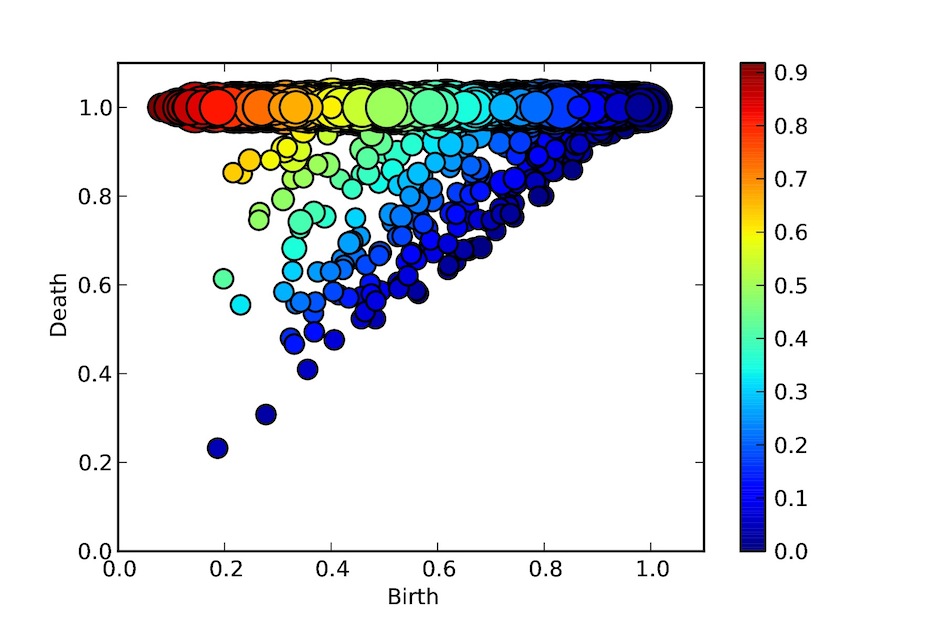}}
\subfigure[]{\includegraphics[width=0.32\textwidth]{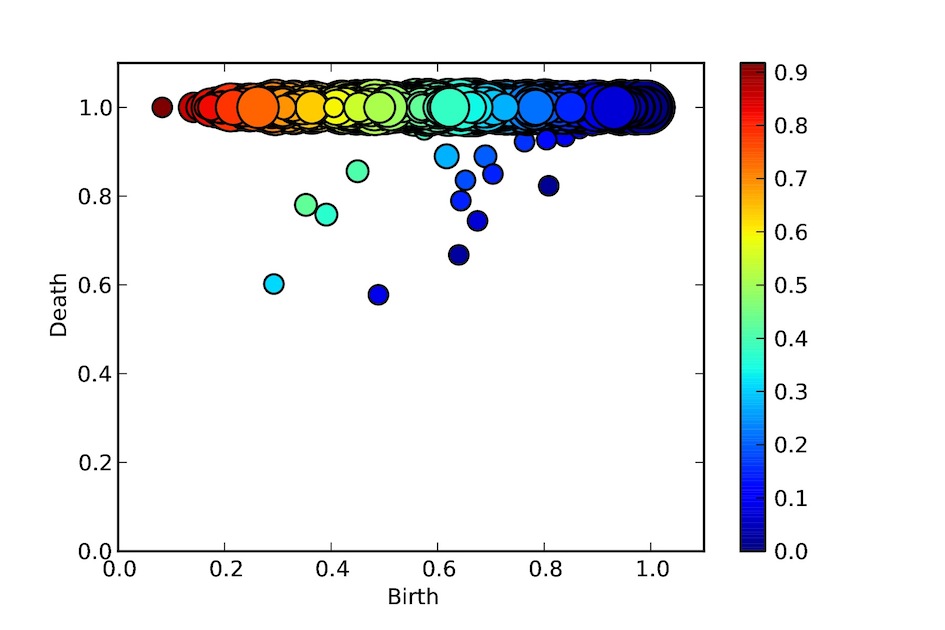}}
\caption{{\bf Summary of $H_1$ persistent homology results for the US airways passenger network for 2002 (Class I).}  }\label{fig::US2002_random_comparison}
\end{figure*}

\begin{figure*}[h]
\centering
\subfigure[]{\includegraphics[width=0.32\textwidth]{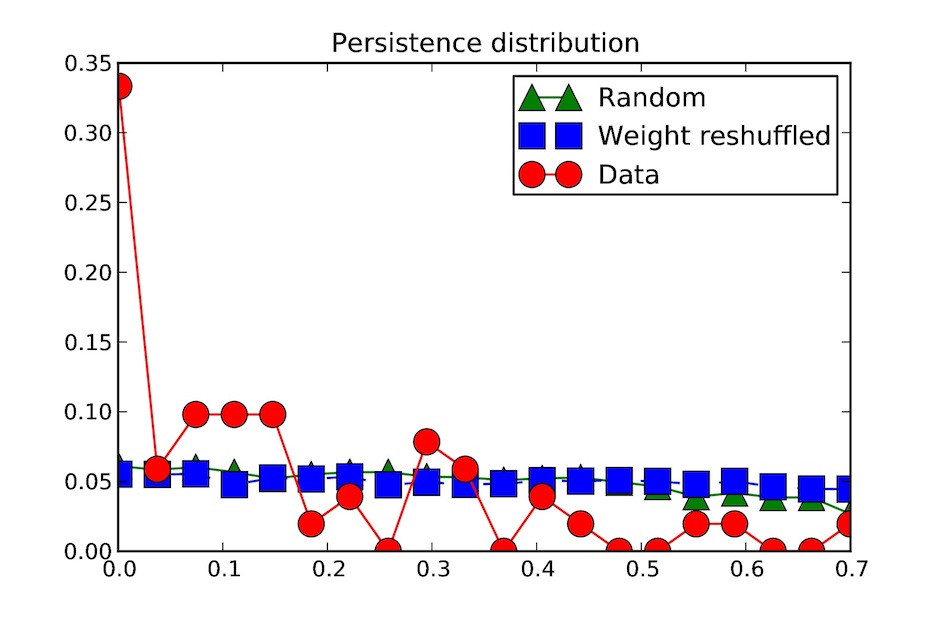}}
\subfigure[]{\includegraphics[width=0.32\textwidth]{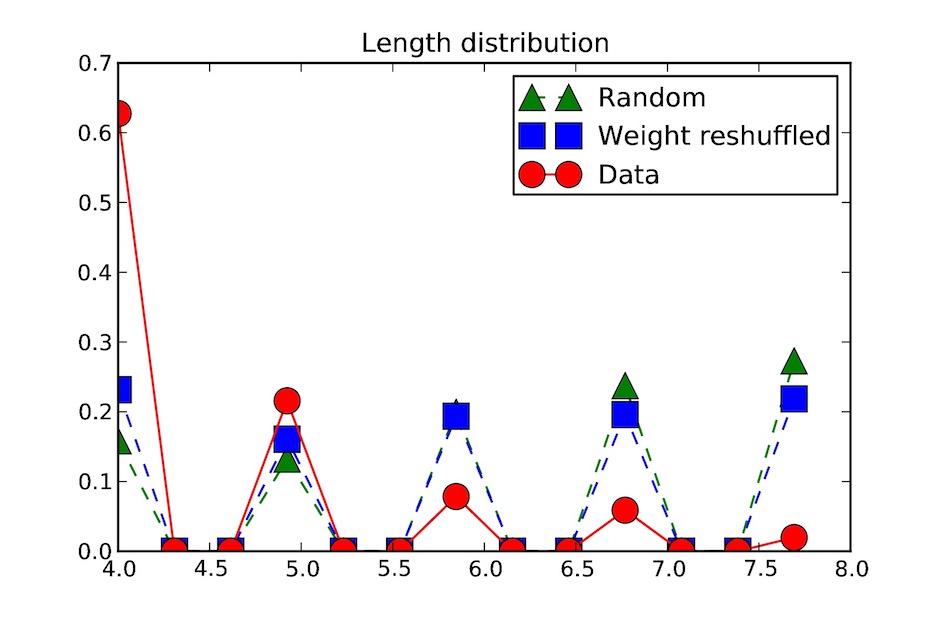}}
\subfigure[]{\includegraphics[width=0.32\textwidth]{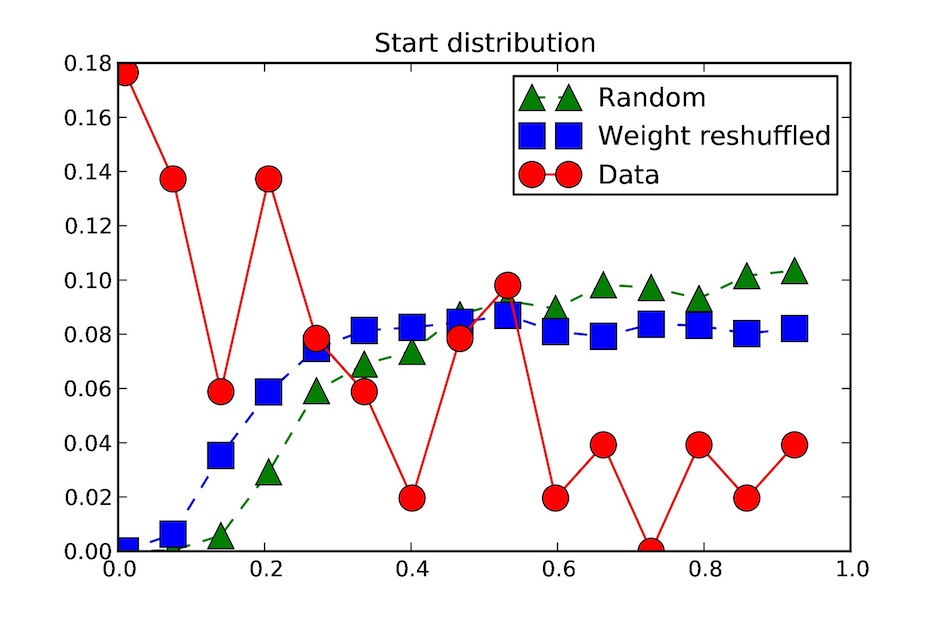}}
\\
\subfigure[]{\includegraphics[width=0.32\textwidth]{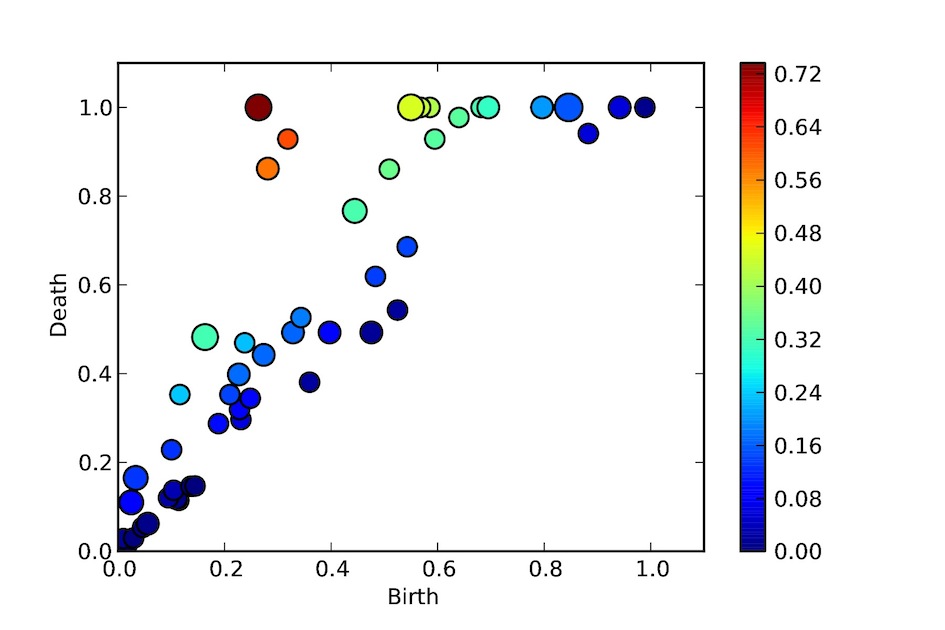}}
\subfigure[]{\includegraphics[width=0.32\textwidth]{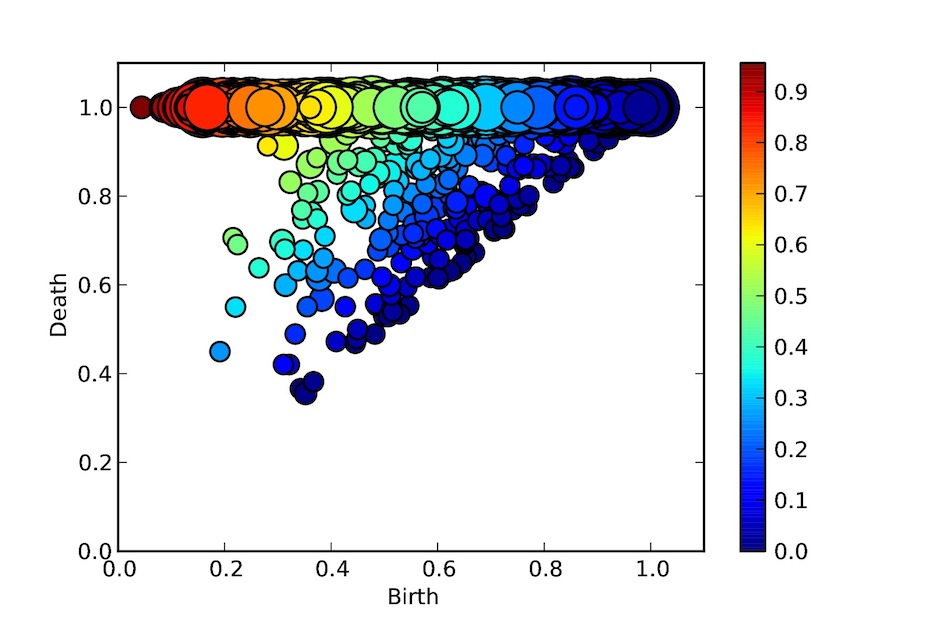}}
\subfigure[]{\includegraphics[width=0.32\textwidth]{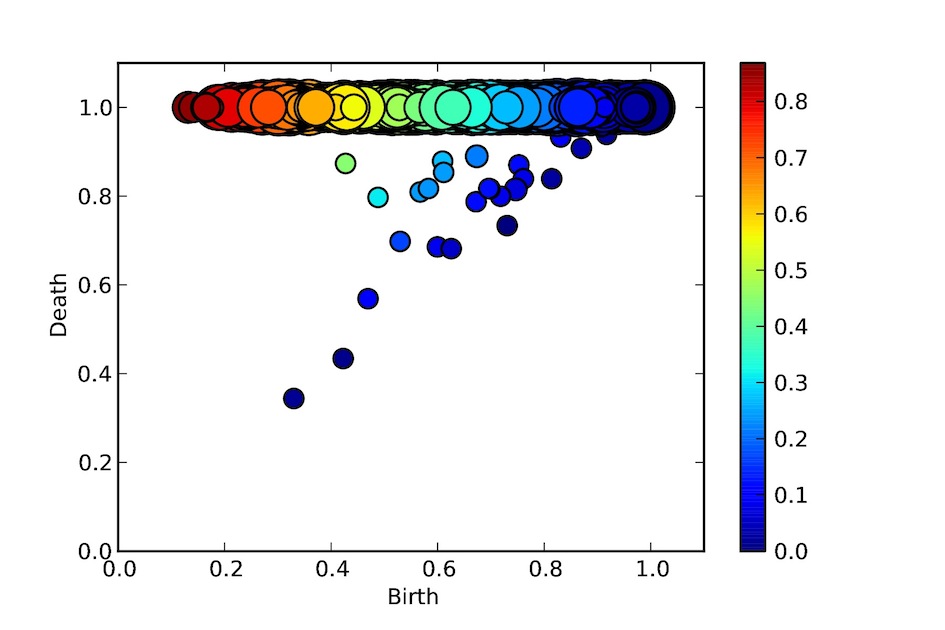}}
\caption{{\bf Summary of $H_1$ persistent homology results for the US airways passenger network for 2006 (Class I).}}\label{fig::US2006_random_comparison}
\end{figure*}

\begin{figure*}[h]
\centering
\subfigure[]{\includegraphics[width=0.32\textwidth]{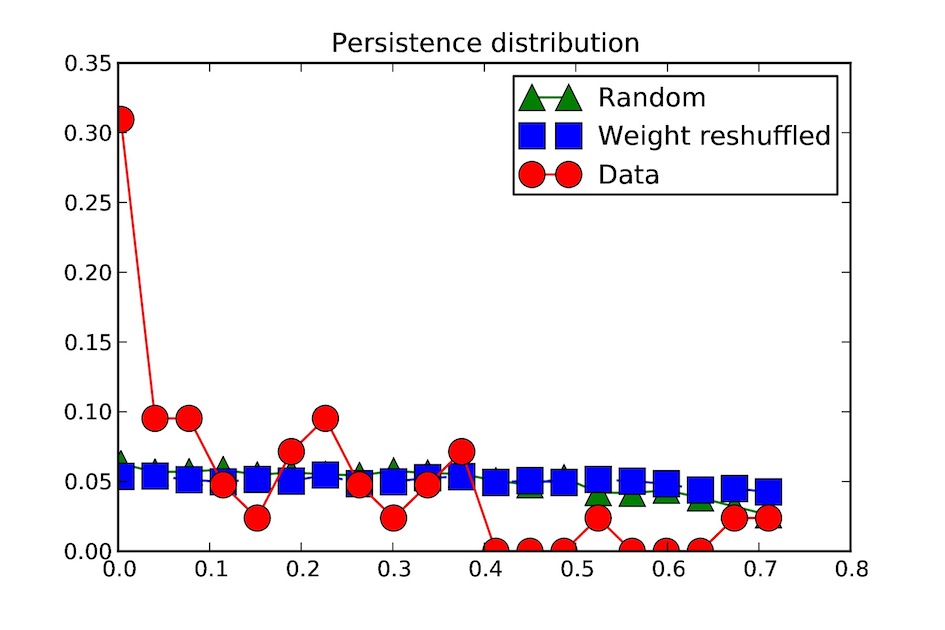}}
\subfigure[]{\includegraphics[width=0.32\textwidth]{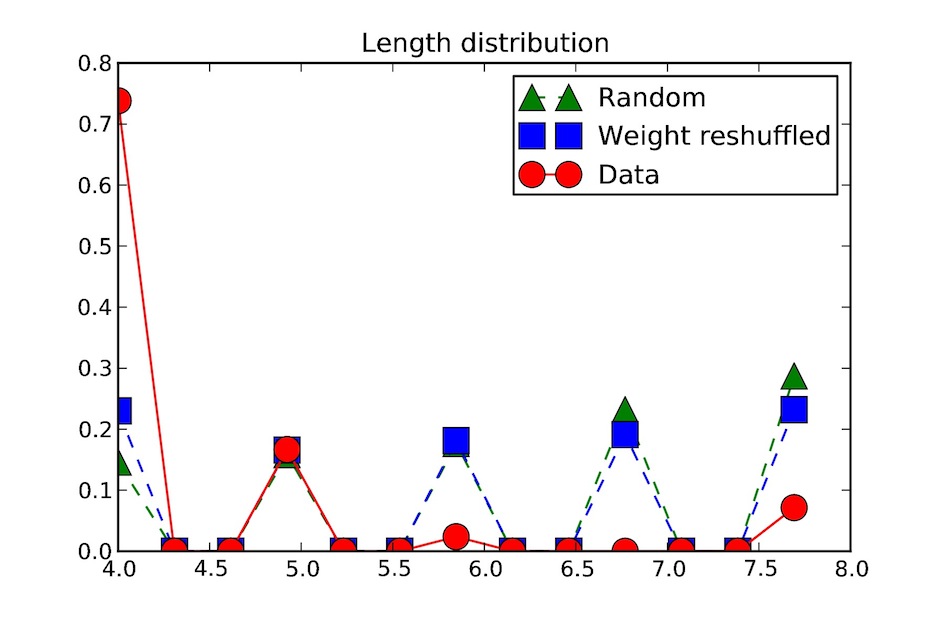}}
\subfigure[]{\includegraphics[width=0.32\textwidth]{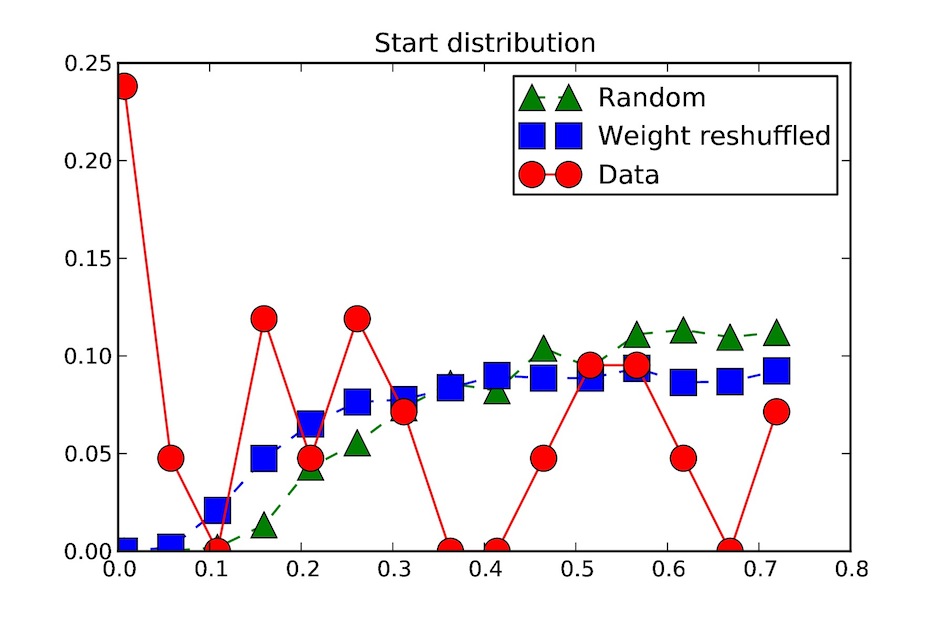}}
\\
\subfigure[]{\includegraphics[width=0.32\textwidth]{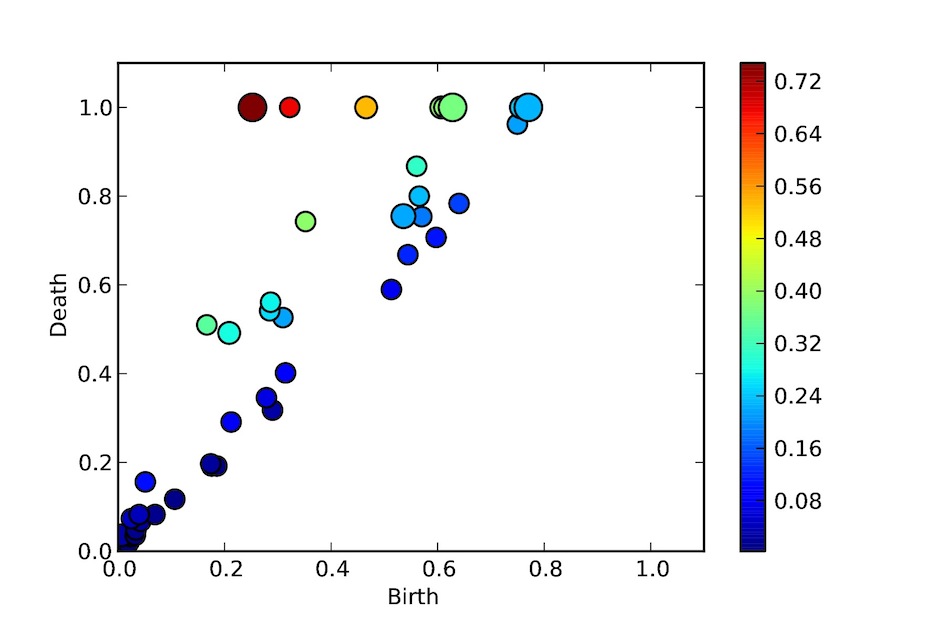}}
\subfigure[]{\includegraphics[width=0.32\textwidth]{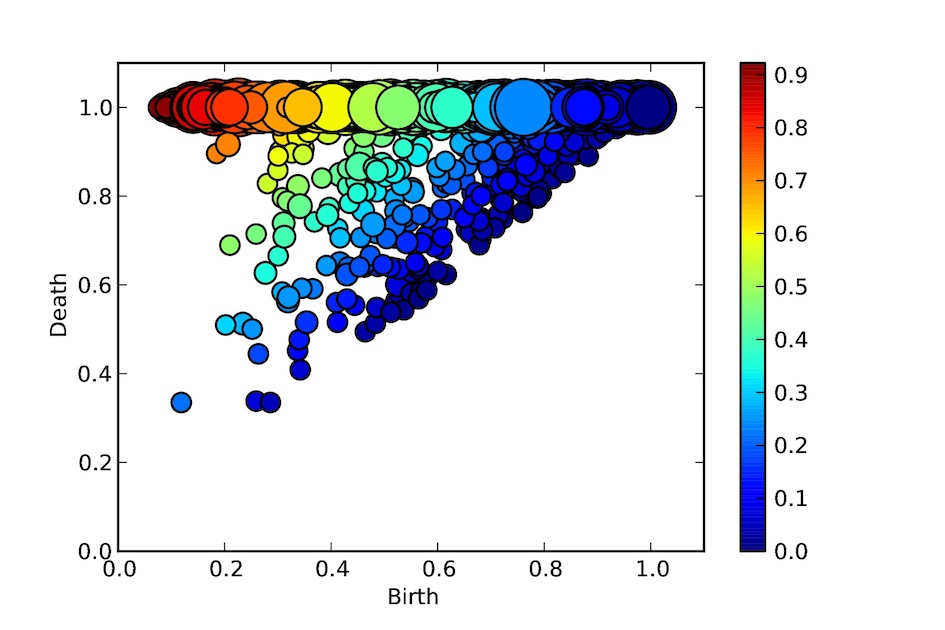}}
\subfigure[]{\includegraphics[width=0.32\textwidth]{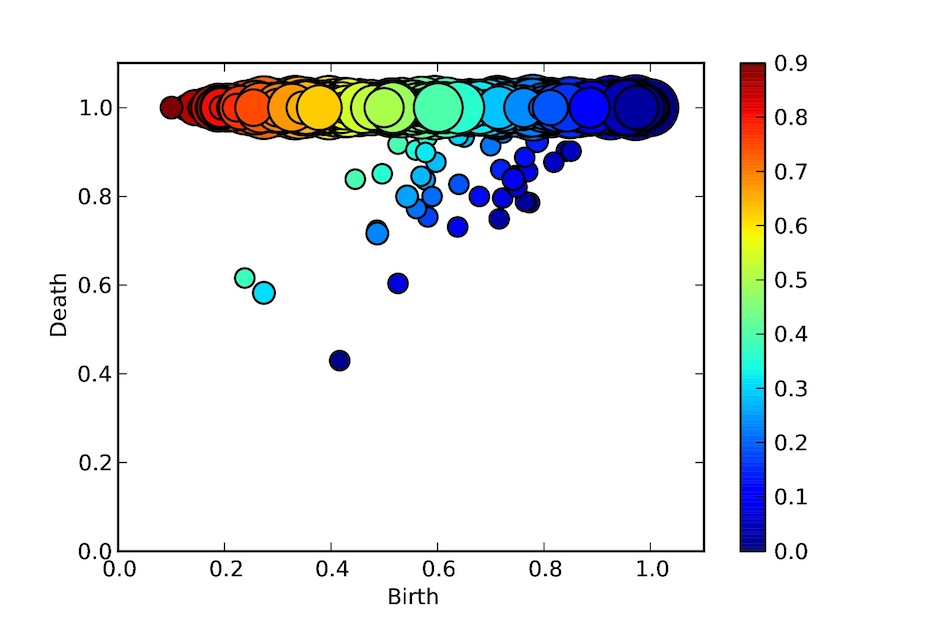}}
\caption{{\bf Summary of $H_1$ persistent homology results for the US airways passenger network for 2011(Class I).}}\label{fig::US2011_random_comparison}
\end{figure*}

\begin{figure*}[h]
\centering
\subfigure[]{\includegraphics[width=0.32\textwidth]{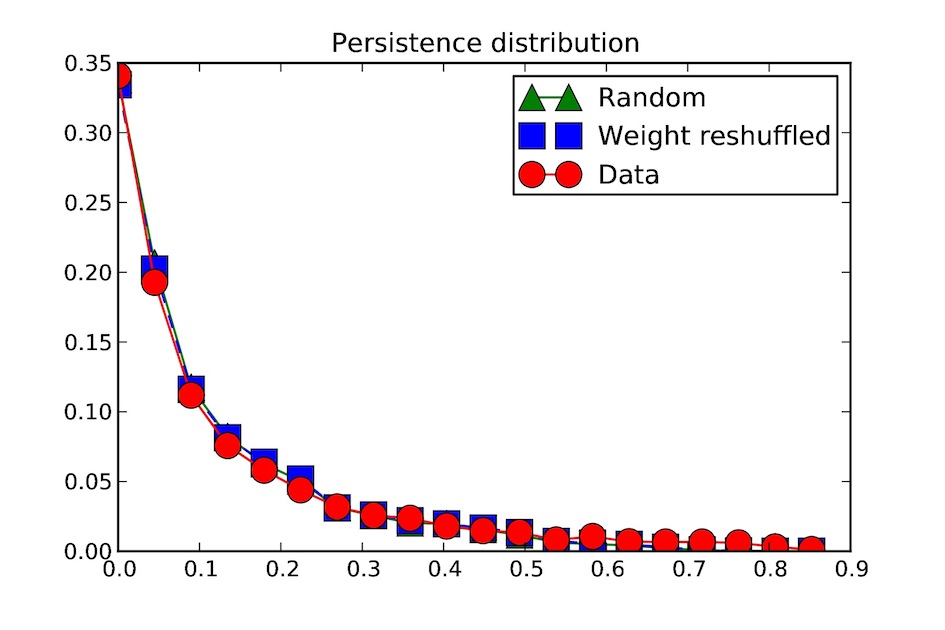}}
\subfigure[]{\includegraphics[width=0.32\textwidth]{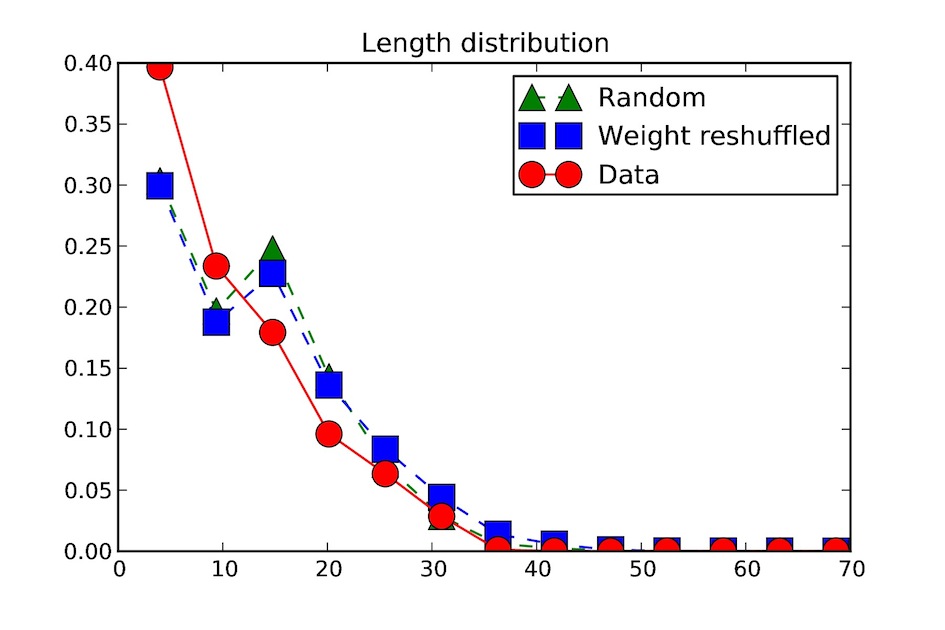}}
\subfigure[]{\includegraphics[width=0.32\textwidth]{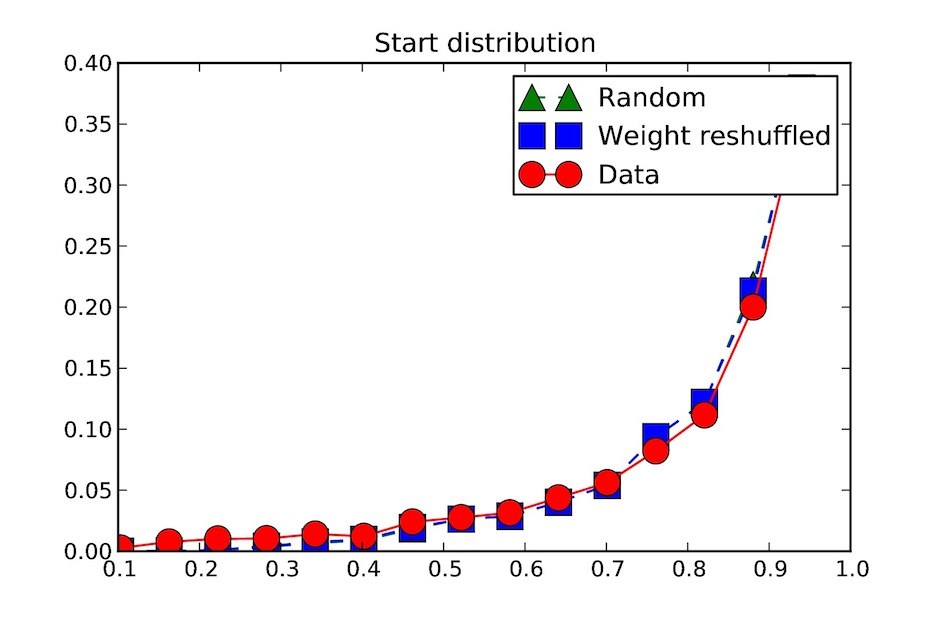}}
\\
\subfigure[]{\includegraphics[width=0.32\textwidth]{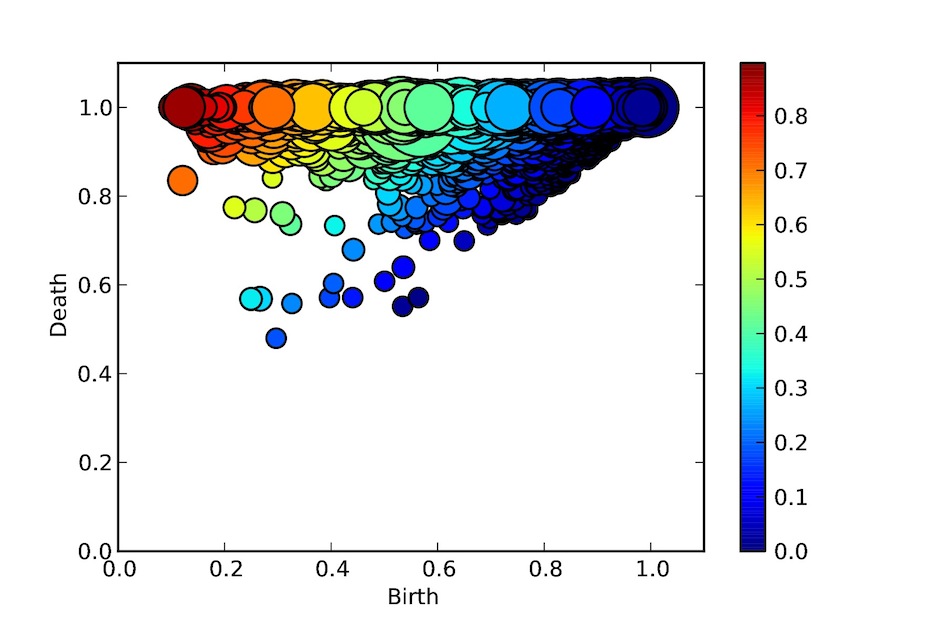}}
\subfigure[]{\includegraphics[width=0.32\textwidth]{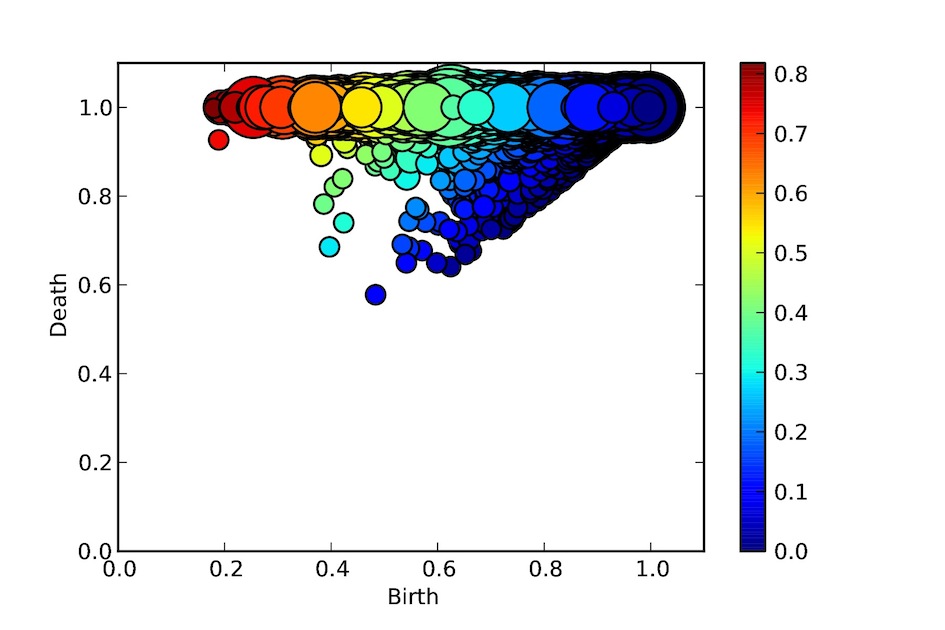}}
\subfigure[]{\includegraphics[width=0.32\textwidth]{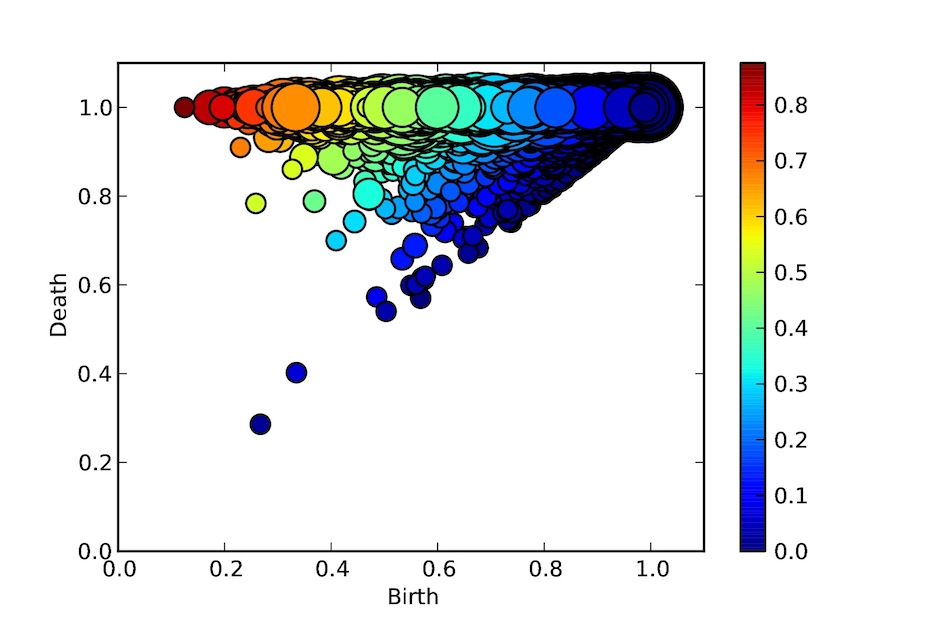}}
\caption{{\bf Summary of $H_1$ persistent homology results for the online messages network of \cite{opsahl1} (Class I).}}\label{fig::messages_random_comparison}
\end{figure*}

\begin{figure*}[h]
\centering
\subfigure[]{\includegraphics[width=0.32\textwidth]{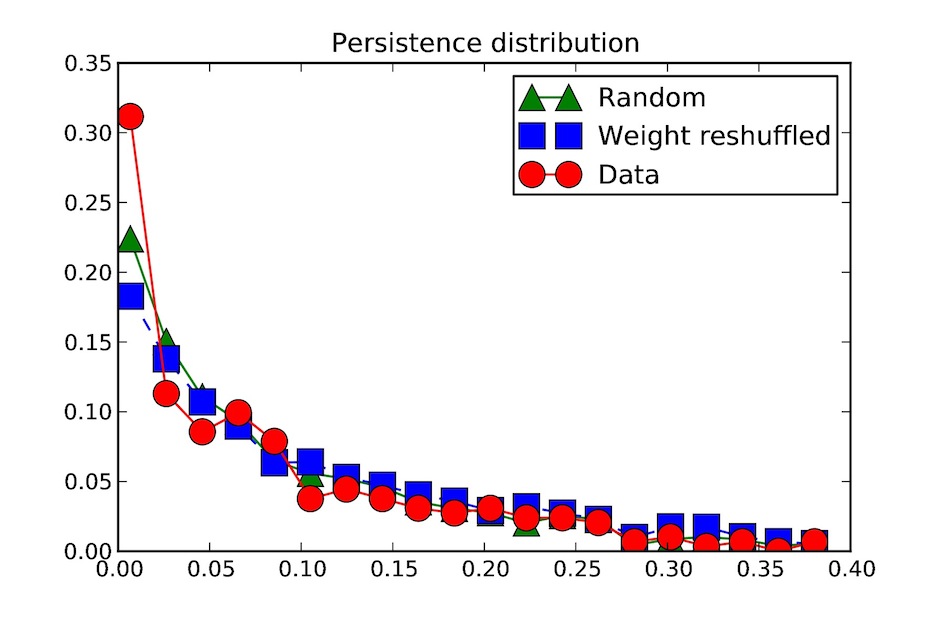}}
\subfigure[]{\includegraphics[width=0.32\textwidth]{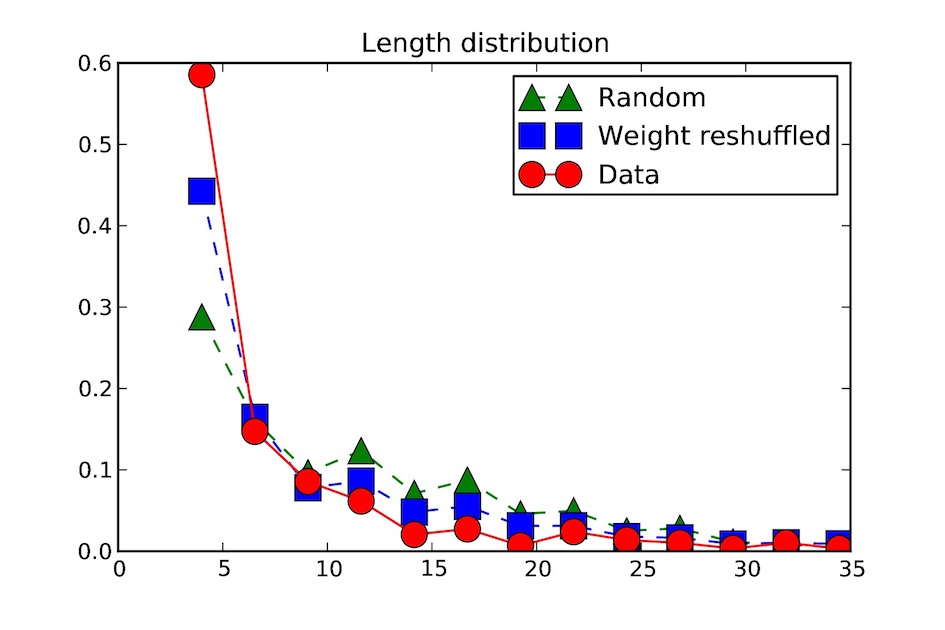}}
\subfigure[]{\includegraphics[width=0.32\textwidth]{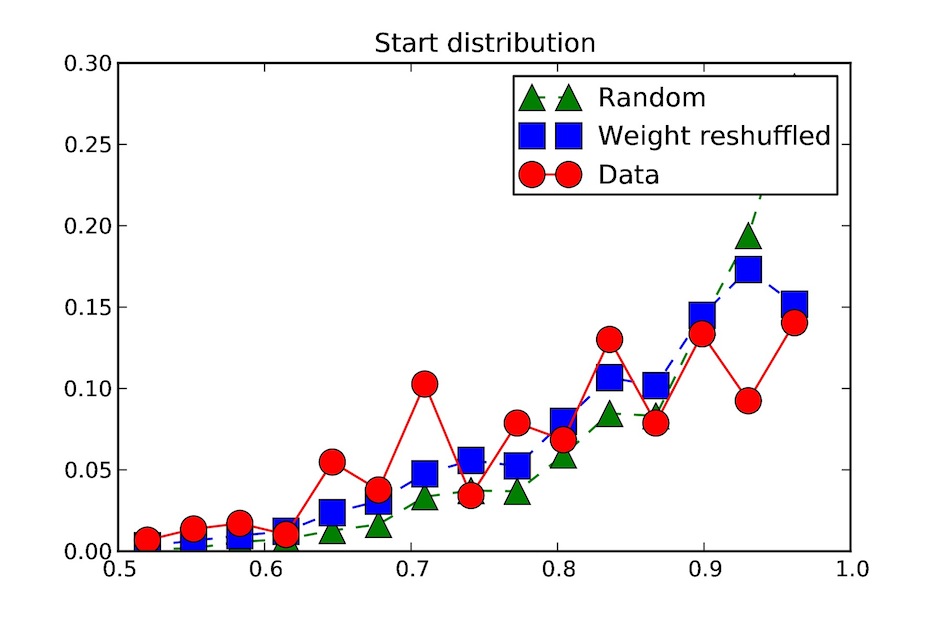}}
\\
\subfigure[]{\includegraphics[width=0.32\textwidth]{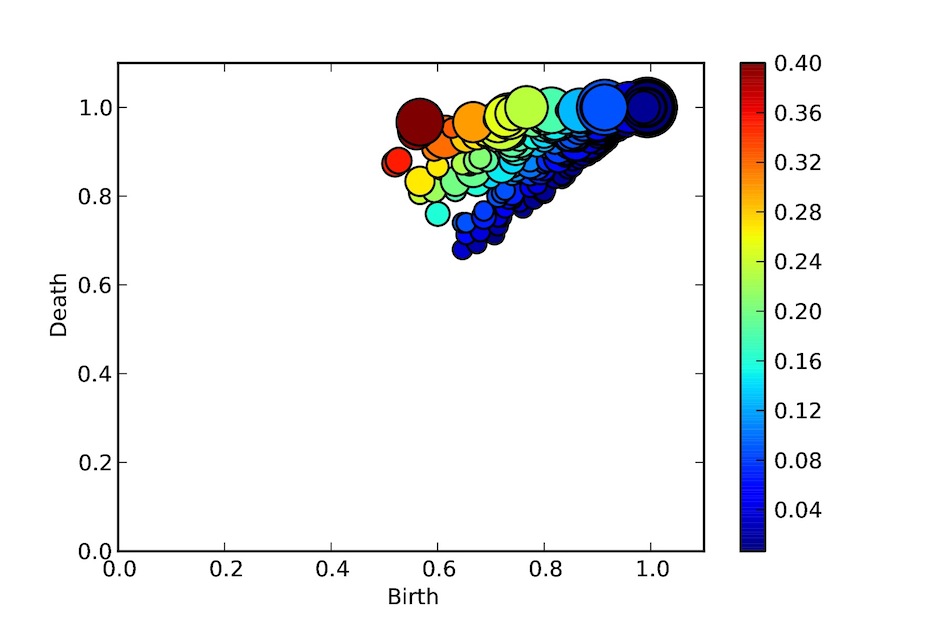}}
\subfigure[]{\includegraphics[width=0.32\textwidth]{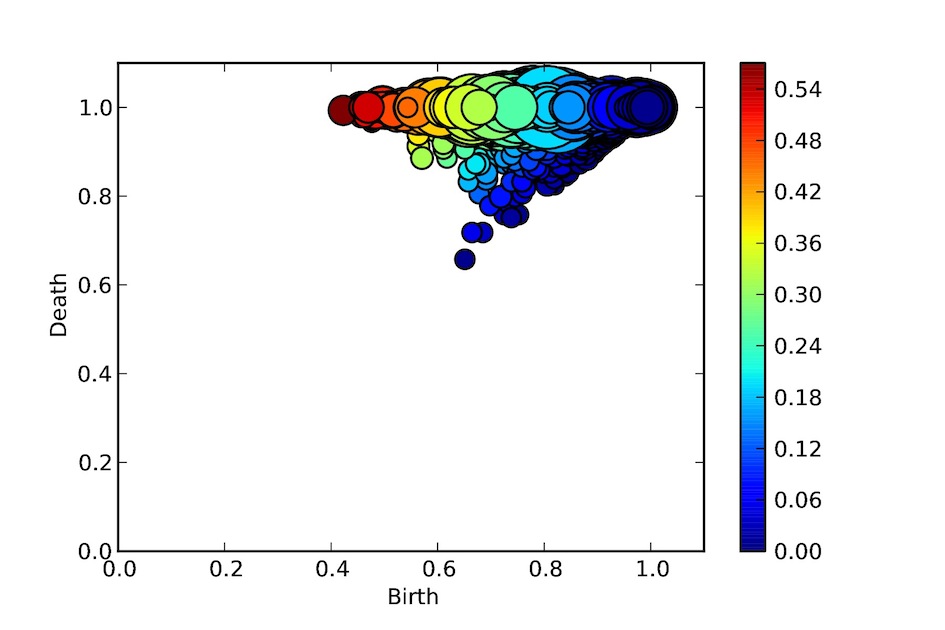}}
\subfigure[]{\includegraphics[width=0.32\textwidth]{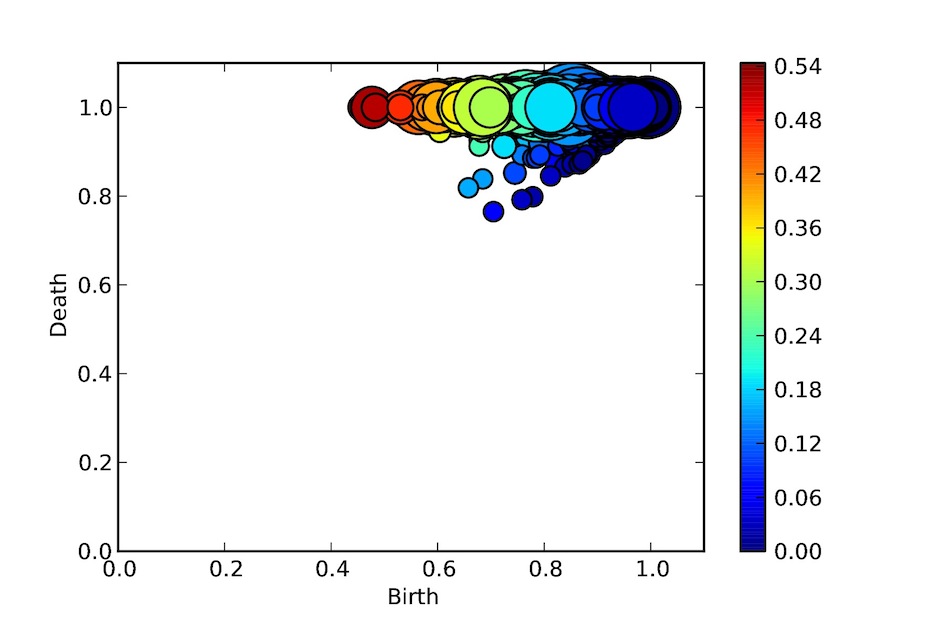}}
\caption{ {\bf Summary of $H_1$ persistent homology results for the day 1 face-to-face contact duration network of children of \cite{juliette} (Class II).}} \label{fig::school_day1_random_comparison}
\end{figure*}

\begin{figure*}[h]
\centering
\subfigure[]{\includegraphics[width=0.32\textwidth]{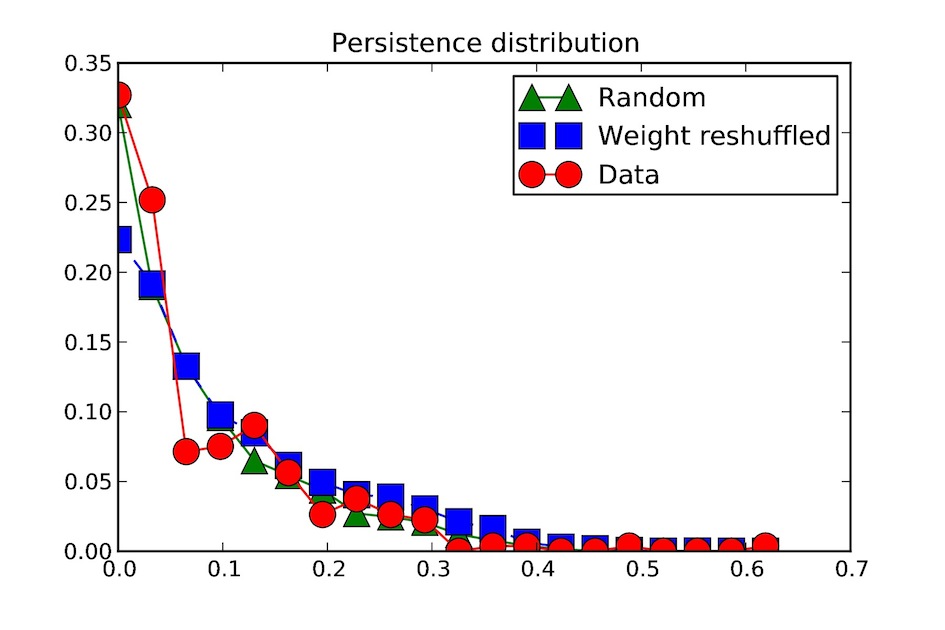}}
\subfigure[]{\includegraphics[width=0.32\textwidth]{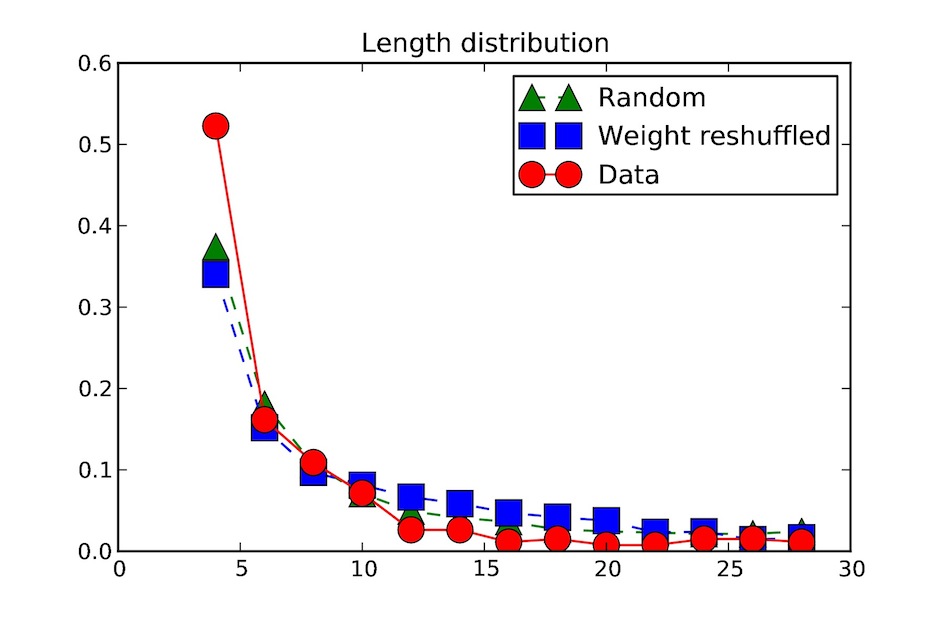}}
\subfigure[]{\includegraphics[width=0.32\textwidth]{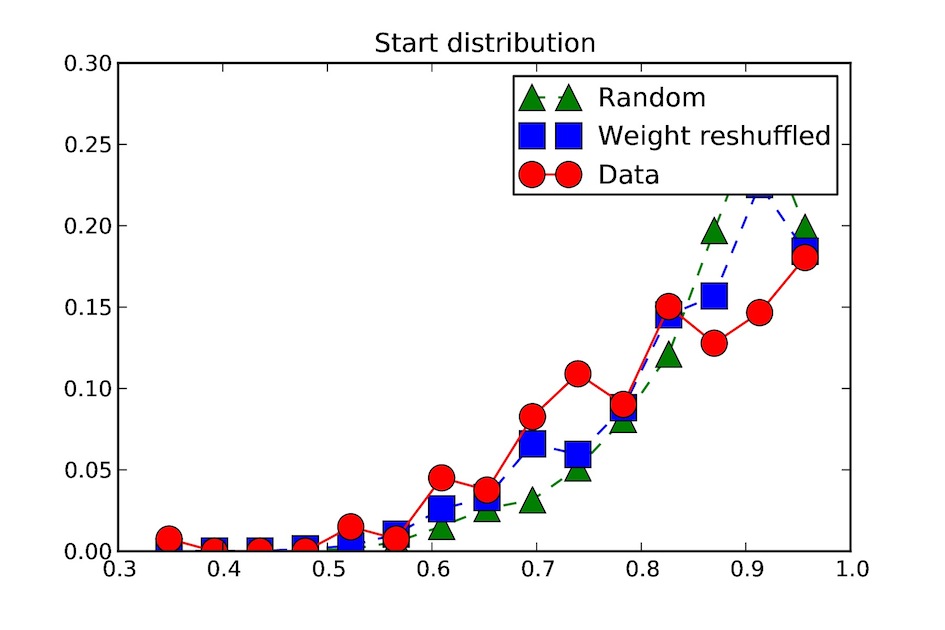}}
\\
\subfigure[]{\includegraphics[width=0.32\textwidth]{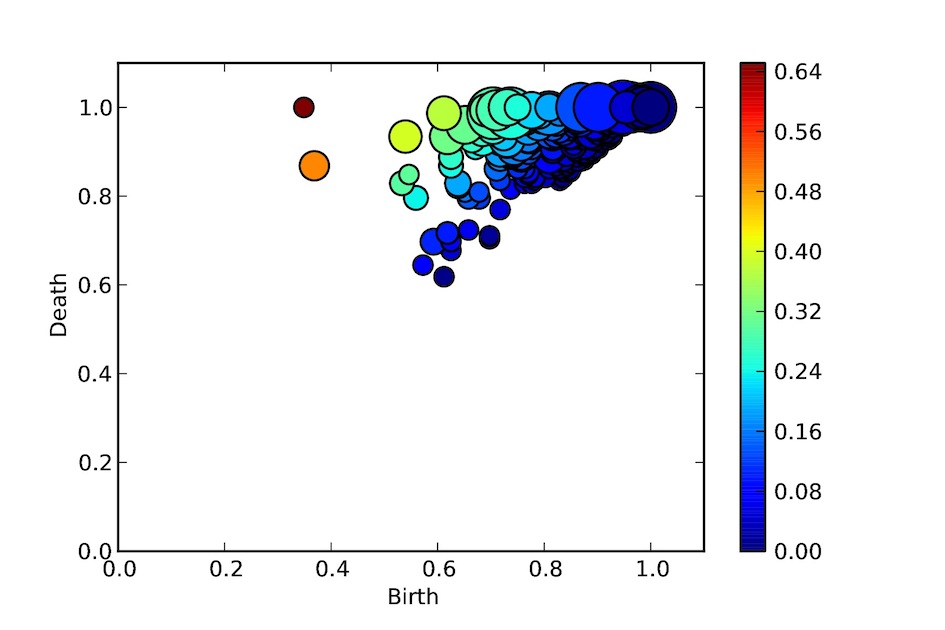}}
\subfigure[]{\includegraphics[width=0.32\textwidth]{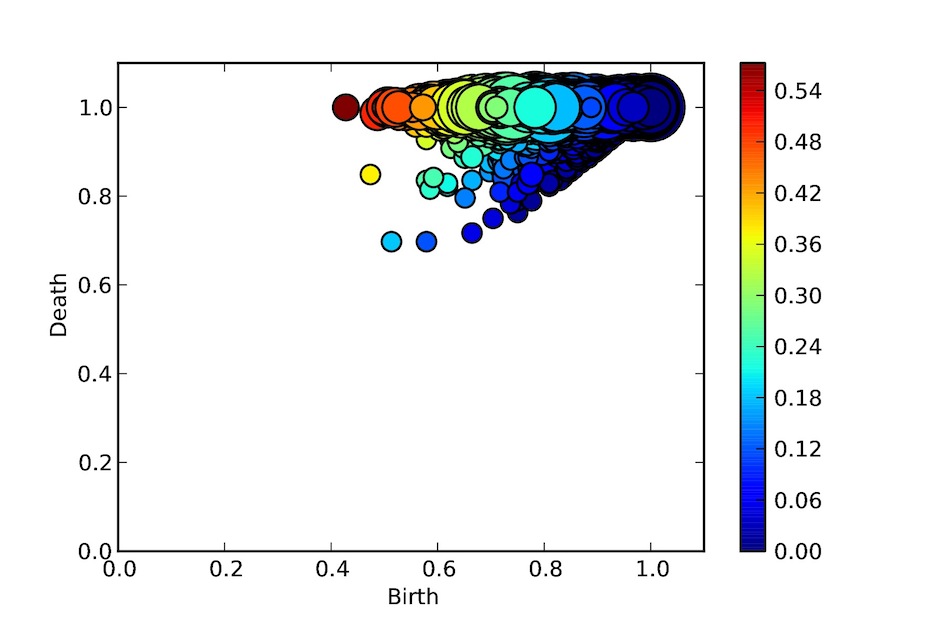}}
\subfigure[]{\includegraphics[width=0.32\textwidth]{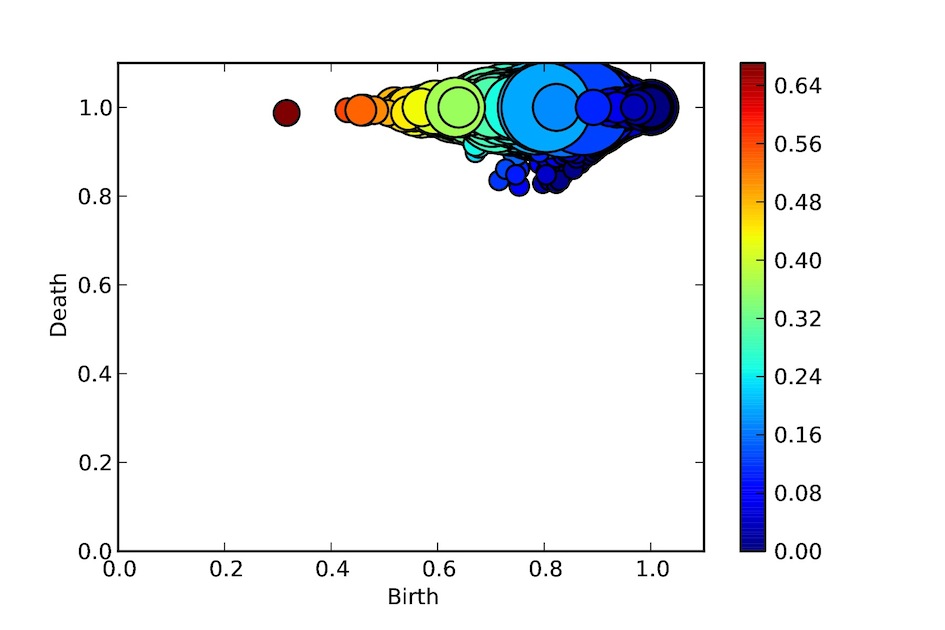}}
\caption{{\bf Summary of $H_1$ persistent homology results for the day 2 face-to-face contact  duration network of children of \cite{juliette} (Class II)}}\label{fig::school_day2_random_comparison}
\end{figure*}

\begin{figure*}[h]
\centering
\subfigure[]{\includegraphics[width=0.31\textwidth]{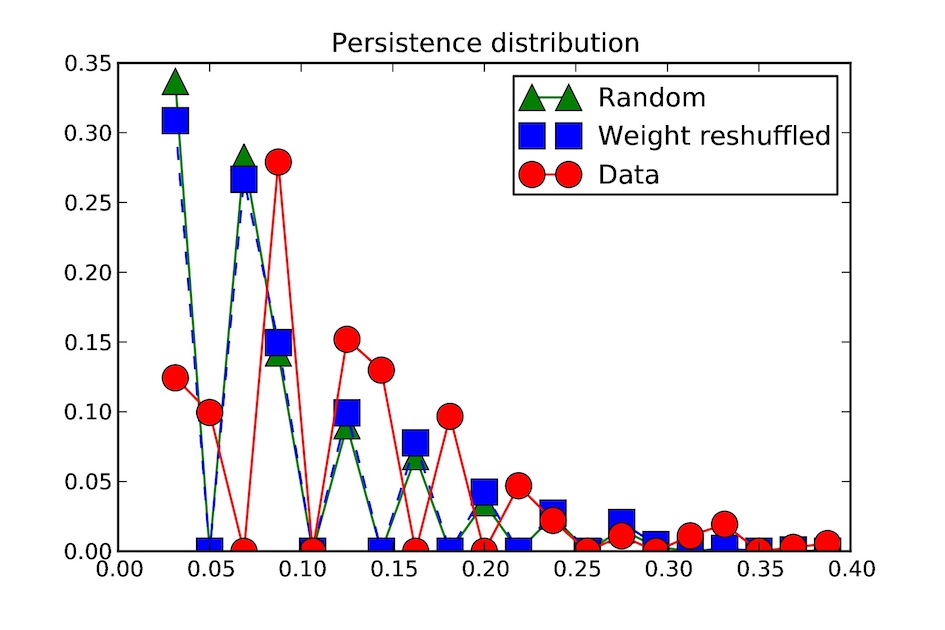}}
\subfigure[]{\includegraphics[width=0.31\textwidth]{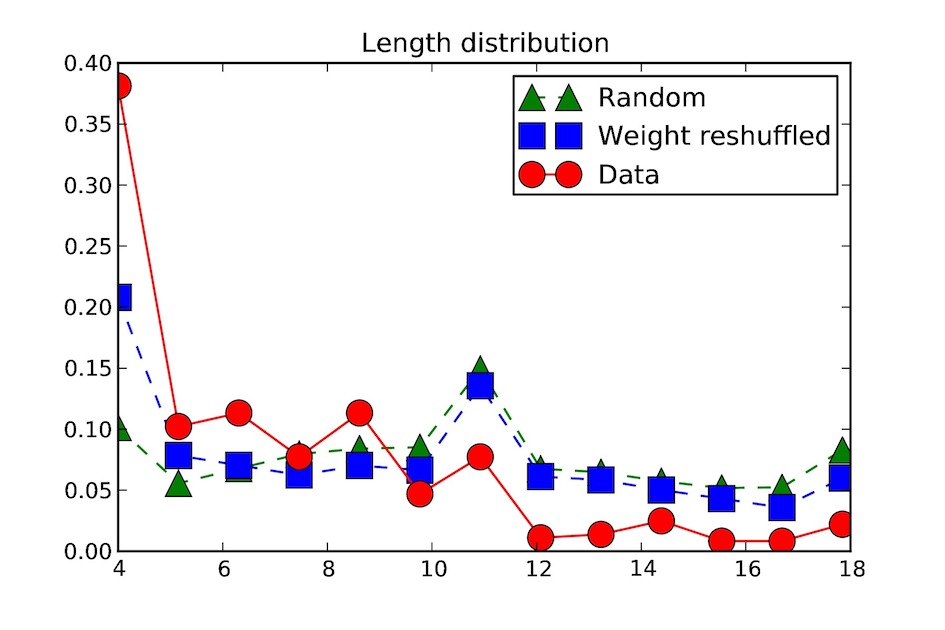}}
\subfigure[]{\includegraphics[width=0.31\textwidth]{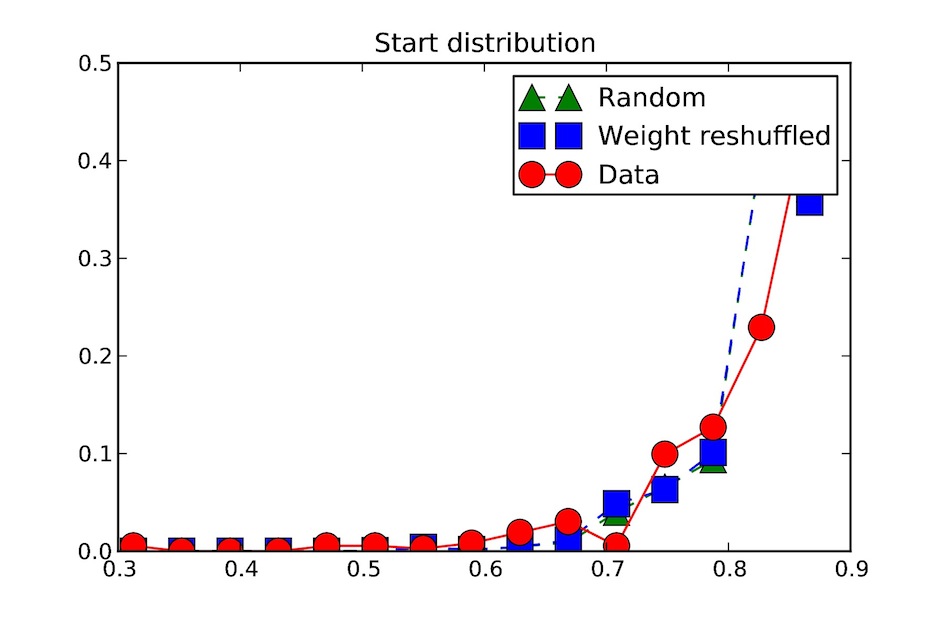}}
\\
\subfigure[]{\includegraphics[width=0.32\textwidth]{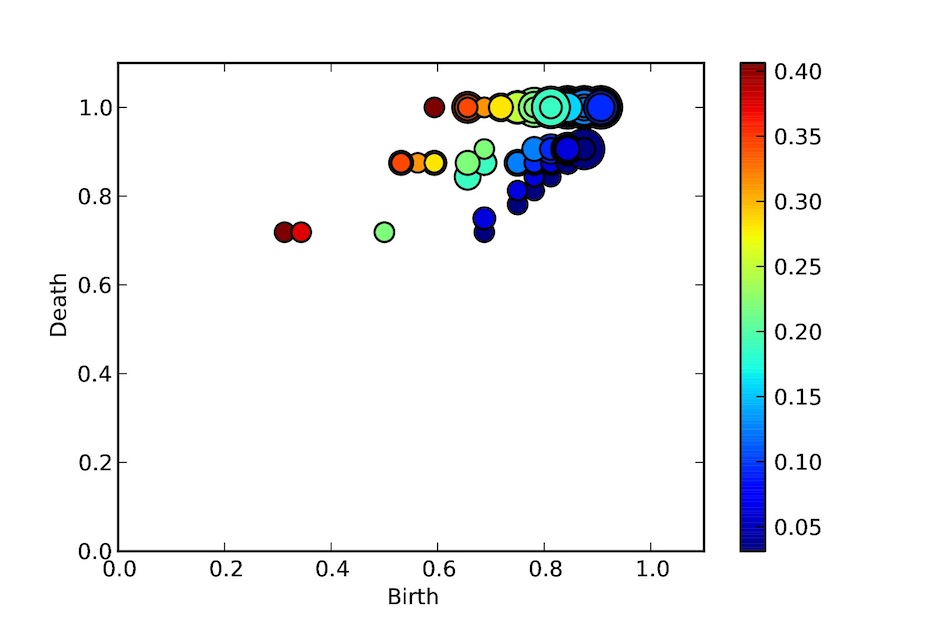}}
\subfigure[]{\includegraphics[width=0.32\textwidth]{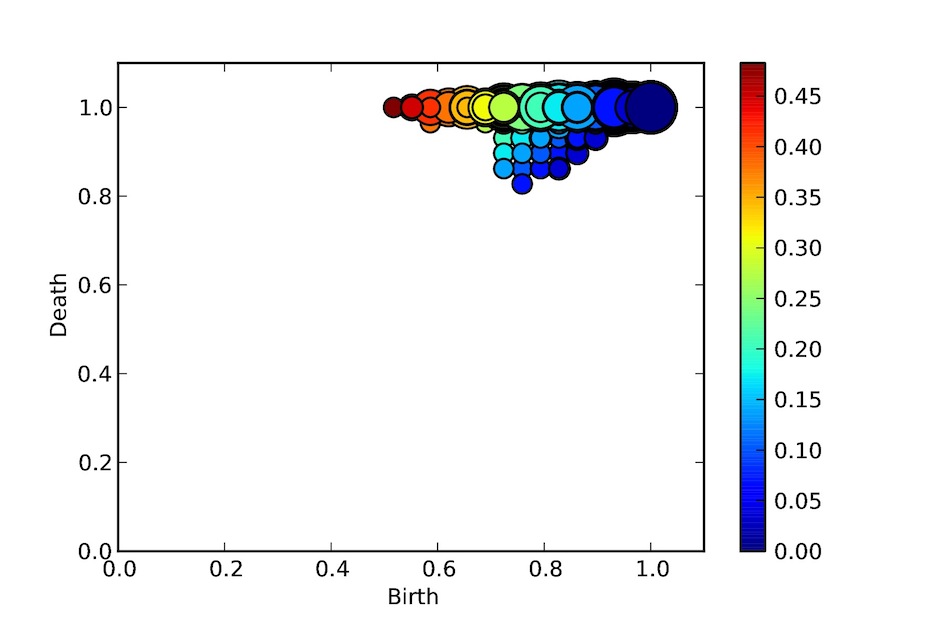}}
\subfigure[]{\includegraphics[width=0.32\textwidth]{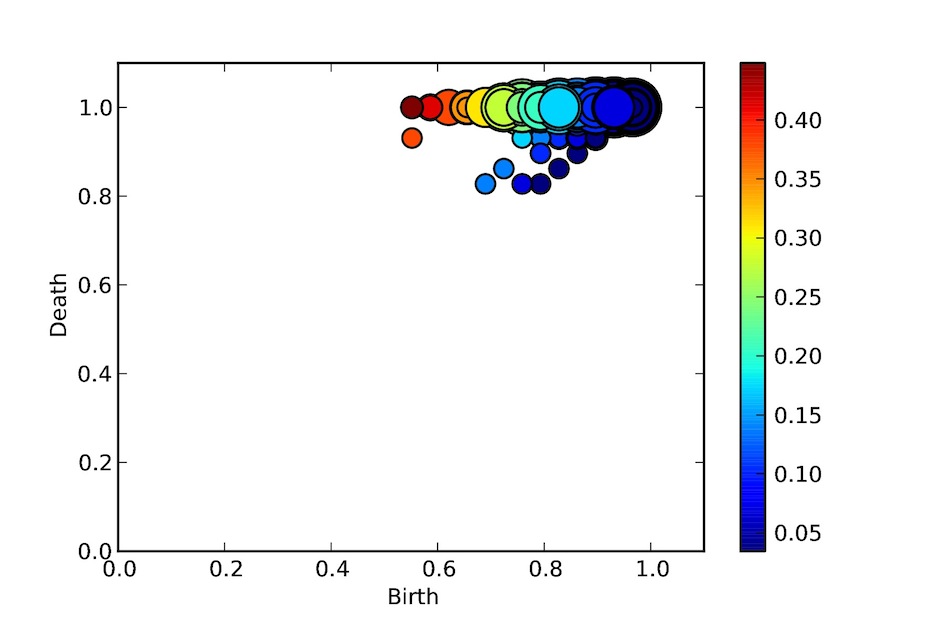}}
\caption{{\bf Summary of $H_1$ persistent homology results for the neural network of the {\it C. elegans} (Class II).}}\label{fig::celegans_random_comparison}
\end{figure*}

\begin{figure*}[h]
\centering
\subfigure[]{\includegraphics[width=0.32\textwidth]{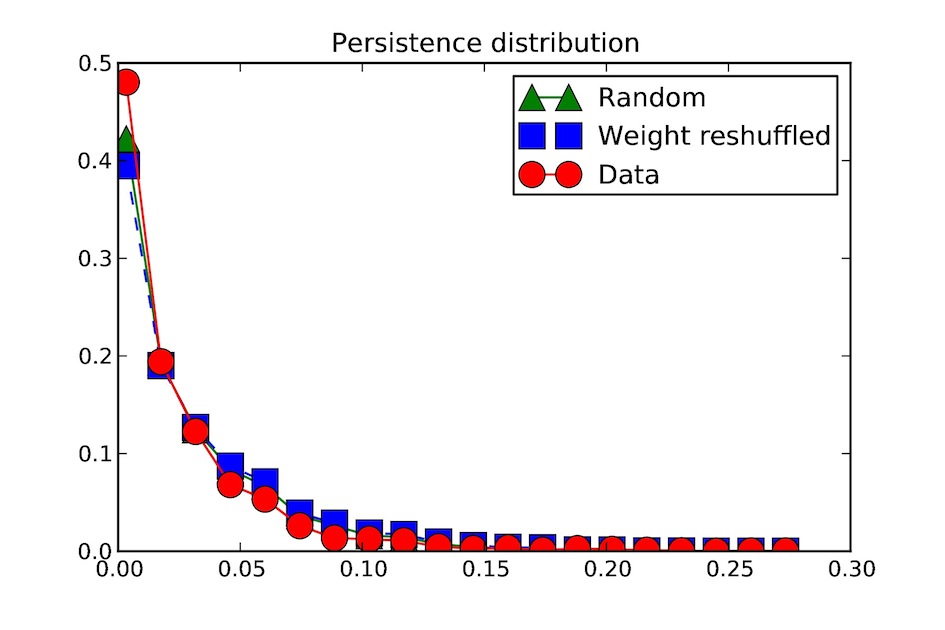}}
\subfigure[]{\includegraphics[width=0.32\textwidth]{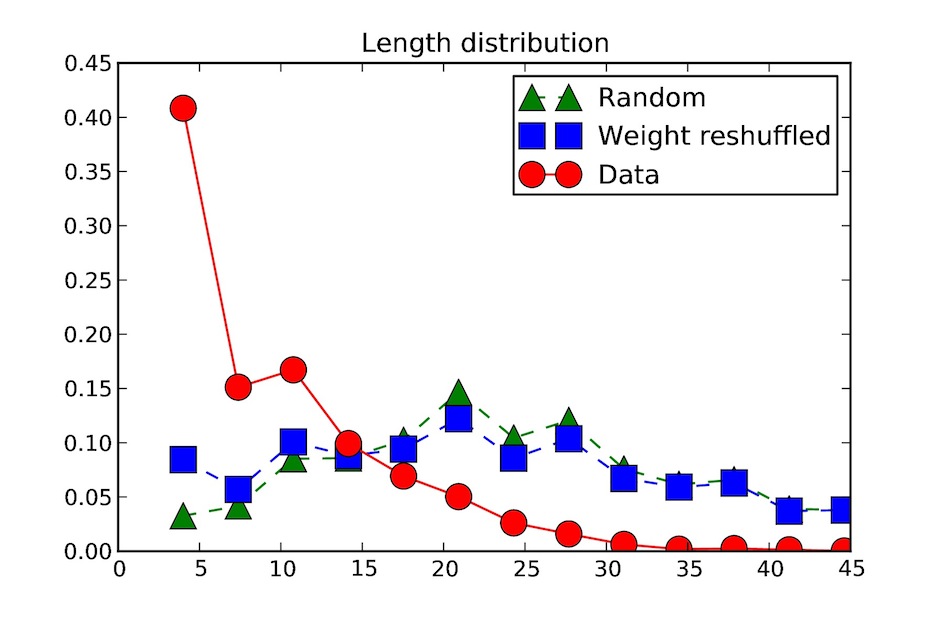}}
\subfigure[]{\includegraphics[width=0.32\textwidth]{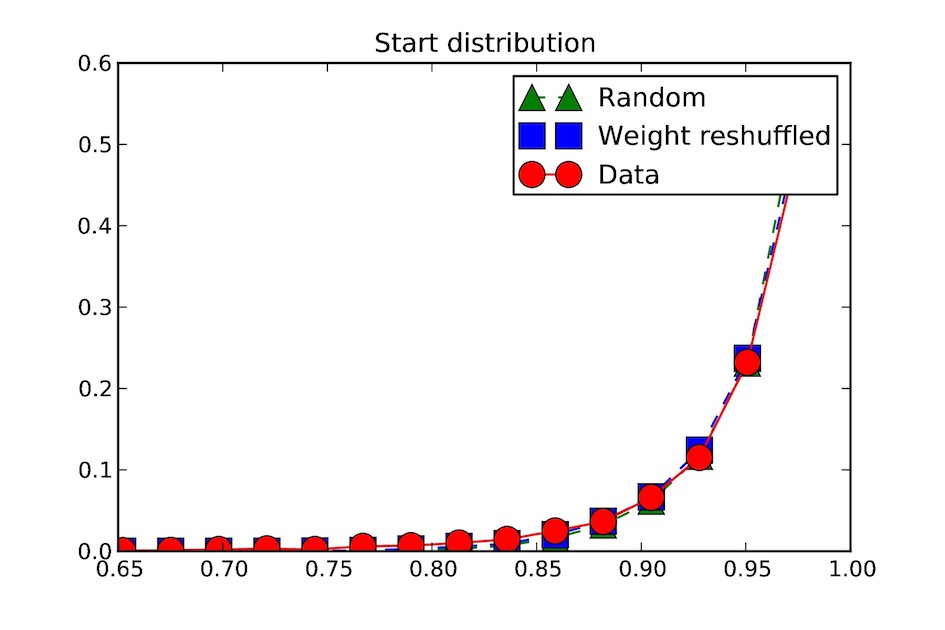}}
\\
\subfigure[]{\includegraphics[width=0.32\textwidth]{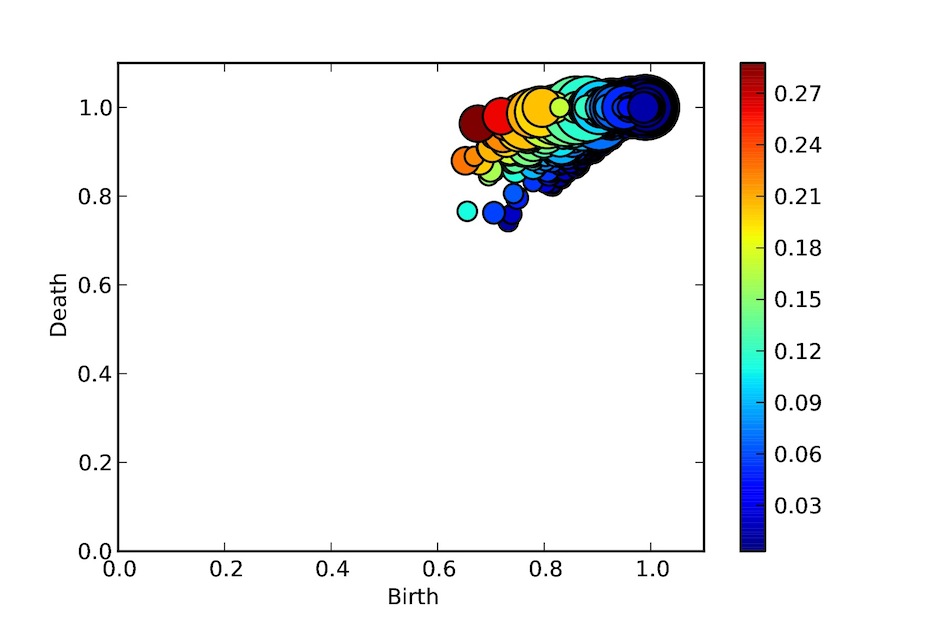}}
\subfigure[]{\includegraphics[width=0.32\textwidth]{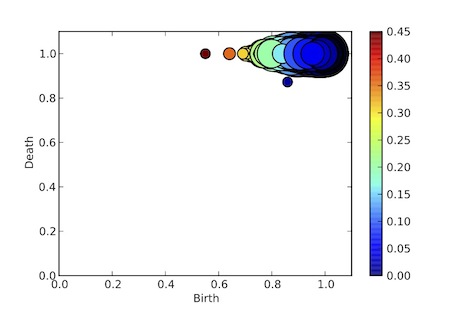}}
\subfigure[]{\includegraphics[width=0.32\textwidth]{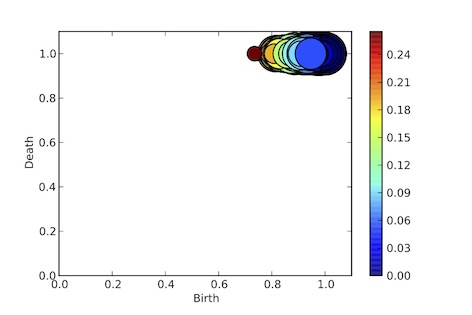}}
\caption{{\bf Summary of $H_1$ persistent homology results for a network of mentions and retweets of a part of the Twitter network (Class II).}}\label{fig::twitter_random_comparison}
\end{figure*}

\begin{figure*}[h]
\centering
\subfigure[]{\includegraphics[width=0.32\textwidth]{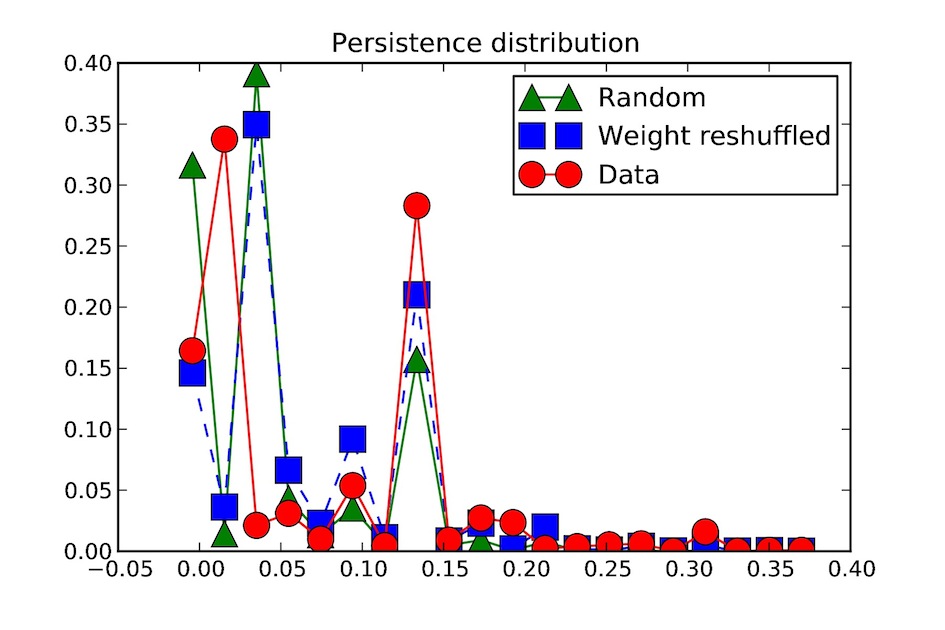}}
\subfigure[]{\includegraphics[width=0.32\textwidth]{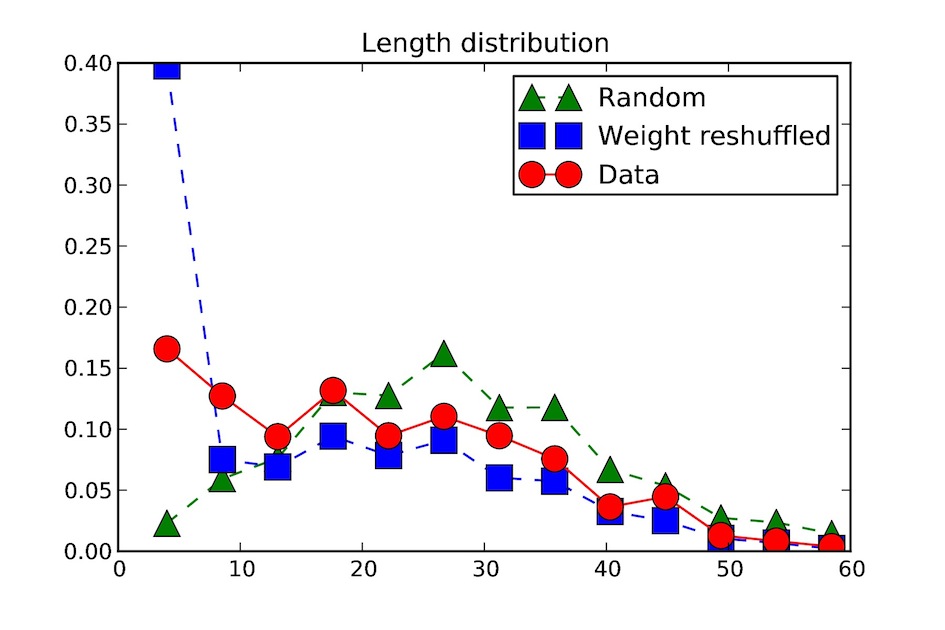}}
\subfigure[]{\includegraphics[width=0.32\textwidth]{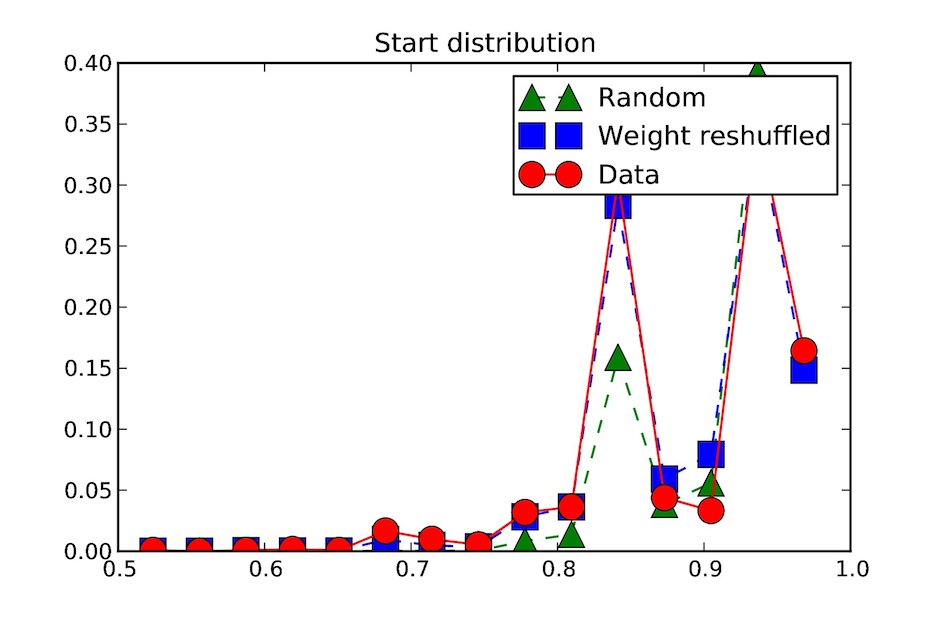}}
\\
\subfigure[]{\includegraphics[width=0.32\textwidth]{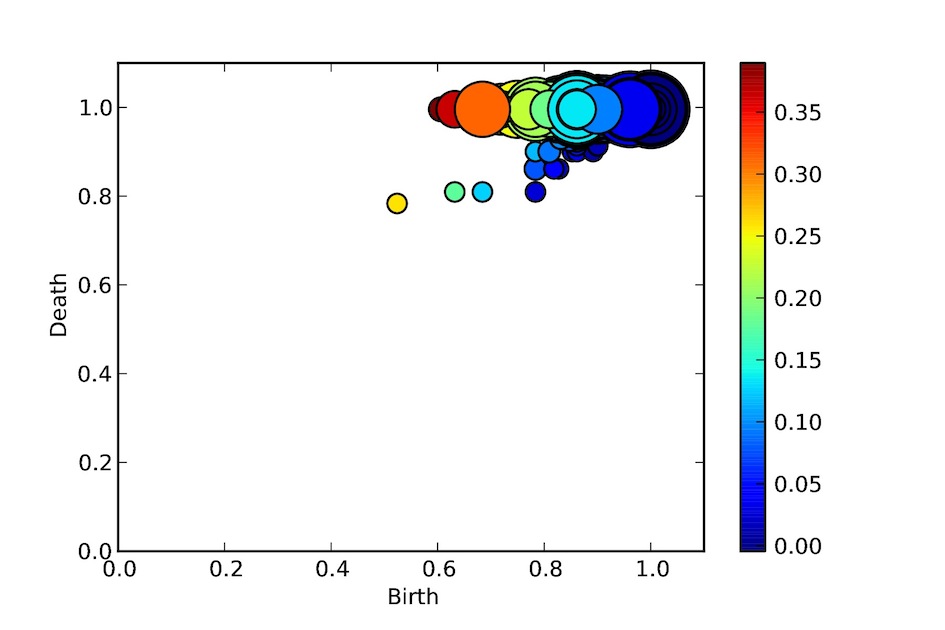}}
\subfigure[]{\includegraphics[width=0.32\textwidth]{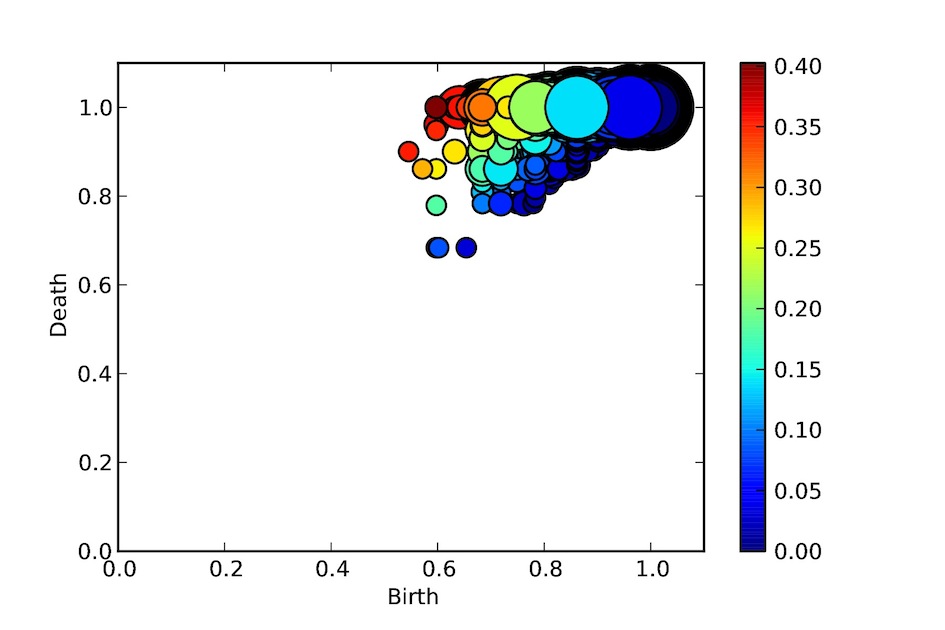}}
\subfigure[]{\includegraphics[width=0.32\textwidth]{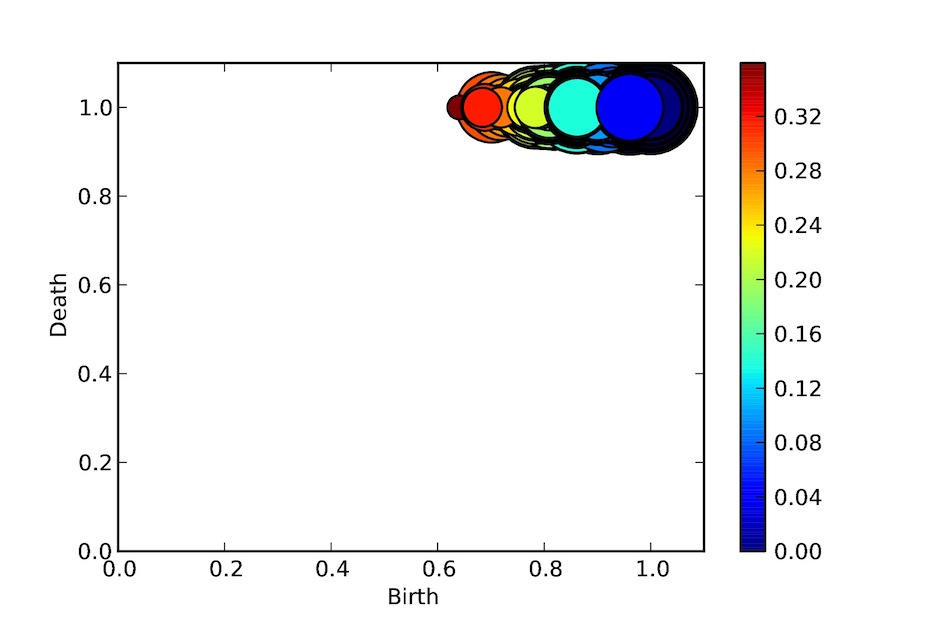}}
\caption{{\bf Summary of $H_1$ persistent homology results for the Hep-th arxiv.....(Class II)}}\label{fig::hepp_random_comparison}
\end{figure*}

\begin{figure*}[h]
\centering
\subfigure[]{\includegraphics[width=0.32\textwidth]{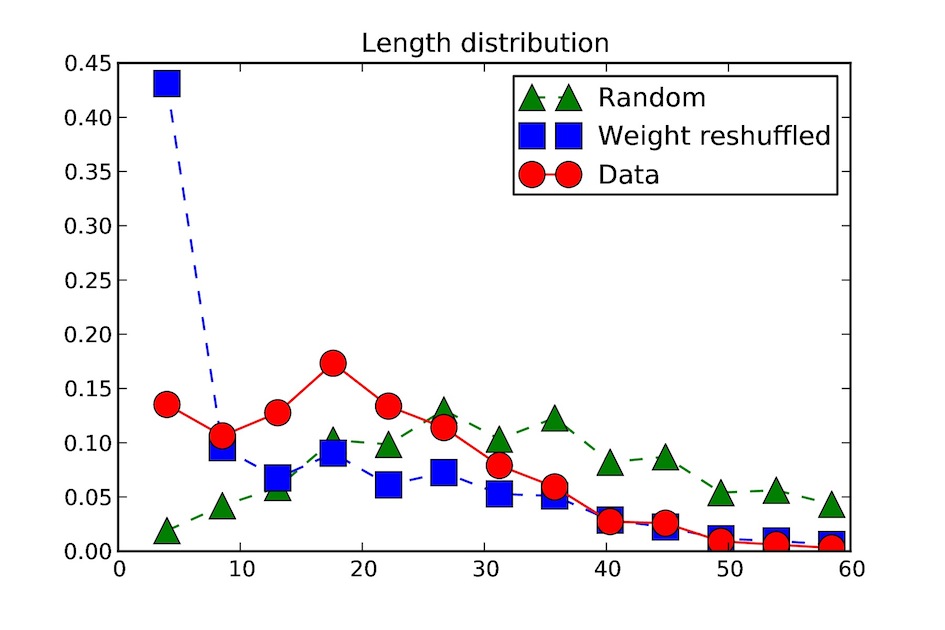}}	
\subfigure[]{\includegraphics[width=0.32\textwidth]{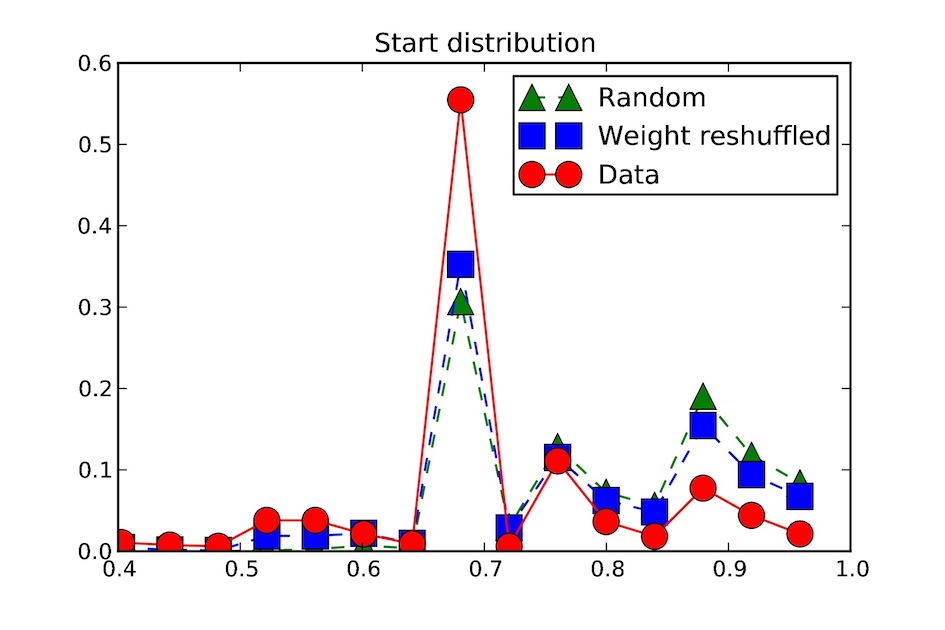}} 
\subfigure[]{\includegraphics[width=0.32\textwidth]{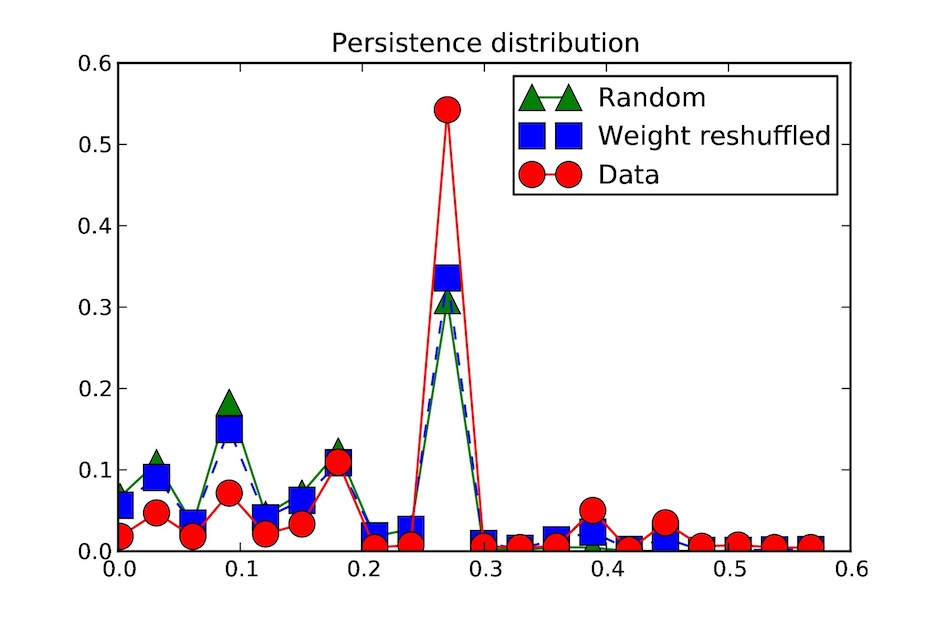}}
\\
\subfigure[]{\includegraphics[width=0.32\textwidth]{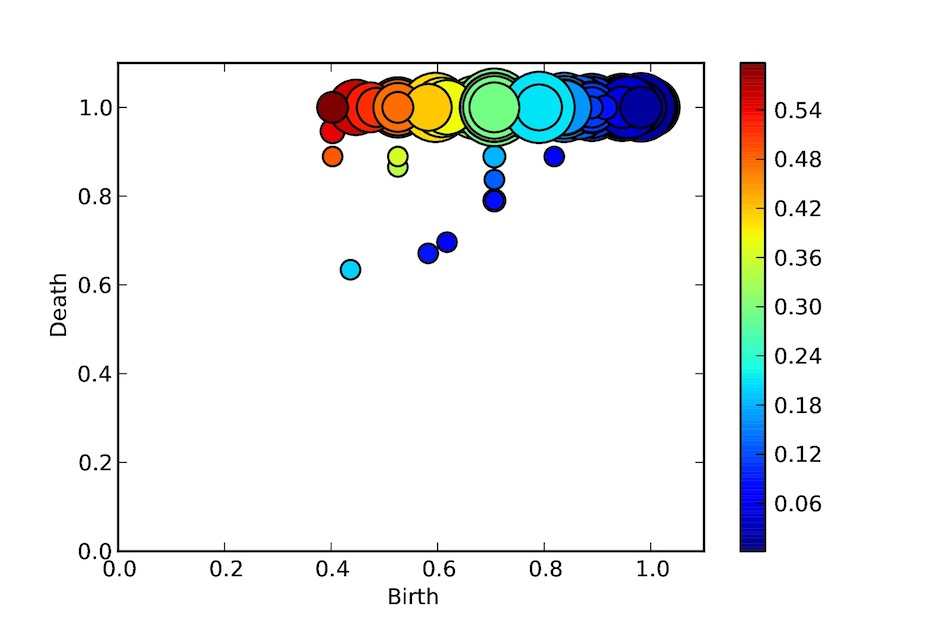}}
\subfigure[]{\includegraphics[width=0.32\textwidth]{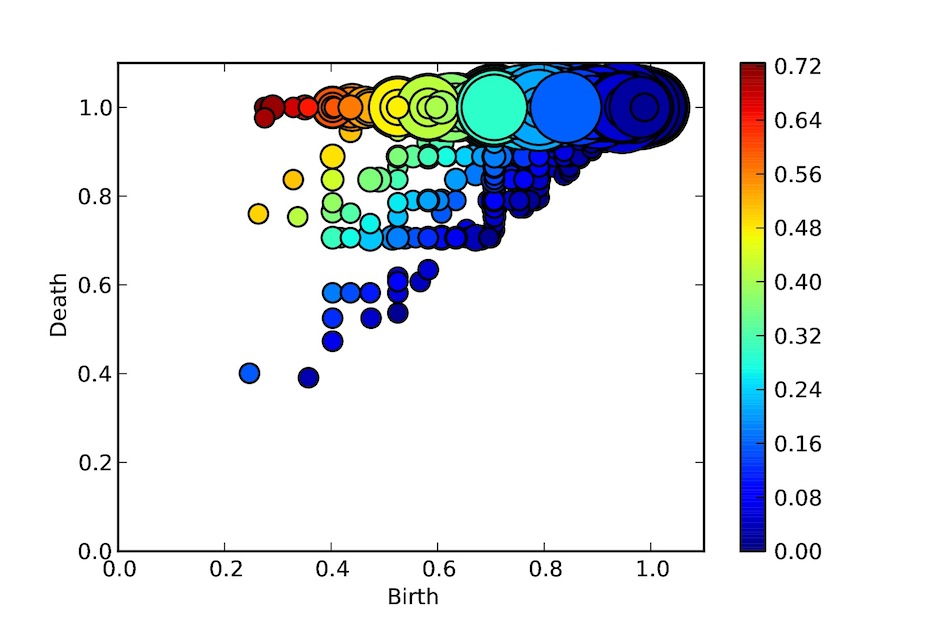}}
\subfigure[]{\includegraphics[width=0.32\textwidth]{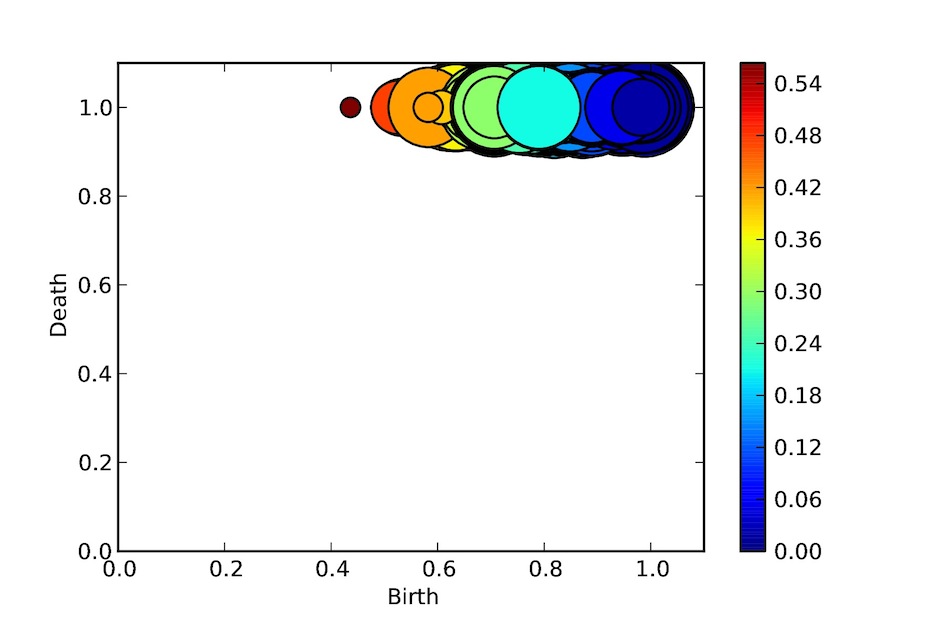}}
\caption{{\bf Summary of $H_1$ persistent homology results for the cond-mat (Class II).}}\label{fig::condmat_comparison}
\end{figure*}

\begin{figure*}
\centering
\subfigure[]{\includegraphics[width=0.32\textwidth]{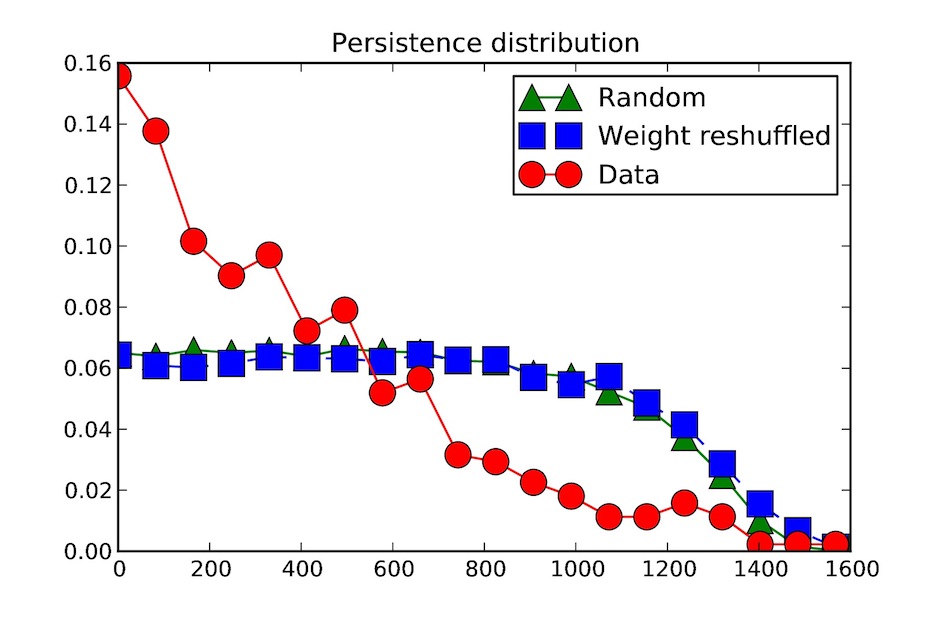}}
\subfigure[]{\includegraphics[width=0.32\textwidth]{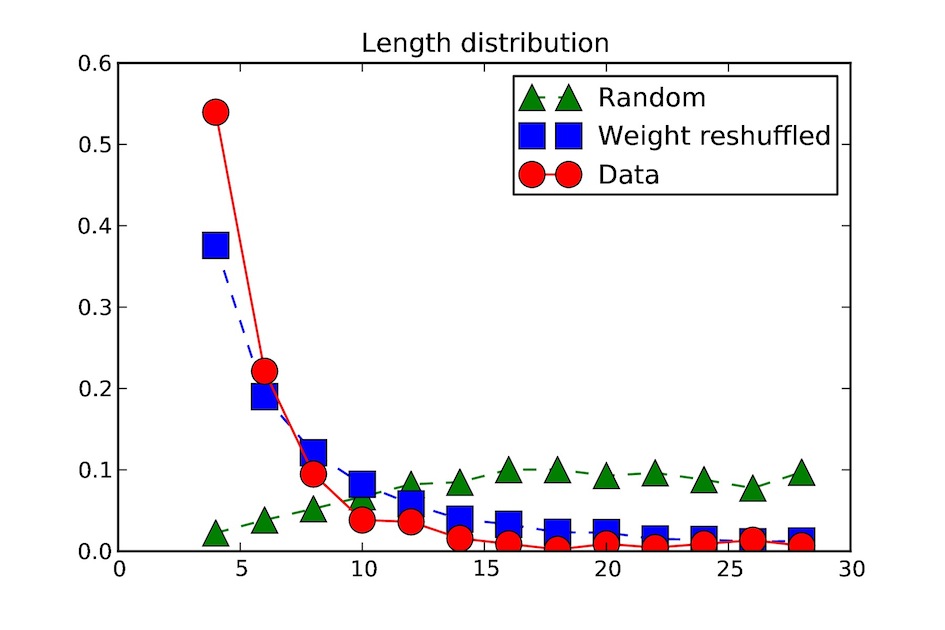}}
\subfigure[]{\includegraphics[width=0.32\textwidth]{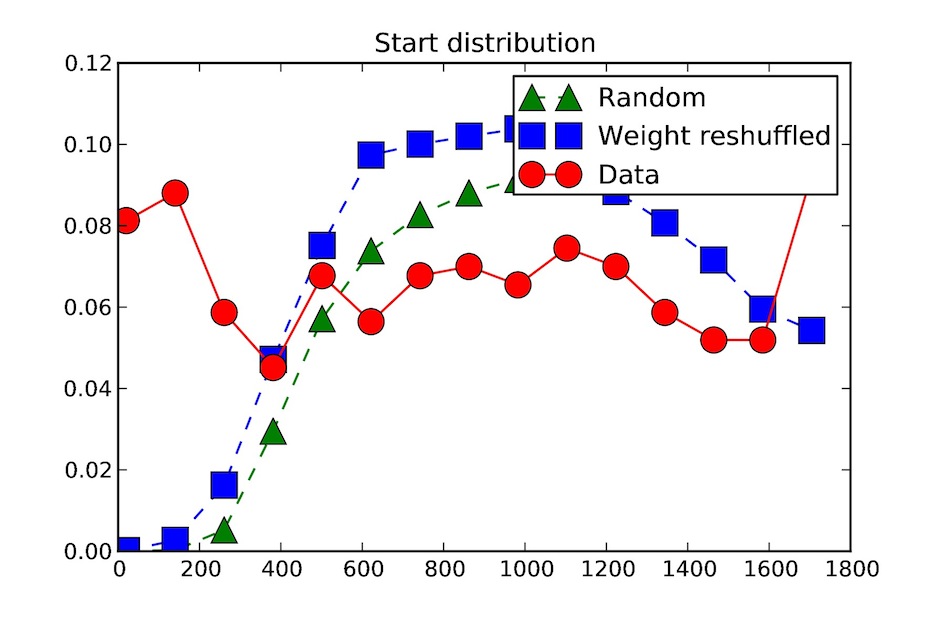}}
\\
\subfigure[]{\includegraphics[width=0.32\textwidth]{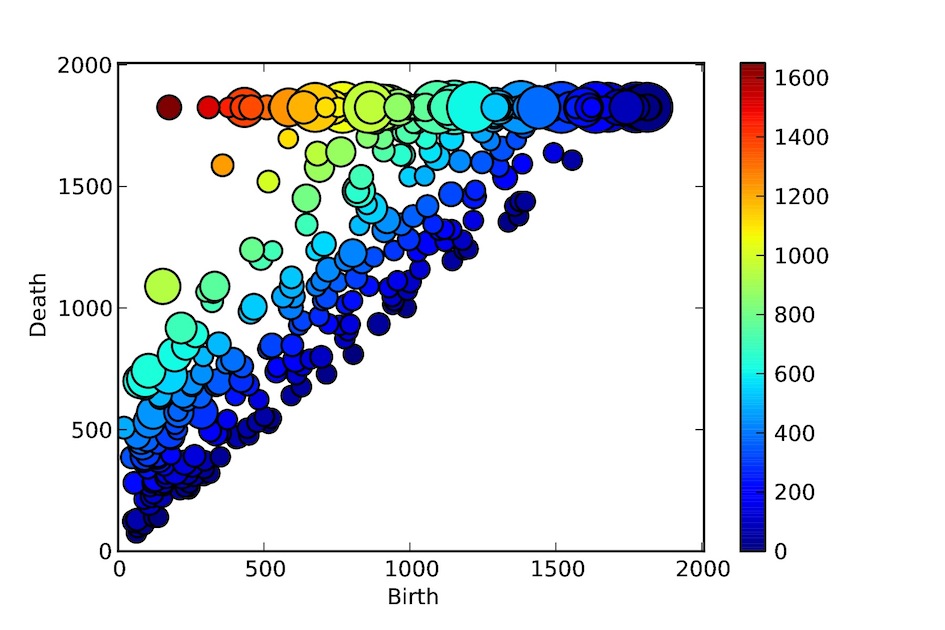}}
\subfigure[]{\includegraphics[width=0.32\textwidth]{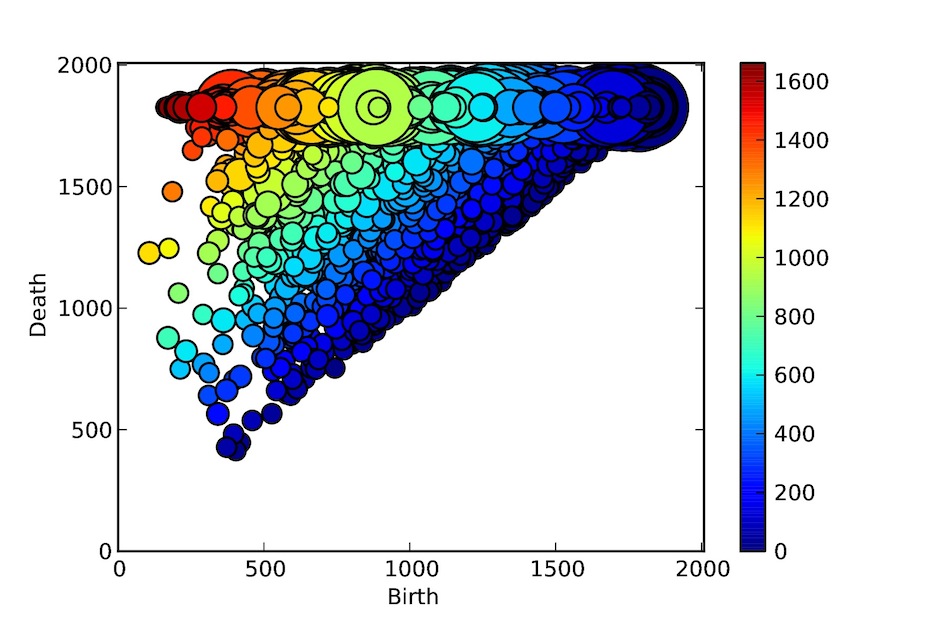}}
\subfigure[]{\includegraphics[width=0.32\textwidth]{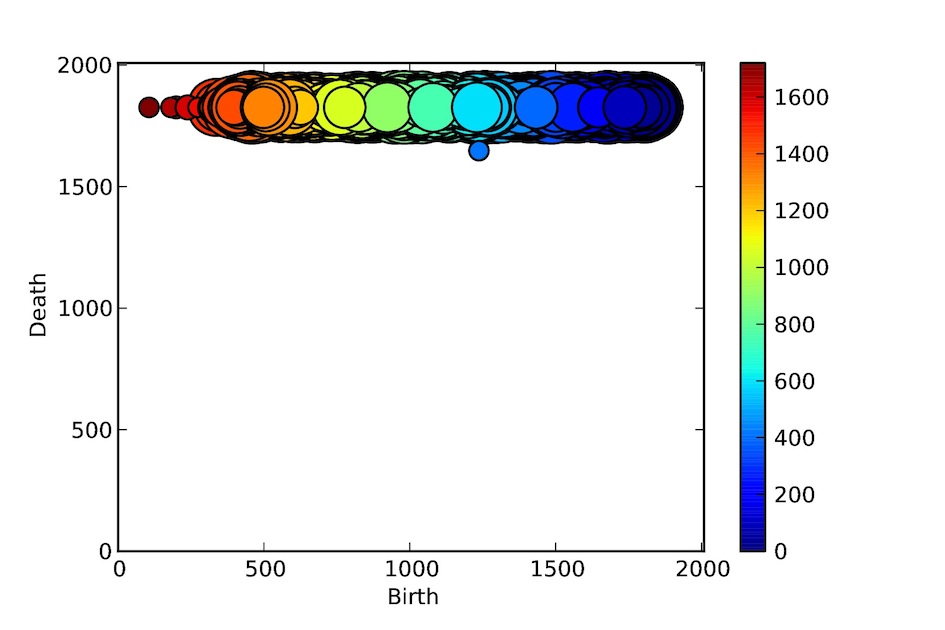}}
\caption{{\bf Summary of $H_1$ persistent homology results for the Random Geometric Graph model with linear weight-degree correlations (Class I).} 
The graph has $N=600$ nodes and a linking distance $d=0.01$. The weight of a link between nodes $i$ and $j$ was set according to $\omega_{ij} \sim (k_i k_j)^\theta X$, where $\theta=1$ and $X$ is a uniform random variable in $(0,1)$.}\label{fig::linearRG_random_comparison}
\end{figure*}

\begin{figure*}
\centering
\subfigure[]{\includegraphics[width=0.32\textwidth]{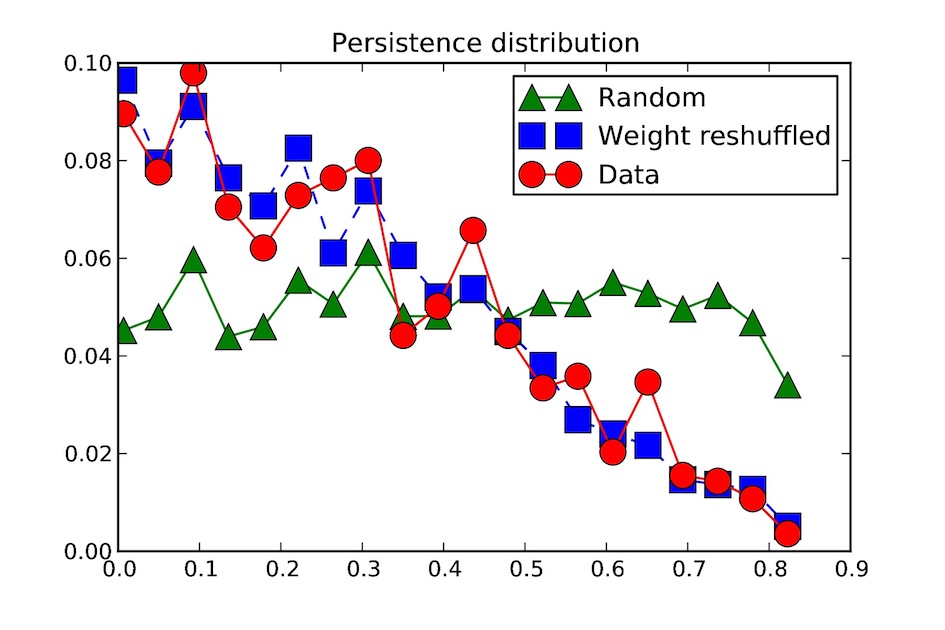}}
\subfigure[]{\includegraphics[width=0.32\textwidth]{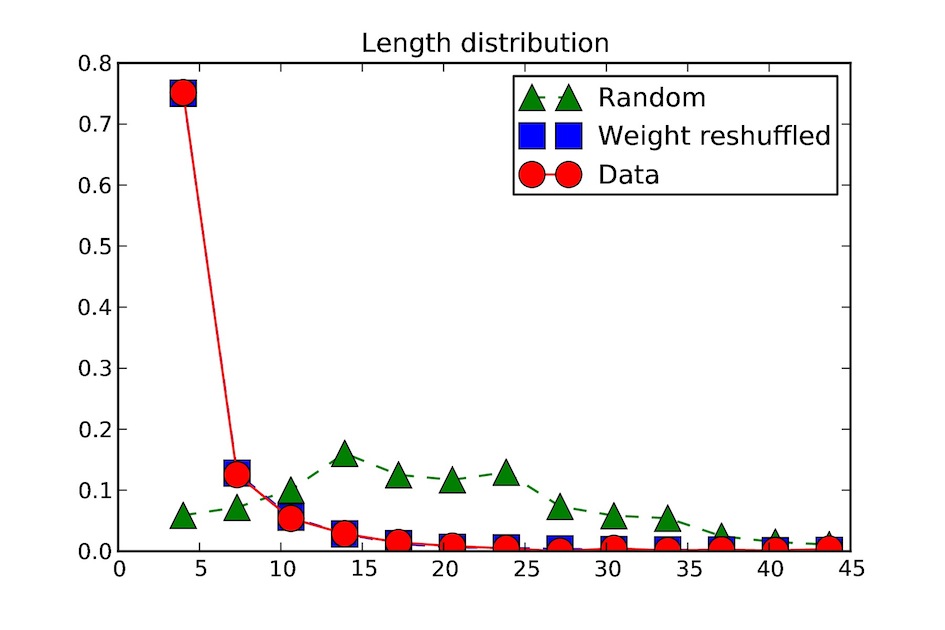}}
\subfigure[]{\includegraphics[width=0.32\textwidth]{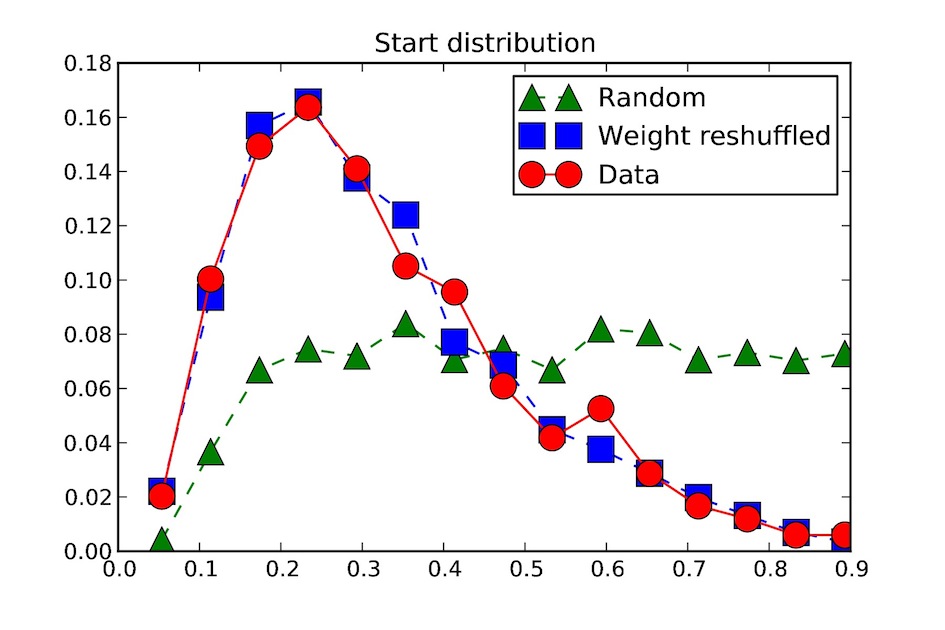}}
\\
\subfigure[]{\includegraphics[width=0.32\textwidth]{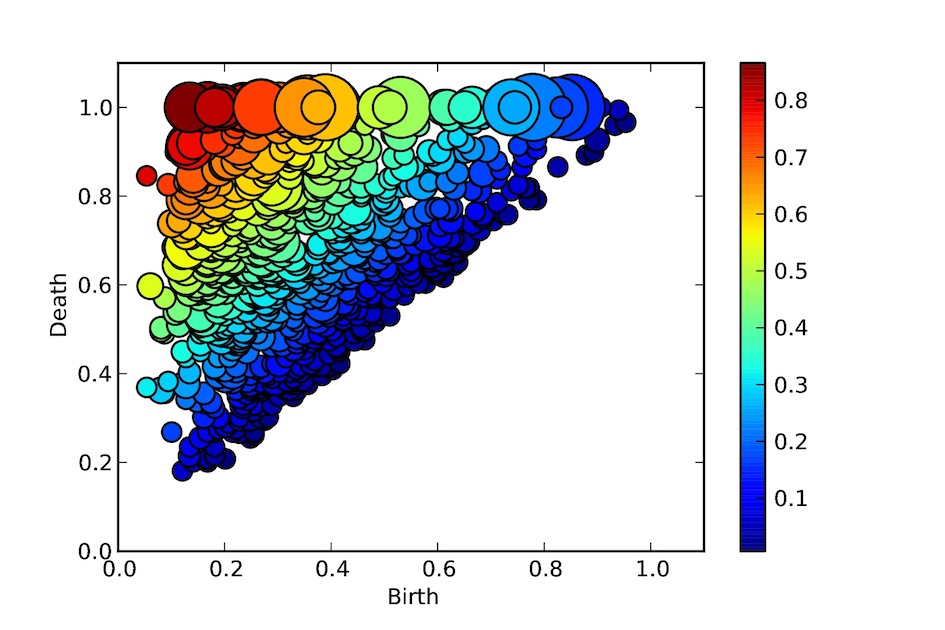}}
\subfigure[]{\includegraphics[width=0.32\textwidth]{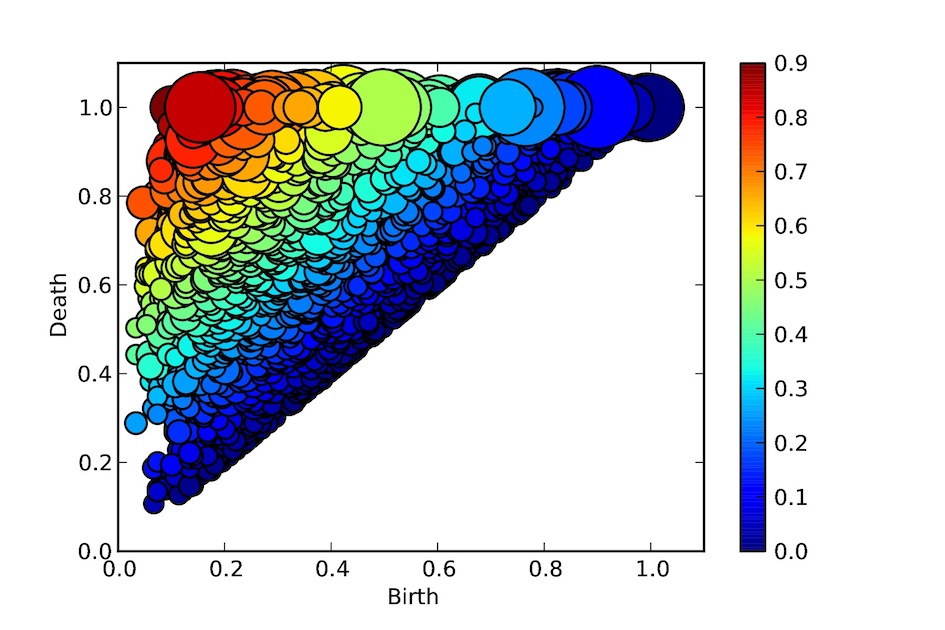}}
\subfigure[]{\includegraphics[width=0.32\textwidth]{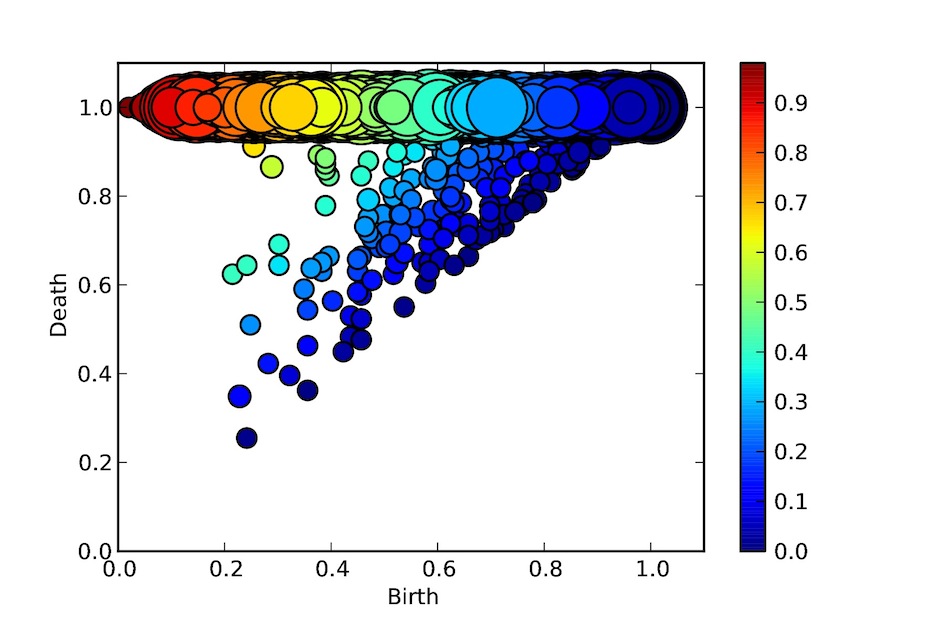}}
\caption{{\bf Summary of $H_1$ persistent homology results for the Random Geometric Graph model with linear weight-degree correlations (Class I).} 
The graph has $N=600$ nodes and a linking distance $d=0.01$. The weight of a link between nodes $i$ and $j$ was set with random uniform weights .}\label{fig::randomRG_random_comparison}
\end{figure*}

We recall that given a network G on $N$ nodes, we consider the weight clique rank filtration on $G$. Let $T$ be the length of the filtration, $\{ g_i \}$  the set of generators of the $i-$th persistence homology module of the filtration and $N_{g_i}$ the cardinality of $\{g_i\}$. For every generator $g_i$, the index $p_{g_i}$ is it's persistence interval, the index $\lambda_{g_i}$ is it's length and $\beta_{g_i}$ is it's birth index. For brevity, $H_1$ generators will be denoted by $g$ rather than $g_1$.\\
There is a conceptual difference in interpreting $H_1$ persistent homology of data with the Rips-Vietoris filtration and $H_1$ persistent homology of weighted networks with the weight rank clique filtration.
While in the first case persistent generators are relevant and considered features of the data, short cycles are more interesting for networks.
This is because random networks, or randomisations of real networks, display one dimensional persistent genrators at all scales, while short lived generators testify the presence of local organisation properties on different scales.\\  
As stated in the main text, the complex networks we considered fall in two main groups.\\
Networks in group I display clear departures from the null counterparts, while class II networks show homological features that are much closer to the randomized versions. 
We collected the complete information about the indices $p_g,\, \lambda_g $ and $\beta_g$ for persistent $H_1$ generators within a series of tableaux (Figures \ref{fig::gene_random_comparison} to \ref{fig::randomRG_random_comparison}). In every figure, panel $a)$ represents the distribution of persistence $p_g$, panel $b)$ the distribution of length $\lambda_g$ and panel $c)$ the distribution of birth index $\beta_g$. 
These quantities are studied for the homology generators in the real world network(red circles), after weight reshuffling of the network(blue squares) and in the network randomisation(green triangles).
Panel $d)$ is the persistence diagram of the network under study, panel $e)$ is the persistence diagram of its weight reshuffled null model and panel $f)$ is the persistence diagram of the random null model.\\
From the perspective of persistence diagrams, class I presents a rich structure of nested cycles covering all scales, as opposed to the weight reshuffled null model and random null model where generators are born uniformly along the filtration and tend to be very persistent, producing largely hollow network instances.  \\
The degree and weight sequences are preserved in the randomisations and therefore cannot account for the differences in the homology. 
Another possibility to explain the different behavior of the two classes could be the presence of degree-degree or weight-degree correlations in class I. 
However, networks in the two classes do not show consistent patterns of assortativity: for example, class I includes the gene network (assortative) and the airport networks (disassortative), while class II  includes the assortative co-authorship networks and the disassortative Twitter data.
Also weight-degree correlations do not appear to be decisive: for example, the RGGs generated with random edge weights did not show significant differences from those generated with edge weights correlated positively to the degrees of the end nodes (see Figs.~\ref{fig::linearRG_random_comparison} and \ref{fig::randomRG_random_comparison}).

\bibliographystyle{plain}

\end{bibunit}
\end{document}